\documentclass[12pt,preprint]{emulateapj}

\slugcomment{AJ}

\shorttitle{ALMA observations of LIRGs using dense gas tracers}
\shortauthors{Imanishi et al.}

\begin{document}


\title{ALMA Observations of Nearby Luminous Infrared Galaxies with 
Various AGN Energetic Contributions Using Dense Gas Tracers} 


\author{Masatoshi Imanishi\altaffilmark{1,2}}
\affil{Subaru Telescope, 650 North A'ohoku Place, Hilo, Hawaii, 96720,
U.S.A.} 
\email{masa.imanishi@nao.ac.jp}

\and

\author{Kouichiro Nakanishi\altaffilmark{1,2}}
\affil{Joint ALMA Observatory, Alonso de C\'{o}rdova 3107, Vitacura 
763-0355, Santiago de Chile}

\altaffiltext{1}{National Astronomical Observatory of Japan, 2-21-1
Osawa, Mitaka, Tokyo 181-8588}
\altaffiltext{2}{Department of Astronomy, School of Science, Graduate
University for Advanced Studies (SOKENDAI), Mitaka, Tokyo 181-8588}

\begin{abstract} 
We present the results of our ALMA Cycle 0 observations, using 
HCN/HCO$^{+}$/HNC J=4--3 lines, of six nearby luminous infrared galaxies 
with various energetic contributions from active galactic nuclei (AGNs) 
estimated from previous infrared spectroscopy. These lines are very
effective for probing the physical properties of high-density molecular
gas around the hidden energy sources in the nuclear regions of these
galaxies. We find that HCN to HCO$^{+}$ J=4--3 flux ratios tend to be
higher in AGN-important galaxies than in starburst-dominated regions,
as was seen at the J=1--0 transition, while there is no clear difference
in the HCN-to-HNC J=4--3 flux ratios among observed sources.   
A galaxy with a starburst-type infrared spectral shape and very 
large molecular line widths shows a high HCN-to-HCO$^{+}$ J=4--3 flux 
ratio, which could be due to turbulence-induced heating. 
We propose that enhanced HCN J=4--3 emission relative to HCO$^{+}$ 
J=4--3 could be used to detect more energetic activity than 
normal starbursts, including deeply buried AGNs, in dusty galaxy 
populations. 
\end{abstract} 
 

\section{Introduction} 
 
Luminous infrared galaxies (LIRGs) and ultraluminous infrared galaxies 
(ULIRGs) emit very strong infrared (8--1000 $\mu$m) radiation with 
luminosity of $L_{\rm IR}$ $>$ 10$^{11}L_{\odot}$ and 
$>$10$^{12}L_{\odot}$, respectively \citep{sam96}. 
The strong infrared emission indicates that (U)LIRGs contain powerful
energy sources hidden behind dust. 
The dust-obscured energy sources of (U)LIRGs may be either nuclear 
fusion reactions inside rapidly formed stars (starburst) and/or
radiative energy generation by an accreting compact supermassive 
black hole (SMBH) with mass of $>$10$^{6}$M$_{\odot}$ (active
galactic nucleus; AGN). 
Understanding the (U)LIRG's hidden energy sources is indispensable to 
clarify the nature of the (U)LIRG population. However, this is not easy 
because the powerful compact nuclear starbursts found in the bulk of 
(U)LIRGs are not clearly distinguishable from very compact AGN activity 
based on imaging observations alone at the distance of many (U)LIRGs 
of interest.
Since (U)LIRGs are the dominant population at $z >$ 1 in terms 
of the cosmic infrared radiation density 
\citep{cap07,got10,mag11,mur11}, 
establishing a reliable method to differentiate the hidden energy 
sources of dusty (U)LIRG populations is useful to 
unravel the history of star formation and SMBH mass growth in the 
dust-obscured side of galaxy formation in the early universe. 
 
If AGNs are surrounded by toroidally distributed (torus-shaped) dusty
medium, the so-called narrow line regions, which are photoionized by AGN 
radiation, should develop at 10--1000 pc along the torus
axis, above the torus scale height \citep{ant93}.  
Since emission from the narrow line regions in AGNs is visible
from all directions, such classical AGNs obscured by torus-shaped dusty
medium are classified optically as Seyfert 2s, and are therefore
distinguishable from starbursts through optical spectroscopic
classification, based on optical emission line flux ratios
\citep{vei87,kew01,kau03}.  
However, (U)LIRGs are major mergers of gas-rich galaxies and have large
amounts of concentrated molecular gas and dust in their nuclei  
\citep{sam96}. 
The putative compact AGNs in (U)LIRG nuclei can be easily 
obscured by dust and gas in virtually all directions in the inner 
part of the surrounding obscuring material. Hence, optical detection of 
AGN signatures becomes very difficult, because the narrow line regions 
can be significantly underdeveloped. 
Understanding the energetic importance of such optically elusive 
{\it buried} AGNs is crucial to clarify the true nature of 
the (U)LIRG population, as well as the SMBH mass growth process during 
gas-rich galaxy mergers \citep{hop06}. 
 
To investigate buried AGNs in dusty (U)LIRG nuclei, it is essential 
to perform observations at wavelengths where dust extinction effects are
small.  
The infrared 2.5--35 $\mu$m band is one such region. Systematic infrared 
2.5--35 $\mu$m spectroscopy of nearby (U)LIRGs has allowed elucidation
of the energetic role of buried AGNs in a quantitative manner  
\citep{ima06a,ima06c,ima07a,ima08,ima10a,ima10b,ima09a,nar08,nar09,nar10,vei09,lee12}.
However, our understanding of distant ($z > 1$) (U)LIRGs is incomplete,
because at $z > 1$, application of this infrared spectroscopy is  
limited to infrared 24 $\mu$m very bright sources only 
\citep{wee06,saj07,das09}, which are strongly biased toward AGNs 
\citep{pap07,don08,lee10}. 
No significant progress is expected until the SPICA satellite, which has
high sensitivity in the infrared 10--200 $\mu$m range, becomes operational  
after 2020 \citep{nak12}. 
 
Starbursts (nuclear fusion) and buried AGNs (mass accretion 
onto SMBHs) have very different energy generation mechanisms. 
First, the radiative energy generation efficiency of a nuclear 
fusion reaction in a starburst is only $\sim$0.7\% of Mc$^{2}$ (where M
is the mass of material used in the nuclear fusion reaction, and c is
the speed of light).  
Thus, the emission surface brightness of a starburst region is modest 
and has both observational \citep{wer76,soi00} and 
theoretical \citep{tho05} upper limits of 
$\sim$10$^{13}$L$_{\odot}$ kpc$^{-2}$. 
An AGN, however, achieves high radiative energy generation 
efficiency (6\%--42\% of Mc$^{2}$, where M is the mass of accreting 
material) \citep{bar70,tho74}, and can produce high luminosity from a 
very compact region. A high emission surface brightness with 
$>$10$^{13}$L$_{\odot}$ kpc$^{-2}$ \citep{soi00} can be achieved in an 
AGN, and a large amount of dust in the close vicinity of the AGN can be
heated to high temperatures (several 100 K). 
Second, while UV is the dominant energetic radiation in 
a starburst, an AGN emits strong X-rays in addition to UV radiation 
\citep{ran03,sha11}. 
Since it is predicted that these AGN-origin high dust temperatures 
and strong X-ray irradiation cause different chemical reactions for 
molecules compared to starbursts \citep{mei05,lin06,har10,har13}, 
different flux ratios of molecular rotational J-transition lines 
could be observed between AGNs and starbursts in the (sub)millimeter 
wavelength range. This difference may be used to distinguish the hidden
energy sources of dusty galaxy populations, because of the negligible
effects of dust extinction at (sub)millimeter wavelengths.  
 
Observational results of possibly different effects and feedback to the
surrounding molecular gas between AGNs and starbursts have been  
reported, based on CO, HCN, HCO$^{+}$, and HNC rotational J-transition 
line observations of nearby bright starburst and Seyfert galaxies 
(modest luminosity AGNs) \citep{koh05,per07,kri08,cos11}. 
For example, it has been argued that AGNs show HCN J=1--0 (rotational 
transition) flux
enhancement, relative to HCO$^{+}$ J=1--0 \citep{koh05,kri08}.  
However, the interpretation of these results is controversial, 
because observational data are sparse for sources for which the
energetic contributions of AGNs are quantitatively well-constrained.  
In nearby Seyfert galaxies, modestly luminous AGNs are surrounded by 
starburst activity in nuclear regions 
\citep{ima02,ima03,rod03,iw04,oi10} and circum-nuclear regions in host 
galaxies \citep{cla00,wat08,woo12}. 
The AGN and starburst energetic contributions can vary markedly 
depending on the aperture size employed. Hence, a comparison among 
various observations at different wavelengths with different aperture 
sizes is not straightforward, which makes the discussion highly uncertain. 
 
Nearby ($z <$ 0.4) ULIRGs are energetically dominated by nuclear compact 
($<$ a few 100 pc in physical scale, or $<$0$\farcs$5 at $z >$ 0.04)
energy sources \citep{soi00,ima11a}, so that various  
observational data at different wavelengths (with different aperture 
sizes) must reflect the properties of nuclear emission, with minimum 
contamination from spatially extended ($>$kpc scale) star formation 
activity in host galaxies. 
The relative energetic contributions from starbursts and AGNs have also 
been estimated {\it quantitatively} and {\it consistently} for many 
nearby ULIRG nuclei through systematic infrared 2.5--35 $\mu$m 
spectroscopy \citep{vei09,ima10a,ima10b,nar10}. 
Nearby ULIRGs are thus excellent laboratories in which to create a 
template of molecular line flux ratio from AGNs and starbursts, and to 
scrutinize AGN effects and feedback on molecular rotational (J)-transition 
line flux ratios in a quantitatively reliable manner. 
 
Since ULIRG nuclear regions are dominated by dense molecular 
gas \citep{gao04}, observations with molecular lines with high 
critical densities (e.g., HCN, HCO$^{+}$, HNC), rather than the widely
used low-J CO lines, are needed to properly probe AGN effects in  
ULIRG nuclei. 
Pre-ALMA interferometric observations using HCN and HCO$^{+}$ J=1--0 
were performed for nearby bright ULIRGs at $z<$ 0.06 to probe 
their nuclear molecular gas properties. It was found that 
ULIRG nuclei classified as AGN-important based on infrared spectra tend
to show higher HCN-to-HCO$^{+}$ J=1--0 flux ratios than
starburst-classified nuclei \citep{ima04,ima06b,in06,ima07b,ima09b},
supporting the suggestion  
that enhanced HCN emission can be used to detect AGNs. 
With the advent of highly sensitive ALMA, this study is in principle 
applicable to fainter sources. 
However, at $z>$ 0.06, these J=1--0 lines will be redshifted beyond the 
frequency (wavelength) coverage of ALMA band 3, and hence will not be
observable with the current specification of ALMA.  
It is particularly important to establish an energy diagnostic method 
using higher J-transition lines at higher frequencies (shorter
wavelengths), which can be applied to distant ULIRGs using ALMA. 
 
For nearby sources, J=4--3 and J=3--2 transition lines of HCN, 
HCO$^{+}$, and HNC were observable in ALMA band 7 (275--373 GHz) and 6 
(211--275 GHz), respectively, during ALMA Cycle 0, while J=2--1 lines
were not. 
Earth's atmospheric background emission is smaller in band 6 than in band 
7, and so observations in band 6 to cover J=3--2 transition lines of 
HCN, HCO$^{+}$, and HNC are easier. 
However, if excitation is thermal at up to J=4--3, the flux of J=4--3 
is higher by a factor of 16/9 than that of J=3--2, largely compensating 
for the higher atmospheric background noise of Earth at J=4--3 in band 7. 
Given reasonable assumptions about the observing conditions, the ALMA
sensitivity calculator showed that the required on source exposure times
for the same detection significance (same S/N ratio) were similar
between the J=3--2 and J=4--3 lines of HCN, HCO$^{+}$, and HNC, if the
excitation is thermal at up to J=4--3. 
Given that the J=4--3 line energy diagnostic is applicable to higher 
redshift sources than that of J=3--2 using ALMA, we selected J=4--3 line 
as our initial choice, with the caution that the J=3--2 line could be
better for detection if the excitation is significantly sub-thermal. 
 
We performed HCN J=4--3 (rest-frame frequency $\nu_{\rm rest}$ = 
354.505 GHz), 
HCO$^{+}$ J=4--3 ($\nu_{\rm rest}$ = 356.734 GHz), and HNC J=4--3 
($\nu_{\rm rest}$ = 362.630 GHz) observations 
of nearby ULIRGs with well-calibrated energy sources. 
Since HCN, HCO$^{+}$, and HNC have similar dipole moments 
($\mu$ = 3.0, 3.9, and 3.1 Debye, respectively) and similar 
frequencies for individual J transitions, it is likely that 
similar dense gas phases are probed by these molecular lines, 
making the interpretation straightforward. 
Throughout this paper, we adopt H$_{0}$ $=$ 71 km s$^{-1}$ Mpc$^{-1}$, 
$\Omega_{\rm M}$ = 0.27, and $\Omega_{\rm \Lambda}$ = 0.73 
\citep{kom09}. 
 
\section{Targets} 
 
Our primary interest is to determine whether AGN-important galaxies
display noticeably different molecular line flux ratios from
starburst-dominated galaxies. Therefore, we targeted galaxies for which
the energetic contributions from AGNs are reasonably well estimated
through infrared spectroscopy and/or other observations. 
Since we expected that the ALMA molecular gas observations of many 
nearby well-studied starburst galaxies and Galactic active star-forming 
regions would be performed with other programs, we mainly selected 
AGN-dominated ULIRGs as the first priority to probe the molecular line 
flux ratios from almost pure AGNs. 

AGN and starburst activity generally coexist in galaxies, so it is not
easy to find such pure AGNs.    
The SMBH and spheroidal stellar mass correlation 
\citep{mag98,fer00,gul09,mcc13} means that the SMBH to stellar mass 
ratio does not change markedly among different galaxies. 
However, the {\it luminosity} ratio between AGNs and starbursts can
change depending on the mass accretion rate onto SMBHs.  
Galaxies containing highly mass-accreting SMBHs have relatively high AGN 
contributions to the total energy budgets, and hence are good targets to 
investigate the effects of AGNs on molecular line flux ratios. 
The energetic importance of AGNs increases with 
the galaxy infrared luminosity. Buried AGN-dominated sources with no
detectable starburst activity were  
found in ULIRGs with L$_{\rm IR}$ $>$ 10$^{12}$L$_{\odot}$,
\citep{vei09,ima09a,ima10a,nar10,ima10b}.  
These sources were excellent targets for our investigation of AGN
effects on molecular line emission.  
While many PG QSOs \citep{sch83} (higher luminosity type-1 unobscured 
AGN populations than Seyfert galaxies), are present at similar distances 
to nearby ULIRGs ($z$=0.05--0.4), and their luminosities largely come from 
AGNs \citep{vei09}, starbursts are still often detectable in their
nuclear regions (with similar nuclear starburst to AGN luminosity
ratios to Seyfert galaxies; Imanishi et al. 2011b), as well as in
spatially extended host galaxies \citep{sch06,net07,shi07,vei09,wat09}.  
Although some PG QSOs do not show detectable starburst activity, we 
chose buried AGN-dominated ULIRGs for the following reasons. 
First, for the same AGN luminosity, we can expect higher molecular 
emission line fluxes in buried AGNs surrounded by a large amount of 
molecular gas and dust than in type-1 unobscured AGNs with a smaller 
dust/gas covering factor. 
Hence, higher S/N ratios can be obtained and more distant sources can be 
observed for buried AGNs than for type-1 unobscured AGNs. 
Second, the primary energy sources of type-1 unobscured AGNs can be 
investigated through optical spectroscopic classification 
\citep{vei87,kew01,kau03}. There is 
no strong need to observe such unobscured AGNs in the (sub)millimeter 
wavelength range just for the purposes of energy diagnostics. 
 
Among buried AGN-dominated ULIRGs with no detectable starburst activity, 
we first picked up those with the highest expected HCN fluxes. This
allowed us to obtain the highest quality spectra with the limited
capabilities of ALMA Cycle 0. Given the correlation between infrared and
HCN J=1--0 line luminosity in galaxies \citep{gao04}, higher HCN J=1--0
emission flux is expected in a galaxy with higher infrared flux.  
IRAS 08572+3915, 12127$-$1412, and 00183$-$7111 were selected as our 
buried AGN-dominated ULIRG sample. 
 
IRAS 08572+3915 ($z =$ 0.058) is a ULIRG (L$_{\rm IR}$ $\sim$ 
10$^{12.1}$L$_{\odot}$) (Table 1) classified optically as a LINER 
\citep{vei99} or a Seyfert 2 in some combination of optical emission 
lines \citep{yua10}. It consists of two merging nuclei with $\sim$5$''$
separation (NW and SE) \citep{sco00,kim02,ima14}.  
The NW nucleus (IRAS 08572+3915 NW) is one of the strongest buried 
AGN candidates in the local universe, because (1) the energy source is 
suggested to be very compact and is more centrally concentrated than the 
nuclear dust, as is expected for a buried AGN 
\citep{dud97,soi00,ima06a,ima07a}, and (2) the infrared polycyclic 
aromatic hydrocarbons (PAH) emission (a good starburst indicator) is 
extremely weak \citep{imd00,ima06a,spo06,arm07,ima07a,vei09,nar10}. 
The detection of many high-J transition CO absorption lines in 
the infrared spectrum is also suggestive of the presence of a luminous 
AGN \citep{shi13}. 
 
IRAS 12127$-$1412 ($z =$ 0.133) is our second buried AGN-dominated ULIRG 
sample (L$_{\rm IR}$ $\sim$ 10$^{12.1}$L$_{\odot}$) (Table 1). 
It consists of two main nuclei (NE and SW) with a separation of 
$\sim$10$''$ \citep{kim02}, and is classified optically as a non-Seyfert 
(LINER or HII-region) \citep{vei99,yua10}. 
The NE nucleus is brighter in the infrared \citep{kim02,ima14}, and is 
diagnosed as buried AGN-dominated based on the suggested 
compact energy source, which is more centrally concentrated than the 
nuclear dust, and (2) non-detectable starburst-originated PAH emission 
features in the infrared \citep{ima06a,ima07a,vei09,ima10a,nar10}. 
 
IRAS 00183$-$7111 ($z =$ 0.327), which is classified optically as a LINER 
\citep{arm89}, is a third almost pure buried AGN candidate, based on its 
buried AGN type infrared spectral shape \citep{spo04,spo09,ima10a}. 
X-ray observations also support the AGN-dominant scenario for this 
ULIRG \citep{nan07}. Although its distance is greater than the former
two buried AGN candidates, the high infrared luminosity (L$_{\rm IR}$
$\sim$ 10$^{12.9}$L$_{\odot}$) keeps the infrared {\it flux} relatively
high, making this a good target for our study.  
 
As a comparison of molecular emission line flux ratios for these buried 
AGN-dominated ULIRGs, we selected a starburst-dominated galaxy. 
Since spatially extended emission with $>$6$''$ could be missed by ALMA 
Cycle 0 observations, we chose the nuclear compact starburst-dominated 
galaxy NGC 1614. NGC 1614 has infrared luminosity with only L$_{\rm IR}$
$\sim$ 10$^{11.6}$L$_{\odot}$ (Table 1), and so is not classified as an
ULIRG. Our ALMA Cycle 0 observational results of NGC 1614 were published
previously by \citet{ima13a}, and detailed descriptions about the
properties of NGC 1614 are found in that paper.  
 
Since starbursts in ULIRGs could be more powerful per unit 
volume than those in LIRGs, molecular gas excitation could differ, 
and so molecular line flux ratios from starburst galaxies could 
vary, depending on the infrared luminosity. Therefore, we also included an 
ULIRG with a starburst-type infrared spectral shape. 
Among many such ULIRGs \citep{ima07a,vei09,ima10a}, to achieve 
the highest S/N ratios with a given ALMA Cycle 0 observing time, we 
searched for sources with high infrared fluxes, at declinations between 
$-$40$^{\circ}$ and 0$^{\circ}$, which were best observable from the
ALMA site in Chile.  
We picked up IRAS 22491$-$1808 ($z$=0.076; L$_{\rm IR}$ $\sim$ 
10$^{12.1}$L$_{\odot}$) as our ALMA Cycle 0 starburst ULIRG sample. 
IRAS 22491$-$1808 has an infrared spectrum dominated by large equivalent 
width PAH emission features, suggesting that the {\it observed} infrared 
flux is dominated by starburst activity \citep{ima07a,vei09,ima10a}. 
However, the starburst-originated PAH emission to infrared luminosity 
ratios are a factor of 3--8 smaller than the typical ratios for 
less-infrared-luminous starburst-dominated galaxies for the 3.3 $\mu$m, 
6.2 $\mu$m, and 11.3 $\mu$m features \citep{ima07a,ima10a}. This 
leaves room for an additional energy source beyond the modestly obscured
detected starburst activity (i.e., a deeply buried AGN that does not
show PAH emission and/or an extremely highly obscured starburst whose PAH
emission is highly flux attenuated). This uncertainty is
unavoidable for the majority of starburst-classified ULIRGs based on
infrared spectra, which show deficiency in PAH to infrared luminosity
ratios. Based on the infrared spectrum of IRAS 22491$-$1808 at $\lambda$
$>$ 5 $\mu$m, \citet{vei09} classified this ULIRG as
starburst-dominated, with the energetic contribution from a buried AGN
of $\sim$18\%, but \citet{nar10} failed to determine the relative energetic
contribution between the starburst and the AGN.  
 
In addition to these five sources, we added a buried AGN 
{\it significant} galaxy, where both AGN and starburst energetically 
contribute to roughly half of the total luminosity. We did this to
determine whether  
such a source shows molecular line flux ratios between AGN-dominated and 
starburst-dominated galaxies. For this class, we chose IRAS 20551$-$4250
($z$=0.043, L$_{\rm IR}$ $\sim$ 10$^{12.0}$L$_{\odot}$). We refer readers
to \citet{ima13b} for the details of our ALMA Cycle 0  
results of IRAS 20551$-$4250. 
 
In total, six sources with various AGN energetic contributions were
observed. Table 1 presents a summary of their basic properties.  
 
\section{Observations and Data Analysis} 
 
All observations were made during ALMA Cycle 0 within the program 
2011.0.00020.S (PI = M. Imanishi). Observation details of NGC 1614 and
IRAS 20551$-$4250 are presented in \citet{ima13a,ima13b}, and are not
repeated here. Table 2 describes the observation details for the
remaining four ULIRGs.  
HCN J=4--3 ($\nu_{\rm rest}$ = 354.505 GHz) and HCO$^{+}$ J=4--3 
($\nu_{\rm rest}$ = 356.734 GHz) lines were simultaneously observable 
with ALMA, but we needed separate observations to cover the HNC 
J=4--3 line ($\nu_{\rm rest}$ = 362.630 GHz). 
Band 7 was used for all sources except IRAS 00183$-$7111 ($z=$ 0.327),
for which band 6 was used due to its high redshift. We adopted the
widest 1.875 GHz band width mode. In ALMA Cycle 0, four different
frequencies can be configured with some restrictions.  
HCN and HCO$^{+}$ J=4--3 lines were covered by two separate correlator 
spectral windows. 
The other two spectral windows were also used to measure the continuum 
flux levels. 
For HNC J=4--3 observations, the HNC line was covered with one 
spectral window, and an additional spectral window was used to probe 
the continuum emission. 
The central frequencies used in each correlator spectral window are 
summarized in Table 3. 
According to the observing log of ALMA Cycle 0, observations of IRAS 
08572$+$3915 at the HNC J=4--3 line were performed, but the data were 
not provided to us because they did not pass the most basic 
quality assessment checks. 
 
We started data analysis from calibrated data provided by the Joint ALMA 
Observatory, using CASA, except the IRAS 22491$-$1808 HNC data, 
for which the phase center position of the phase calibrator
(J2258$-$279) on the UV data delivered to us was
incorrect and was slightly shifted by $\sim$1$\farcs$1.
We corrected for the phase center by using the CASA task
'fixvis' and re-calibrated the IRAS 22491$-$1808 HNC data. 
We first checked the visibility plots to determine whether the signatures of 
targeted emission lines were recognizable. 
For most of the detected emission lines, the presence of the lines 
was evident in the visibility plot. 
We then selected channels that were free from strong emission lines to 
estimate the continuum flux level. 
As will be discussed in the following section, in addition to the targeted 
HCN, HCO$^{+}$, and HNC J=4--3 emission lines, clear signatures of other 
emission lines were detected in a few sources in the spectral windows 
originally set to measure the continuum emission. 
Those channels that were affected by strong emission lines were excluded
from the continuum extraction.  
We then subtracted the estimated continuum emission from the spectra, 
and performed the task ``clean'' for continuum-subtracted molecular line 
data, as well as for the continuum data themselves. 
To create spectra, we binned 40 channels (17--18 km s$^{-1}$ in band 7 
and 22--23 km s$^{-1}$ in band 6). 
For spatial binning, we employed 0$\farcs$3 pixel$^{-1}$ or 0$\farcs$1 
pixel$^{-1}$, depending on the beam size of the obtained data. 
The resulting beam sizes of continuum and individual molecular line
data of the observed sources are  shown in Tables 4 and 5, respectively. 
The energetically-dominant nuclear regions ($<$a few 100 pc) 
of the observed (U)LIRGs are fully covered with our ALMA Cycle 0 data.
 
\section{Results} 
 
Our ALMA Cycle 0 observational results of NGC 1614 and IRAS 20551$-$4250 
have been published in \citet{ima13a} and \citet{ima13b}, respectively. 
Here, we present the results of the remaining four ULIRGs. 
Figure 1 shows the continuum-a (taken with HCN and HCO$^{+}$ J=4--3 
observations) and continuum-b (taken with HNC J=4--3) maps. 
The continuum peak fluxes and peak coordinates are summarized in Table
4.  
The peak coordinates generally agree between continuum-a and -b.
 
The spectra at the continuum peak positions within the beam size are 
shown in Figure 2. 
Although we targeted HCN, HCO$^{+}$, and HNC J=4--3 emission 
line observations, the signatures of other emission lines were seen at the 
rest frame frequency of $\nu_{\rm rest}$ $\sim$ 369.1 GHz and 
$\sim$349.4 GHz, which we ascribed to H$_{2}$S 3(2,1)--3(1,2) 
($\nu_{\rm rest}$ $\sim$ 369.101 GHz) and CH$_{3}$CN v=0 19(3)--18(3) 
($\nu_{\rm rest}$ $\sim$ 349.393 GHz) with possible contamination from 
CCH v=0 N=4--3 J=7/2--5/2 ($\nu_{\rm rest}$ $\sim$ 349.400 GHz), 
respectively. 
For IRAS 12127$-$1412, the signature of the CS J = 7--6 line 
($\nu_{\rm rest}$ = 342.883 GHz) may be present. 

Figures 3, 4, 5, and 6 are integrated intensity (moment 0) maps and 
spectra at the continuum peak, within the beam size, of 
possibly detected emission lines for IRAS 08572$+$3915, IRAS 
12127$-$1412, IRAS 00183$-$7111, and IRAS 22491$-$1808, respectively. 
In the spectra (right), the continuum level was found to be
well-subtracted, so that the moment 0 maps (left) should reflect the
properties of molecular emission lines rather than the residual of
continuum subtraction.  
The peak flux values and detection significance in the moment 0 maps are 
summarized in Table 5. 
The peak coordinates of the molecular emission lines agree within 1 
pixel with those of the simultaneously observed continuum-a (for 
HCN, HCO$^{+}$, and H$_{2}$S) and continuum-b (for HNC and CH$_{3}$CN), 
with the exception of the HNC line of IRAS 00183$-$7111, the emission 
peak of which is shifted from the continuum-b peak by 
two pixels (0$\farcs$6) to the west. 
Given the beam size of 2$\farcs$2 $\times$ 1$\farcs$2 
(PA = 173$^{\circ}$; almost north-south direction) and the detection 
significance of 3.6$\sigma$ for the HNC moment 0 map of IRAS 
00183$-$7111, the positional uncertainty of the HNC emission is on the
order of (beam-size)/(signal-to-noise ratio) $\sim$0$\farcs$33
in the east-west direction. 
The east-west offset could be larger than the statistical uncertainty,
and the marginal (3.6$\sigma$) detection of the HNC emission in IRAS  
00183$-$7111 requires further confirmation.  
 
To estimate the fluxes of individual molecular emission lines detected 
in the spectra, the lines were fitted with a Gaussian profile. 
The results of the Gaussian fits for individual molecular lines are 
summarized in Table 5. 
The detection significance of the possible CS J=7--6 line in IRAS 
12127$-$1412 is $<$3$\sigma$ both in the moment 0 map 
(peak flux value relative to the rms noise level derived from the
standard deviation of sky signals) and the Gaussian fit 
in the spectrum at the emission peak, within the beam size (Table 5). 
The molecular line luminosities are derived from the fluxes based on 
Gaussian fits, and are summarized in Table 6, where we adopted equations
(1) and (3) from \citet{sol05}.  
 
Emission lines are generally detected at the expected frequency from 
optically derived redshifts (Table 1 and downward arrows in Figure 2). 
However, for IRAS 22491$-$1808, the detected molecular lines are 
appreciably offset from the expected frequency of $z=$ 0.076 
\citep{soi87,kim98}. 
Our molecular line data of dense gas tracers indicate $z =$ 0.0776, 
$\sim$500 km s$^{-1}$ redshifted. 
Our molecular line redshift is comparable to the optical redshift 
of $z =$ 0.078 derived by \citet{str92}.
 
For the bright HCN and HCO$^{+}$ J=4--3 emission lines of IRAS 08572$+$3915 
and 22491$-$1808, the intensity-weighted mean velocity (moment 1) and 
intensity-weighted velocity dispersion (moment 2) maps are shown in 
Figures 7 and 8, respectively. 
For IRAS 08572$+$3915, no signature of rotating motion was seen, 
suggesting that the dense molecular gas is very compact and 
not clearly spatially resolved. 
For IRAS 22491$-$1808, a rotation sign is barely seen such that the 
north-eastern (south-western) molecular gas is redshifted (blueshifted) 
with respect to the galactic nucleus. This suggests that the dense
molecular gas is just spatially resolved.  
A noticeable feature of IRAS 22491$-$1808 is that HCN and HCO$^{+}$ 
J=4--3 emission show considerably larger velocity dispersion 
($>$150 km s$^{-1}$) in the bulk of the signal detected region than 
IRAS 08572$+$3915 ($\lesssim$100 km s$^{-1}$) in the moment 2 maps. 
 
IRAS 08572$+$3915 and 12127$-$1412 have distinct fainter secondary 
nuclei, SE and SW, with apparent separation of $\sim$5$''$ and 
$\sim$10$''$ from the brighter nuclei, respectively \citep{kim02,ima14}. 
For IRAS 08572+3915 SE, we detected no significant ($>$3$\sigma$)
emission of any of the observed molecular lines (HCN, HCO$^{+}$, and
H$_{2}$S) in the moment 0 maps within $<$1'' of the near-infrared
$K$-band (2.2 $\mu$m) peak coordinate at (09 00 25.67, $+$39 03 50.2) in
J2000 \citep{kim02}.  
For 12127$-$1412 SW, significant ($>$3$\sigma$) emission was not seen 
in the moment 0 maps of HCN, HCO$^{+}$, HNC, CH$_{3}$CN, and CS, 
within $<$1'' of the near-infrared $K$-band peak coordinate at 
(12 15 18.64, $-$14 29 48.8) in J2000 \citep{kim02}. 
The upper limits are $<$3$\sigma$ of the rms noise of individual moment 
0 maps, shown in Table 5, column 5. 
We also searched for continuum emission within $<$1$''$ of the
near-infrared $K$-band peak positions of IRAS 08572$+$3915 SE and
12127$-$1412 SW, but no signs were found. The upper limits are
$<$3$\sigma$ of the rms noise, as shown in Table 4, column 5.  
 
The vibrationally excited (v$_{2}$=1f) HCN J=4--3 emission 
line ($\nu_{\rm rest}$ = 356.256 GHz) was detected in two external 
galaxies, NGC 4418 \citep{sak10} and IRAS 20551$-$4250 \citep{ima13b}, 
both of which are thought to contain luminous buried AGNs. 
The expected frequency of this emission line is shown in Figure 9 for 
the four ULIRGs, but the line is not clearly recognizable, probably 
because (1) the flux of the HCN J=4--3 (v$_{2}$=1f) emission line itself is 
not high, and (2) the line width of the nearby, much stronger HCO$^{+}$ 
J=4--3 emission at the vibrational ground level (v=0) is not as small 
as in NGC 4418 and IRAS 20551$-$4250. 
 
\section{Discussion} 
 
\subsection{Molecular Line Flux Ratio} 
 
Figure 10 shows a plot of the HCN-to-HCO$^{+}$ J=4--3 and HCN-to-HNC J=4--3 
flux ratios in the four ULIRGs, together with the starburst-dominated 
LIRG, NGC 1614 \citep{ima13a}, and the AGN-starburst 
composite ULIRG, IRAS 20551$-$4250 \citep{ima13b}. 
The buried AGN-dominated ULIRGs, IRAS 08572$+$3915, 12127$-$1412, 
and 00183$-$7111, show higher HCN-to-HCO$^{+}$ flux ratios at J=4--3 than 
the starburst-dominated LIRG, NGC 1614. 
The AGN and starburst composite ULIRG, IRAS 20551$-$4250, is located 
between the buried AGN-dominated ULIRGs and a starburst-dominated LIRG, 
NGC 1614, on the ordinate, as expected ($\S$2). 
The starburst-classified ULIRG in the infrared spectrum, IRAS 
22491$-$1808, shows a high HCN-to-HCO$^{+}$ J=4--3 flux ratio, which 
will be discussed in the next subsection. 
 
On the other hand, the HCN-to-HNC J=4--3 flux ratios in the abscissa are 
not clearly different between AGN-dominated and starburst-dominated 
sources. A high HCN-to-HCO$^{+}$ J=4--3 flux ratio could be an indicator
of a luminous buried AGN, and could be used to distinguish AGNs from
starbursts \citep{ima10c,ion13}.  
 
\subsection{Interpretation} 
 
We consider three possible mechanisms for the strong HCN J=4--3 emission 
in AGNs, relative to HCO$^{+}$ J=4--3. First, the HCN abundance
enhancement is a natural explanation for the enhanced HCN emission. 
In an AGN, due to the high radiative energy generation efficiency of a
mass-accreting SMBH, a larger amount of dust is heated to high
temperatures (several 100 K) than in a starburst ($\S$1).  
It is predicted that the HCN abundance could be enhanced, relative to 
HCO$^{+}$ in high dust temperature chemistry \citep{har10}. 
In an AGN, the HCN abundance enhancement by strong X-ray radiation 
is also calculated by \citet{lin06}, but the HCN-to-HCO$^{+}$ abundance
ratio under X-ray irradiation is also shown to be highly
model-dependent \citep{mei05,har13}.  
 
Second, vibrationally excited HCN emission lines were detected in
AGN-hosting galaxies \citep{sak10,ima13b}.  
Infrared radiative pumping \citep{aal95} by absorbing infrared $\sim$14 
$\mu$m photons is the most natural mechanism for vibrational excitation,
because vibrationally excited levels of HCN with $>$1000  
K are very difficult to achieve with collisions \citep{sak10}. 
Since an AGN emits infrared 14 $\mu$m continuum emission much more
strongly due to AGN-heated hot dust than a starburst, this infrared
radiative pumping is expected to work more effectively in an AGN.  
This mechanism can increase the HCN J=4--3 flux at v=0 through a cascade 
process \citep{ran11} in comparison to collisional excitation alone. 
On the other hand, detection of the vibrationally excited HCO$^{+}$
emission line has not been reported to date in any AGNs.
In fact, the HCN v$_{2}$=1--0 absorption features at infrared 14 $\mu$m
were detected in several obscured AGNs, while the HCO$^{+}$ v$_{2}$=1--0
absorption features at infrared 12.1 $\mu$m were not \citep{lah07,vei09}.
This may indicate that the necessary condition for the infrared
radiative pumping is fulfilled more effectively for HCN than HCO$^{+}$.
Thus, this infrared radiative pumping may be the reason for the enhanced
HCN J=4--3 fluxes relative to HCO$^{+}$ J=4--3 observed in AGNs.

Third, the higher HCN-to-HCO$^{+}$ J=4--3 flux ratios in AGNs compared to 
starbursts could be simply due to an excitation effect. As the critical
density of HCN J=4--3 (n$_{\rm crit}$ $\sim$ 2 $\times$ 10$^{7}$
cm$^{-3}$) is higher than that of HCO$^{+}$ J=4--3 (n$_{\rm crit}$
$\sim$ 4 $\times$ 10$^{6}$ cm$^{-3}$) \citep{mei07}, HCO$^{+}$ J=4--3 is
more easily excited than HCN J=4--3. The HCN-to-HCO$^{+}$ flux ratio at
J=4--3 should be lower than that at J=1--0.  
AGN-heated hot dust could create large amounts of hot molecular gas, 
and more HCN J=4--3 excitation is expected in an AGN than a starburst. 
This can result in a higher HCN-to-HCO$^{+}$ J=4--3 flux ratio in an AGN 
than in a starburst, even without invoking HCN abundance enhancement 
and/or the flux enhancement of the HCN J-transitions at v=0 through the 
infrared radiative pumping mechanism. 
 
HCN and HNC have comparable critical densities \citep{mei07}, so their 
excitation is expected to be similar, as long as the same molecular gas 
phase is probed. 
The observational result that AGNs tend to show higher
HCN-to-HCO$^{+}$ J=4--3 flux ratios than starbursts but similar
HCN-to-HNC J=4--3 flux ratios suggests that excitation is an important
factor.  
That is, the larger amount of hot molecular gas in an AGN may boost 
HCN J=4--3 excitation, relative to HCO$^{+}$ J=4--3, increasing 
the HCN-to-HCO$^{+}$ J=4--3 flux ratios compared to a starburst, but
also increase the excitation of HCN and HNC up to J=4--3 in a similar
manner, without changing the HCN-to-HNC J=4--3 flux ratio.  
 
It is also possible that the HCN abundance is enhanced relative to 
HCO$^{+}$ in AGNs due to AGN-related phenomena (high dust temperature 
and/or strong X-ray irradiation), but that the HNC abundance is similarly 
enhanced. 
Or, it may be that the HNC J=4--3 flux at v=0 is also enhanced through
vibrational excitation by infrared radiative pumping \citep{cos13}, in a
similar way to HCN, but the HCO$^{+}$ J=4--3 flux at v=0 is not.  
 
IRAS 22491$-$1808 shows an infrared spectrum whose observed flux can be  
explained solely by starburst activity, and yet shows a high 
HCN-to-HCO$^{+}$ J=4--3 flux ratio. As mentioned in $\S$2, the presence
of a buried AGN in IRAS 22491$-$1808 is not precluded, due to the
smaller  
observed PAH-to-infrared luminosity ratio compared to the known 
starburst-dominated, less infrared-luminous galaxies. 
However, the direct signature of such an AGN is also lacking. 
A notable feature of IRAS 22491$-$1808 is the very large molecular line
widths, with $>$400--500 km s$^{-1}$ in full width at half maximum
(FWHM)  
or $>$700 km s$^{-1}$ in full width at zero intensity (FWZI) for HCN and 
HCO$^{+}$ J=4--3 (Figure 6). 
These values are much larger than those for normal ULIRGs whose 
HCN and HCO$^{+}$ line widths are typically $\sim$200--300 km s$^{-1}$
in FWHM \citep{gao04,in06,ima07b}, and are as high as the values of the
merger-induced shock-dominated highly turbulent luminous infrared galaxy
NGC 6240 \citep{nak05}. The intensity-weighted velocity dispersion
(moment 2) maps of HCN and HCO$^{+}$ J=4--3 of IRAS 22491$-$1808 in
Figure 8 also show much larger peak values in the bulk of galactic areas
than those of IRAS 08572$+$3915 (Figure 7), NGC 1614 \citep{ima13a}, and
IRAS 20551$-$4250 \citep{ima13b}. This suggests that 
the high-density molecular gas in IRAS 22491$-$1808 is particularly
turbulent compared to the majority of ordinary ULIRGs.  
A turbulent heating mechanism \citep{pan09} may produce large amounts 
of hot dust and molecular gas, and could (1) boost HCN J=4--3 excitation 
compared to normal starburst galaxies, (2) enhance HCN abundance under
high dust temperature chemistry (see also Papadopoulos 2007), and 
(3) increase the effect of the  
infrared radiative pumping mechanism due to an increase in infrared 14 
$\mu$m photons. 
%Adopting the optical redshift of $z$=0.076 \citep{kim98}, the
%significantly ($\sim$500 km s$^{-1}$) redshifted dense molecular gas  
%emission in IRAS 22491$-$1808 (Figure 2) may be due to outflow or inflow 
%in the direction away from the Earth, and may be related to the high
%observed turbulence levels.  
 
If sufficient amounts of hot dust and molecular gas are created by a 
mechanism other than a luminous AGN, and if the enhanced 
HCN-to-HCO$^{+}$ J=4--3 flux ratios are due to phenomena related 
to hot dust and molecular gas (i.e., increased HCN J=4--3 excitation,
HCN abundance increase by hot dust chemistry, and increased efficiency
of infrared radiative pumping), then the high HCN-to-HCO$^{+}$ flux
ratios could pick up not only luminous buried AGNs, but also starbursts
with highly turbulent dense molecular gas. Although we obtained the
molecular line data of the starburst-dominated LIRG NGC 1614
\citep{ima13a}, data on ordinary starburst-dominated ULIRGs that show 
usual dense molecular gas properties are lacking. Starbursts in
ULIRGs could be more intense per unit volume than those in LIRGs, and
could create different  
molecular gas excitation in such a way that HCN J=4--3 is excited to a
greater extent  
in starbursts in ULIRGs than in the less infrared-luminous LIRGs. 
Molecular line data on starburst-dominated ULIRGs with normal dense
molecular gas properties, if obtained in the future, would provide 
insight into the origin of the observed molecular line flux ratios in 
the highly turbulent ULIRG, IRAS 22491$-$1808. 
Observations of further well-studied (U)LIRGs are needed to 
better understand how molecular emission line flux ratios 
depend on the relative AGN and starburst energetic contributions, 
and other galaxy properties.
 
Finally, we briefly mention some other possibilities that could 
affect the observed molecular line flux ratios. 
Emission line fluxes can be reduced by dust extinction. 
Since the frequencies of HCN, HCO$^{+}$, and HNC J=4--3 lines are 
$\sim$350 GHz ($\sim$850 $\mu$m) in the submillimeter wavelength range, 
flux attenuation of these molecular lines by dust extinction is usually not 
significant.
\citet{mat09} estimated that in the nearby well-studied ULIRG, Arp 220, 
dust extinction  may not be negligible even in the submillimeter wavelength.
Even if dust extinction at $\sim$850 $\mu$m is not negligible in some of the 
observed (U)LIRGs, flux attenuation of HCN, HCO$^{+}$, and HNC J=4--3 
lines should be comparable, due to their similar frequencies (wavelengths).
Thus, molecular line flux ratios will not change significantly by possible 
dust extinction in the submillimeter wavelength.

Line opacity could also have an effect on the observed molecular line 
fluxes. 
According to the wide-accepted standard scenario for molecular gas 
clouds in the Galaxy and external galaxies, molecular gas clouds 
consist of clumps with a small volume filling factor \citep{sol87}.
The line opacity of each clump is thought to be larger than unity for 
CO \citep{sol87}.
Emission from the other side of each clump is not completely probed 
from CO observations.
Each clump has small molecular line widths by thermal broadening, but 
shows large random motion inside molecular gas clouds.
Due to the velocity difference of each clump, line absorption by 
foreground clumps is insignificant, and clumps at the other side 
of the molecular gas clouds are observationally detectable. 
This model (the so-called mist model) can explain observed molecular 
gas properties very well \citep{sol87}.
The detailed physical parameters of each clump are observationally 
not well-constrained.
Thus, the physical properties of each clump are assumed to be uniform 
inside a molecular cloud, as well as among different galaxies. 
The optical depths of HCN J=4--3 lines in external galaxies could be 
larger than unity \citep{ngu92,sak10,jia11}.
Even if each clump is optically thick for HCN, as long as the uniform
physical properties of each gas clump is assumed, the enhanced HCN
abundance results in an increased area filling factor of HCN emission in
a molecular gas cloud, and so a higher HCN J=4--3 flux from a molecular
gas cloud \citep{ima07b}, not changing our previous discussion.
We should note that the unconstrained physical properties
of each molecular gas clump could cause the largest uncertainty about
the interpretation of the observed molecular line flux ratios.

Summarizing, we regard that the proposed three scenarios 
(HCN abundance enhancement, infrared radiative pumping, 
and excitation effect) are the plausible mechanisms that could explain 
the variation of the observed molecular line flux ratios in (U)LIRGs with 
different AGN energetic contributions. 
If the excitation effect is important, the HCN-to-HCO$^{+}$ flux ratios 
are expected to systematically decrease at higher J-transition.
On the other hand, in the case of HCN abundance enhancement, 
the HCN-to-HCO$^{+}$ flux ratios are not sensitive to J-transitions. 
Multiple J-transition molecular line observations are needed to
differentiate physical mechanisms behind the observed molecular line
flux ratios, with the aid of modeling \citep{van07}. 
For (U)LIRGs with small molecular line widths, vibrationally-excited 
HCN emission lines may be detectable, by separating from the nearby,
much stronger HCO$^{+}$ emission lines at v=0 \citep{sak10,ima13b}.
The strengths of the detected vibrationally-excited HCN emission lines 
can be used to understand the general role 
of the infrared radiative pumping for HCN excitation in AGN-hosting 
(U)LIRGs. 
 
\subsection{Implications} 
 
As mentioned in $\S$1, although the energy diagnostic method developed
using the J=1--0 transition lines of HCN and HCO$^{+}$
\citep{koh05,ima04,in06,ima06b,ima07b,kri08,ima09b}  
is powerful, using ALMA it is applicable only to nearby galaxies with $z
<$ 0.06. If we can establish a reliable energy diagnostic method using a
higher J transition at higher frequency (shorter wavelength), the method
could potentially be applied to more distant galaxies using ALMA.  
Furthermore, the HCN J=4--3 lines selectively trace dense 
molecular gas in the nuclear regions of galaxies where AGNs are 
expected to be present, with even less contamination from 
spatially extended galaxy-wide molecular gas than at lower
J-transitions. 
Hence, the energy diagnostic method using HCN J=4--3 could be very
sensitive to the presence of luminous AGNs. These were the primary
reasons why we conducted our ALMA Cycle 0 program.  
 
In the case of thermal excitation, the flux of high-J transition lines
increases proportional to the square of the frequency
($\propto$$\nu^{2}$), and so the J=4--3 flux can be 16 times higher  
than the J=1--0 flux. 
If realized, the high J=4--3 flux can keep the detection significance of 
molecular emission lines high, even under the higher Earth background
noise at the frequency of J=4--3 compared to J=1--0.  
Given that the bulk of the infrared luminosity of nearby (U)LIRGs
usually comes from the nuclear regions ($\S$1), we may be able to
relate the nuclear molecular gas emission properties with the
IRAS-measured whole galactic infrared emission.
Based on our pre-ALMA interferometric HCN J=1--0 observations of nearby 
(U)LIRG nuclei, the nuclear HCN J=1--0 peak flux and whole galactic
infrared flux are found to be correlated (Figure 11; Left), where the
actual data are summarized in Table 7.  
Therefore, we can empirically predict the emission peak flux of HCN
J=1--0 line in the (U)LIRG nuclei from the IRAS-measured infrared flux,
and can convert to the HCN J=4--3 emission peak flux assuming thermal 
excitation. 
Figure 11 (right) compares the HCN J=1--0 luminosities at (U)LIRG nuclei
based on our pre-ALMA interferometric observations, with the
IRAS-measured infrared luminosities from the whole galactic regions of  
(U)LIRGs. Our plots, particularly for the nuclei of
ULIRGs with L$_{\rm IR}$ $\gtrsim$ 10$^{12}$L$_{\odot}$, roughly follow
the ratio established for ULIRGs based on single-dish telescope data
\citep{gra08}, supporting the scenario that the bulk of high-density
molecular gas in ULIRGs is concentrated in the energetically-dominant
nuclear regions and is recovered with our pre-ALMA interferometric
data with $\sim$1 to several arcsec spatial resolution.   
 
Table 8 presents a comparison of the expected HCN J=4--3 peak flux for
thermal excitation with the observed peak flux. It can be seen that the
HCN J=4--3 emission peak is substantially smaller than that expected from
thermal excitation, indicating that HCN is sub-thermally excited at
J=4--3 even in our active ULIRG sample, often hosting luminous buried
AGNs.  
Thus, lower J transitions (J=3--2 or 2--1) may be better in terms of the 
detection significance of HCN J-transition line per given observing time. 
It is necessary to find the best J transition lines with which we can 
reliably and efficiently separate deeply buried AGNs from 
the surrounding starbursts in a dusty ULIRG population. 
 
\section{Summary} 
We conducted HCN, HCO$^{+}$, and HNC J=4--3 observations of nearby 
dusty (U)LIRGs with different levels of AGN energetic contributions 
to the observed infrared luminosities, in ALMA Cycle 0, to probe 
the physical properties of dense molecular gas around the dust-obscured 
energy sources in the nuclei. Six sources were observed in total, and
the results of two sources had been published previously
\citep{ima13a,ima13b}. We mainly discussed the results of the remaining
four sources in this paper, and made the following summary.
 
\begin{enumerate} 
 
\item HCN, HCO$^{+}$, HNC J=4--3 lines were significantly ($>$3$\sigma$) 
detected in most of the observed ULIRGs, suggesting the presence of
a large amount of dense molecular gas that is sufficiently excited to
J=4--3.  
 
\item The three buried AGN-dominated ULIRGs, IRAS 00183$-$7111, 
08572$+$3915, and 12127$-$1412, showed higher HCN-to-HCO$^{+}$ J=4--3 
flux ratios than the less infrared-luminous starburst-dominated galaxy, 
NGC 1614, supporting the previously reported trend that HCN emission is 
enhanced relative to HCO$^{+}$, in AGNs compared to starbursts. 
We saw no clear differences in the HCN-to-HNC J=4--3 flux ratios between 
AGN-important ULIRGs and the starburst-dominated galaxy, NGC 1614. 
 
\item Three possible mechanisms were proposed for the origin of the 
enhanced HCN J=4--3 emission relative to HCO$^{+}$ J=4--3 in AGNs: (1) 
HCN abundance enhancement under high dust temperature chemistry and/or 
X-ray induced chemistry, (2) infrared radiative pumping of HCN, and 
(3) more HCN J=4--3 excitation. Multiple J-transition data of HCN,
HCO$^{+}$, and HNC, both at vibrational ground and excited levels, are
needed to better separate these scenarios.  
 
\item The ULIRG, IRAS 22491$-$1808, whose observed infrared spectral shape 
is explained by starburst activity, showed a high HCN-to-HCO$^{+}$ J=4--3 
flux ratio, similar to that of buried AGN-dominated ULIRGs. This ULIRG
also showed remarkably high turbulent motion. 
Turbulence-induced heating may heat gas and dust, and enhance
the HCN J=4--3 flux. Data on other ordinary starburst-dominated ULIRGs
are needed to better understand the origin of the strong HCN emission in
IRAS 22491$-$1808, and to investigate mechanisms other than AGN activity
that can enhance HCN emission.  

\item The vibrationally excited HCN J=4--3 (v$_{2}$=1f) lines were 
covered with our ALMA observations, but were not clearly detected in any 
of the four ULIRGs. 
 
\item In addition to the targeted molecular lines, 
H$_{2}$S 3(2,1)--3(1,2) and CH$_{3}$CN v=0 19(3)--18(3) emission lines 
were serendipitously detected in a few sources, demonstrating the high 
sensitivity of ALMA. 
 
\item The observed HCN J=4--3 emission peak fluxes were found to be 
significantly smaller than those expected from thermal excitation, 
suggesting that the HCN J=4--3 line is sub-thermally excited 
even in our active galaxy sample. We need to find the best J-transition
lines, with which we can separate ULIRGs hosting luminous buried AGNs
from starbursts more efficiently but with sufficient reliability.  
 
\end{enumerate} 
 
Further observations of a larger sample of well-calibrated (U)LIRGs 
at multiple J-transition lines are clearly necessary to confirm the trend of 
molecular emission line flux ratios, depending on the primary energy 
sources and other galaxy properties, implied from our ALMA Cycle 0 data 
presented in this paper, and to understand the physical origin of the 
observed molecular line flux ratios.
 
\acknowledgments 
 
We thank the referee, Paul Ho, for his invaluable comment.
We are grateful to Drs. E. Mullar, H. Nagai, and K. Saigo for their
helpful advice on ALMA data analysis. 
M.I. is supported by Grants-in-Aid for Scientific Research
(no. 23540273) and the ALMA Japan Research Grant of NAOJ Chile Observatory, 
NAOJ-ALMA-0001. This paper made use of the following ALMA
data: ADS/JAO.ALMA\#2011.0.00020.S. 
ALMA is a partnership of ESO (representing its member states), NSF (USA) 
and NINS (Japan), together with NRC (Canada) and NSC and ASIAA (Taiwan), 
in cooperation with the Republic of Chile. The Joint ALMA Observatory is 
operated by ESO, AUI/NRAO, and NAOJ. 

%\appendix

\clearpage

%%%%%%%%%% Table 1 %%%%%%%%%
\begin{deluxetable}{lccrrrrccl}
\tabletypesize{\scriptsize}
%\rotate
\tablecaption{Basic Properties of Observed Luminous Infrared Galaxies
\label{tbl-1}}
\tablewidth{0pt}
\tablehead{
\colhead{Object} & \colhead{Redshift}   & \colhead{Scale}   & 
\colhead{f$_{\rm 12}$}   & 
\colhead{f$_{\rm 25}$}   & 
\colhead{f$_{\rm 60}$}   & 
\colhead{f$_{\rm 100}$}  & 
\colhead{log L$_{\rm IR}$} &
\colhead{Optical}  & \colhead{Energy} \\
\colhead{} & \colhead{}   & \colhead{[kpc/"]}
& \colhead{[Jy]} & \colhead{[Jy]} 
& \colhead{[Jy]} & \colhead{[Jy]}  & \colhead{[L$_{\odot}$]} 
& \colhead{Class} & \colhead{Source}  \\
\colhead{(1)} & \colhead{(2)} & \colhead{(3)} & \colhead{(4)} & 
\colhead{(5)} & \colhead{(6)} & \colhead{(7)} & \colhead{(8)} & 
\colhead{(9)} & \colhead{(10)}   
}
\startdata
IRAS 08572$+$3915 & 0.058 & 1.1 & 0.32 & 1.70 & 7.43  & 4.59  & 12.1 & LI$^{a}$(Sy2$^{b}$) & B-AGN \\  
IRAS 12127$-$1412 & 0.133 & 2.3 & $<$0.13 & 0.24 & 1.54 & 1.13 & 12.1 & LI$^{a}$ (HII$^{b}$) & B-AGN \\  
IRAS 00183$-$7111 & 0.327 & 4.7 & $<$0.07 & 0.13 & 1.20 & 1.19 & 12.9 & LI$^{c}$ & B-AGN \\  
IRAS 20551$-$4250 & 0.043 & 0.84 & 0.28 & 1.91 & 12.78 & 9.95  & 12.0 & LI(HII)$^{d}$ & B-AGN + SB  \\  
IRAS 22491$-$1808 & 0.076 \tablenotemark{A} & 1.4 & 0.05 & 0.55 & 5.44 & 4.45 & 12.1 &
HII$^{a,b}$ & SB only (?) \\  
NGC 1614 (IRAS 04315$-$0840) & 0.016 & 0.32 & 1.38 & 7.50 & 32.12 & 34.32 &
11.6 & HII $^{b,e,f}$ & SB \\ \hline
\enddata

\tablenotetext{A}{The redshift was estimated to be $z$=0.076 by
\citet{kim98} and \citet{soi87}, but to be $z$=0.078 by \citet{str92}. 
In this paper, we adopt $z$=0.076 \citep{kim98} to be consistent to
other ULIRGs in the IRAS 1Jy sample \citep{kim98}.  
}

\tablecomments{ 
Col.(1): Object name. 
Col.(2): Redshift. 
Col.(3): Physical scale in kpc for 1 arcsec.
Cols.(4)--(7): f$_{12}$, f$_{25}$, f$_{60}$, and f$_{100}$ are 
{\it IRAS} fluxes at 12 $\mu$m, 25 $\mu$m, 60 $\mu$m, and 100 $\mu$m, 
respectively, taken from \citet{kim98}, \citet{san03}, or IRAS faint 
source catalog. 
Col.(8): Decimal logarithm of infrared (8$-$1000 $\mu$m) luminosity 
in units of solar luminosity (L$_{\odot}$), calculated with 
$L_{\rm IR} = 2.1 \times 10^{39} \times$ D(Mpc)$^{2}$ 
$\times$ (13.48 $\times$ $f_{12}$ + 5.16 $\times$ $f_{25}$ + 
$2.58 \times f_{60} + f_{100}$) [ergs s$^{-1}$] \citep{sam96}. 
Col.(9): Optical spectral classification. 
``HII'', ``LI'', and ``Sy2'' denote HII-region, LINER, and Seyfert 2, 
respectively. 
$^{a}$: \citet{vei99}.
$^{b}$: \citet{yua10}.
$^{c}$: \citet{arm89}.
$^{d}$: \citet{duc97}
$^{e}$: \citet{vei95}.
$^{f}$: \citet{kew01}.
Col.(10): Infrared and/or X-ray spectroscopic classification 
of primary energy sources. 
``B-AGN'' and ``SB'' denote buried AGN and starburst, respectively. 
References are given in $\S$2, \citet{ima13a}, and \citet{ima13b}.
}

\end{deluxetable}

%%%%%%%%%% Table 2 %%%%%%%%%
\begin{deluxetable}{lllccc|ccc}
%\tabletypesize{\small}
%\rotate
\tabletypesize{\scriptsize}
\tablecaption{Log of ALMA Cycle 0 Observations \label{tbl-2}} 
\tablewidth{0pt}
\tablehead{
\colhead{Object} & \colhead{Line} & \colhead{Date} & \colhead{Antenna} & 
\colhead{Baseline} & \colhead{Exposure} & \multicolumn{3}{c}{Calibrator} \\ 
\colhead{} & \colhead{} & \colhead{[UT]} & \colhead{Number} & \colhead{[m]} &
\colhead{[min]} & \colhead{Bandpass} & \colhead{Flux} & \colhead{Phase}  \\
\colhead{(1)} & \colhead{(2)} & \colhead{(3)} & \colhead{(4)} &
\colhead{(5)} & \colhead{(6)} & \colhead{(7)}  & \colhead{(8)} & \colhead{(9)}
}
\startdata 
IRAS 08572$+$3915 & HCN/HCO$^{+}$ & 2012 October 21 & 22 & 21--384 & 19 & 3C279 &
Callisto & J0927+390 \\   
% & HNC & & & & & \\ 
IRAS 12127$-$1412 & HCN/HCO$^{+}$ & 2012 May 31 -- June 1 & 20 & 21--402 & 26 & 3C279 &
Titan & J1130$-$148 \\
 &  & 2012 December 14 & 24 & 15--402 & 22 & 3C279 & Titan & J1130$-$148 \\ 
 & HNC & 2012 June 1 & 21 & 21--402 & 42 & 3C279 & Titan & J1130$-$148 \\
 &  & 2012 November 5 & 23 & 21--375 & 40 & 3C279 & Titan & J1130$-$148 \\ 
IRAS 00183$-$7111 & HCN/HCO$^{+}$ & 2011 November 26--27 & 18 & 18--201 &  30 & 3C454.3
& Callisto & J2157$-$694 \\
& & 2011 November 28 & 14 & 18--201 & 29 & 3C454.3 & Neptune & J2157$-$694 \\  
& & 2012 January 10 & 17 & 19--269 & 30 & 3C454.3 & Neptune & J2157$-$694 \\  
& & 2012 January 13 & 17 & 19--269 & 30 & 3C454.3 & Neptune & J2157$-$694 \\  
& HNC & 2011 November 29 & 14 & 18--201 & 31 & 3C454.3 & Neptune & J2157$-$694 \\  
& & 2012 January 13 & 17 & 19--269 & 31 & 3C454.3 & Neptune & J2157$-$694 \\  
& & 2012 January 23 & 17 & 19--269 & 65 & 3C454.3 & Neptune & J2157$-$694 \\ 
%& & 2012 January 23 & 17 & 3C454.3 & Neptune & J2157$-$694 \\  
%IRAS 20551$-$4250 & HCN/HCO$^{+}$ & 2012 June 1 & 18 & 3C454.3 & Neptune
%& J2056$-$472 \\
%& & 2012 July 26 & 17 & 3C454.3 & Neptune & J2056$-$472 \\
%& HNC & 2012 June 2 & 19 & 3C454.3 & Neptune & J2056$-$472 \\
%& & 2012 July 26 & 18 & 3C454.3 & Neptune & J2056$-$472 \\
IRAS 22491$-$1808 & HCN/HCO$^{+}$ & 2012 October 21 & 20 & 21--384 & 26 & 3C454.3 &
Neptune & J2258$-$279 \\
& HNC & 2011 November 27 \tablenotemark{a} & 17 & 18--201 & 17 & B0521$-$365 & Callisto & J2258$-$279 \\
& & 2012 November 7 & 23 & 21--382 & 17 & B0521$-$365 & Callisto & J2258$-$279 \\
% NGC 1614 & HCN/HCO$^{+}$ & 2011 November 15 & 16 & 3C454.3 & Callisto & J0423$-$013\\
% & HNC & 2011 November 15 & 16 & 3C454.3 & Callisto & J0423$-$013\\
\enddata

\tablenotetext{a}{
The HNC data of IRAS 22491$-$1808 taken on 2011 November 27 were much
noisier than those taken on 2012 November 7. 
We used only the latter data to investigate the quantities of HNC 
J=4--3 line and nearby continuum emission from IRAS 22491$-$1808,
because addition of the former data resulted in larger noise.} 

\tablecomments{ 
Col.(1): Object name. 
Col.(2): Observed molecular line. HCN and HCO$^{+}$ J=4--3 lines were 
simultaneously taken. HNC J=4--3 observations were made separately. 
Col.(3): Observing date in UT. 
Col.(4): Number of antennas used for observations. 
Col.(5): Baseline length in meter. Minimum and maximum baseline lengths are 
shown.  
Col.(6): Net on source exposure time in min.
In ALMA Cycle 0, requested sensitivity was defined, and 
observations were performed until this sensitivity was achieved. 
The net on source exposure times agree with the calculated 
exposure times in our ALMA Cycle 0 proposal within a factor of 2. 
Cols.(7), (8), and (9): Bandpass, flux, and phase calibrator for the 
target source, respectively.
}

\end{deluxetable}

%%%%%%%%%% Table 3 %%%%%%%%%
\begin{deluxetable}{lll}
%\tabletypesize{\small}
\tabletypesize{\scriptsize}
\tablecaption{Central Frequency of Individual Correlator Spectral Window
\label{tbl-3}}
\tablewidth{0pt}
\tablehead{
\colhead{Object} & \colhead{Line} & \colhead{Frequency} \\ 
\colhead{} & \colhead{} & \colhead{[GHz]}  \\
\colhead{(1)} & \colhead{(2)} & \colhead{(3)}   
}
\startdata 
IRAS 08572$+$3915 & HCN/HCO$^{+}$ & 335.071 (HCN), 336.989 (HCO$^{+}$),
347.164, 349.055 \\
% & HNC &  342.751 (HNC), 354.064 \\
IRAS 12127$-$1412 & HCN/HCO$^{+}$ & 312.891 (HCN), 314.858 (HCO$^{+}$), 300.971, 302.736\\
 & HNC & 320.062 (HNC), 308.561 \\
IRAS 00183$-$7111 & HCN/HCO$^{+}$ & 267.148 (HCN), 268.828 (HCO$^{+}$),
254.333, 252.449 \\
& HNC & 273.271 (HNC), 258.176 \\
IRAS 22491$-$1808 & HCN/HCO$^{+}$ & 329.466 (HCN), 331.351 (HCO$^{+}$),
341.543, 343.309 \\
& HNC \tablenotemark{a} & 337.017 (HNC) \\
\enddata

\tablenotetext{a}{
Although two spectral windows were used for the HNC observations of 
IRAS 22491$-$1808 on 2012 November 7, data of only one spectral 
window, containing the HNC J=4--3 emission line, were provided to us.
}

\tablecomments{ 
Col.(1): Object name. 
Col.(2): Observed molecular line. 
Col.(3): Central observed frequency of each correlator spectral window 
in GHz. 
For HCN and HCO$^{+}$ J=4--3 observations, four spectral windows were 
used. 
For HNC J=4--3 observations, two spectral windows were used.
} 

\end{deluxetable}

%%%%%%%%%% Table 4 %%%%%%%%%
\begin{deluxetable}{lcrccl}
%\tabletypesize{\small}
\tabletypesize{\scriptsize}
\tablecaption{Continuum Emission \label{tbl-4}} 
\tablewidth{0pt}
\tablehead{
\colhead{Object} & \colhead{Continuum} &
\colhead{Flux} & \colhead{Peak Coordinate} & \colhead{rms} &
\colhead{Beam patten} \\ 
\colhead{} & \colhead{[GHz]} & \colhead{[mJy beam$^{-1}$]}
& \colhead{(RA,DEC)J2000} & \colhead{[mJy beam$^{-1}$]} & 
\colhead{[arcsec $\times$ arcsec] ($^{\circ}$)} \\  
\colhead{(1)} & \colhead{(2)} & \colhead{(3)}  & \colhead{(4)}  &
\colhead{(5)} & \colhead{(6)} 
}
\startdata 
IRAS 08572$+$3915 NW & a (342) &  6.1 (39$\sigma$) & (09 00 25.39, $+$39 03 54.2) &
0.16 & 1.8 $\times$ 1.1 ($-$110$^{\circ}$)\\  
%                  & b & &  & \\  
IRAS 12127$-$1412 NE & a (308) & 2.8 (28$\sigma$) & (12 15 19.13, $-$14 29 41.7)
& 0.10 & 0.6 $\times$ 0.5 (58$^{\circ}$) \\
           & b (314) &  2.6 (20$\sigma$) & (12 15 19.13, $-$14 29 41.8) &
0.13 & 0.6 $\times$ 0.5 (49$^{\circ}$) \\
IRAS 00183$-$7111 & a (261) &  2.2 (26$\sigma$) & (00 20 34.70, $-$70 55
26.4) & 0.084 & 2.5 $\times$ 1.2 (177$^{\circ}$) \\
           & b (266) &  1.5 (23$\sigma$) & (00 20 34.70, $-$70 55 26.4) &
0.067 & 2.2 $\times$ 1.2 (175$^{\circ}$) \\  
IRAS 22491$-$1808 & a (336) &  10.4 (64$\sigma$) & (22 51 49.35, $-$17 52 24.2)
& 0.16 & 0.6 $\times$ 0.5 ($-$8$^{\circ}$) \\ 
           & b (337) & 10.6 (39$\sigma$) & (22 51 49.35, $-$17 52 24.1)
& 0.27 & 1.1 $\times$ 0.5 (99$^{\circ}$) \\   
\enddata

\tablecomments{ 
Col.(1): Object name. 
Col.(2): Continuum-a or -b. Continuum-a data were taken simultaneously with 
HCN and HCO$^{+}$ J=4--3 observations. 
Continuum-b data were taken simultaneously with HNC J=4--3 observations. 
The central frequency in GHz is added in parentheses. 
Col.(3): Flux in [mJy beam$^{-1}$] at the emission peak. 
Detection significance relative to the rms noise (1$\sigma$) is shown 
in parentheses. 
The rms noise level is derived from the standard deviation of sky signals 
in individual continuum maps.
We did not include possible systematic uncertainty, which is difficult 
to quantify. 
Col.(4): The coordinate of the continuum emission peak in J2000. 
For reference, the near-infrared $K$-band (2.2$\mu$m) continuum emission 
peaks in J2000 are (09 00 25.44, $+$39 03 55.0) for IRAS 08572$+$3915 (NW 
nucleus), (12 15 19.14, $-$14 29 41.4) for IRAS 12127$-$1412 (NE 
nucleus), (22 51 49.30, $-$17 52 22.5) for IRAS 22491$-$1808 (E nucleus) 
\citep{kim02}. 
Col.(5): rms noise (1$\sigma$) level in [mJy beam$^{-1}$], estimated 
from the standard deviation of sky signals. 
Col.(6): Beam size in [arcsec $\times$ arcsec] and position angle. 
Position angle is 0$^{\circ}$ along the north-south direction, 
and increases counterclockwise.
}

\end{deluxetable}

\clearpage

%%%%%%%%%% Table 5 %%%%%%%%%
\begin{deluxetable}{ll|llcl|cccc}
%\rotate
%\tabletypesize{\small}
\tabletypesize{\scriptsize}
\tablecaption{Molecular Line Flux \label{tbl-5}} 
\tablewidth{0pt}
\tablehead{
\colhead{Object} & \colhead{Line} & 
\multicolumn{4}{c}{Integrated intensity (moment 0) map} & 
\multicolumn{4}{c}{Gaussian line fit} \\  
\colhead{} & \colhead{} & \colhead{Peak} & \colhead{rms} & 
\colhead{} & 
\colhead{Beam} & \colhead{Velocity} & \colhead{Peak} & \colhead{FWHM} & 
\colhead{Flux} \\ 
\colhead{} & \colhead{} & \multicolumn{2}{c}{[Jy beam$^{-1}$ km s$^{-1}$]} &
\colhead{Sum} & 
\colhead{[$''$ $\times$ $''$] ($^{\circ}$)} &
\colhead{[km s$^{-1}$]} & \colhead{[mJy]} & \colhead{[km s$^{-1}$]} & 
\colhead{[Jy km s$^{-1}$]} \\ 
\colhead{(1)} & \colhead{(2)} & \colhead{(3)} & \colhead{(4)} & 
\colhead{(5)} & \colhead{(6)} & \colhead{(7)} & \colhead{(8)} &
\colhead{(9)} & \colhead{(10)}    
}
\startdata 
IRAS 08572$+$3915 & HCN J=4--3 & 2.8 (18$\sigma$) & 0.15 & 23 & 1.8 $\times$ 1.1
($-$113$^{\circ}$) & 17501$\pm$10 & 9.7$\pm$0.6 & 330$\pm$21 & 3.2$\pm$0.3  \\  
                  & HCO$^{+}$ J=4--3 & 3.5 (29$\sigma$) & 0.12 & 19 & 2.5 $\times$ 1.1
($-$111$^{\circ}$) & 17489$\pm$7 & 14$\pm$0.8 & 290$\pm$17 & 4.2$\pm$0.3 \\  
                  & H$_{2}$S & 1.2 (4.5$\sigma$) & 0.26 & 39 & 2.1 $\times$ 1.1
($-$108$^{\circ}$) & 17494$\pm$41 & 2.4$\pm$0.3 & 520$\pm$99 & 1.3$\pm$0.3 \\  
%                  & HNC & &&&&&&     \\  
IRAS 12127$-$1412 & HCN J=4--3 & 1.1 (8.7$\sigma$) & 0.12 & 40 & 0.6 $\times$ 0.5
(59$^{\circ}$) & 39924$\pm$32 & 2.2$\pm$0.2 & 570$\pm$78 & 1.2$\pm$0.2 \\
                  & HCO$^{+}$ J=4--3 & 0.56 (5.2$\sigma$) & 0.11 & 30 & 0.6 $\times$
0.5 (59$^{\circ}$) & 39985$\pm$42 & 1.5$\pm$0.2 & 570$\pm$110 & 0.77$\pm$0.19     \\  
          & CS J=7--6 & $<$0.34 ($<$3$\sigma$) & 0.11 & 36 & 0.6 $\times$ 0.5 (40$^{\circ}$) & 
          39869$\pm$73 & 1.1$\pm$0.5 & 240 (fix) & 0.26$\pm$0.11 \\ 
                  & HNC J=4--3 & 0.81 (5.3$\sigma$) & 0.15 & 39 & 0.6 $\times$ 0.5
(31$^{\circ}$) & 40039$\pm$43 & 2.0$\pm$0.4 & 480$\pm$86 & 0.91$\pm$0.23 \\   
                  & CH$_{3}$CN & 0.35 (5.1$\sigma$) & 0.068 & 31 & 0.6 $\times$ 0.5
(62$^{\circ}$) & 39984$\pm$95 & 0.7$\pm$0.3 & 480 (fix) & 0.30$\pm$0.12 \\  
IRAS 00183$-$7111 & HCN J=4--3 & 0.70 (3.9$\sigma$) & 0.18 & 41 & 2.4 $\times$ 1.2
(177$^{\circ}$) & 98576$\pm$100 & 1.6$\pm$0.6 & 620$\pm$230 & 0.78$\pm$0.42 \\
                  & HCO$^{+}$ J=4--3 & 0.57 (3.3$\sigma$) & 0.17 & 42 & 2.3 $\times$ 1.2
(177$^{\circ}$) & 98423$\pm$209 & 1.1$\pm$0.4 & 620 (fix) & 0.55$\pm$0.21 \\  
                  & HNC J=4--3 & 0.40 (3.6$\sigma$) & 0.11 & 47 & 
2.2 $\times$ 1.2 (173$^{\circ}$) & 98216$\pm$173 & 0.77$\pm$0.19 & 620(fix) & 0.39$\pm$0.10 \\  
IRAS 22491$-$1808 & HCN J=4--3 & 7.1 (38$\sigma$) & 0.19 & 34 & 0.6 $\times$ 0.6 (29$^{\circ}$) & 
23297$\pm$9 & 18$\pm$1 & 420$\pm$21 & 7.6$\pm$0.5 \\
                  & HCO$^{+}$ J=4--3 & 4.7 (31$\sigma$) & 0.15 & 42 & 0.6 $\times$ 0.5 ($-$6$^\circ$) & 
                  23296$\pm$16 & 9.4$\pm$0.5 & 570$\pm$35 & 5.4$\pm$0.4 \\  
                  & H$_{2}$S & 2.3 (26$\sigma$) & 0.09 & 21 & 0.6 $\times$ 0.5
($-$12$^{\circ}$) & 23275$\pm$22 & 9.7$\pm$0.4 & 360$\pm$44 & 3.5$\pm$0.5 \\  
                  & HNC J=4--3 & 6.5 (39$\sigma$) & 0.17 & 33 & 1.1 $\times$ 0.5
(99$^{\circ}$) & 23298$\pm$5 & 21$\pm$1 & 340$\pm$11 & 7.2$\pm$0.3 \\  
\enddata

\tablecomments{ 
Col.(1): Object name. 
Col.(2): Observed molecular line. 
HCN, HCO$^{+}$, H$_{2}$S, and CS lines were observed simultaneously with 
continuum-a. 
HNC and CH$_{3}$CN lines were taken at the same time as continuum-b. 
Col.(3): Rest-frame frequency of each molecular line in [GHz]. 
Col.(4): Integrated intensity in [Jy beam$^{-1}$ km s$^{-1}$] at the 
emission peak. 
Detection significance relative to the rms noise (1$\sigma$) in the 
moment 0 map is shown in parentheses. 
Col.(5): rms noise (1$\sigma$) level in the moment 0 map in 
[Jy beam$^{-1}$ km s$^{-1}$], derived from the standard deviation 
of sky signals in each moment 0 map. 
Col.(6): The number of spectral elements summed to create moment 0 maps.
Each spectral element ($\sim$19.5 MHz width) consists of 40 correrator
channels binning (see $\S$3).
Col.(7): Beam size in [arcsec $\times$ arcsec] and position angle. 
Position angle is 0$^{\circ}$ along the north-south direction, 
and increases counterclockwise. 
Cols.(8)--(11): Gaussian fits of emission lines in the spectra at the 
continuum peak position, within the beam size. 
Col.(8): Optical velocity (v$_{\rm opt}$) of emission peak in [km s$^{-1}$]. 
Col.(9): Peak flux in [mJy]. 
Col.(10): Observed FWHM in [km s$^{-1}$] in the right panels of 
Figures 3--6.  
Col.(11): Flux in [Jy km s$^{-1}$]. The observed FWHM in 
[km s$^{-1}$] in column 10 is divided by ($1+z$) to obtain the
intrinsic FWHM in [km s$^{-1}$]. 
}

\end{deluxetable}

%%%%%%%%%% Table 6 %%%%%%%%%
\begin{deluxetable}{lll|cc}
%\rotate
%\tabletypesize{\small}
\tabletypesize{\scriptsize}
\tablecaption{Molecular Line Luminosity \label{tbl-6}} 
\tablewidth{0pt}
\tablehead{
\colhead{Object} & \colhead{Line} & \colhead{$\nu_{\rm rest}$} & 
\multicolumn{2}{c}{Luminosity} \\  
\colhead{} & \colhead{} & \colhead{[GHz]} & \colhead{[10$^{4}$ L$_{\odot}$]} & 
\colhead{[10$^{7}$ K km s$^{-1}$ pc$^{2}$]} \\ 
\colhead{(1)} & \colhead{(2)} & \colhead{(3)} & \colhead{(4)} & 
\colhead{(5)}   
}
\startdata 
IRAS 08572$+$3915 NW & HCN J=4--3 & 354.505 & 7.3$\pm$0.7 & 5.1$\pm$0.5 \\  
                  & HCO$^{+}$ J=4--3 & 356.734 & 9.7$\pm$0.7 & 6.6$\pm$0.5 \\
                  & H$_{2}$S & 369.101 & 3.1$\pm$0.7 & 1.9$\pm$0.4 \\
IRAS 12127$-$1412 NE & HCN J=4--3 & 354.505 & 15$\pm$3 & 11$\pm$2\\
                  & HCO$^{+}$ J=4--3 & 356.734 & 9.6$\pm$2.4 & 6.6$\pm$1.6\\
                  & CS J=7--6 & 342.883 & 3.1$\pm$1.3 & 2.5$\pm$1.0 \\
                  & HNC J=4--3 & 362.630 & 12$\pm$3 & 7.6$\pm$2.2\\
                  & CH$_{3}$CN & 349.393 & 3.7$\pm$1.5 & 2.7$\pm$1.1\\
IRAS 00183$-$7111 & HCN J=4--3 & 354.505 & 63$\pm$34 & 44$\pm$24\\
                  & HCO$^{+}$ J=4--3 & 356.734 & 45$\pm$17 & 31$\pm$12\\
                  & HNC J=4--3 & 362.630 & 32$\pm$8 & 21$\pm$5\\
IRAS 22491$-$1808 & HCN J=4--3 & 354.505 & 30$\pm$2 & 21$\pm$1 \\
                  & HCO$^{+}$ J=4--3 & 356.734 & 22$\pm$2 & 15$\pm$1 \\
                  & H$_{2}$S & 369.101 & 14$\pm$2 & 9.0$\pm$1.3 \\
                  & HNC J=4--3 & 362.630 & 29$\pm$1 & 19$\pm$1 \\
IRAS 20551$-$4250 & HCN J=4--3 & 354.505 & 11$\pm$1 & 8.0$\pm$0.2 \\
                  & HCO$^{+}$ J=4--3 & 356.734 & 17$\pm$1 & 12$\pm$1 \\
                  & H$_{2}$S & 369.101 & 4.1$\pm$0.2 & 2.5$\pm$0.2 \\
                  & HNC J=4--3 & 362.630 & 7.1$\pm$0.2 & 4.6$\pm$0.2 \\
                  & CH$_{3}$CN & 349.393 & 5.2$\pm$0.2 & 3.8$\pm$0.2 \\
                  & HCN v$_{2}$=1f J=4--3 & 356.256 & 0.5$\pm$0.1 & 0.3$\pm$0.1 \\
\enddata

\tablecomments{ 
Col.(1): Object name. 
IRAS 20551$-$4250 is included, because molecular line luminosity information 
is not shown in \citet{ima13b}. 
Col.(2): Observed molecular line. 
Col.(3): Rest-frame frequency of each molecular line in [GHz]. 
Col.(4): Molecular line luminosity in [10$^{4}$ L$_{\odot}$], derived from 
the flux based on Gaussian fits (Table 5, column 11). 
Col.(5): Molecular line luminosity in [10$^{7}$ K km s$^{-1}$ pc$^{2}$], 
derived from the flux based on Gaussian fits (Table 5, column 11).
}

\end{deluxetable}

%%%%%%%%%% Table 7 %%%%%%%%%
\begin{deluxetable}{lcccccc}
%\tabletypesize{\small}
\tabletypesize{\scriptsize}
\tablecaption{HCN J=1--0 and Infrared Emission of (U)LIRGs 
Based on Our Pre-ALMA Interferometric Data \label{tbl-7}} 
\tablewidth{0pt}
\tablehead{
\colhead{Object} & \colhead{Redshift} & \colhead{f$_{\rm IR}$} & 
\colhead{log L$_{\rm IR}$} & \colhead{HCN J=1--0 peak} & 
\colhead{HCN J=1--0 flux} & \colhead{HCN J=1--0 luminosity} \\  
\colhead{} & \colhead{} & \colhead{[10$^{6}$ L$_{\odot}$ Mpc$^{-2}$]} &
\colhead{[L$_{\odot}$]} & \colhead{[mJy]} & 
\colhead{[Jy km s$^{-1}$]} & \colhead{[10$^{7}$ K km s$^{-1}$ pc$^{2}$]} \\
\colhead{(1)} & \colhead{(2)} & \colhead{(3)} &
\colhead{(4)} & \colhead{(5)} & \colhead{(6)} & \colhead{(7)} 
}
\startdata 
NGC 4418 & 0.007 & 8.9 & 11.0 & 40 & 10 & 3.7 \\
UGC 5101 & 0.040 & 2.6 & 12.0 &  9 & 6.4 & 77 \\ 
Mrk 273  & 0.038 & 4.2 & 12.2 & 10 & 4.6  & 50 \\
IRAS 17208$-$0014 & 0.042 & 5.7 & 12.4 & 24 & 13 & 180 \\
Arp 220  & 0.018 & 19.1 & 12.2 & 52 & 33 & 81 \\
Mrk 231  & 0.042 & 8.2 & 12.5 & 38 & 10 & 130 \\
IRAS 08572$+$3915 & 0.058 & 1.7 & 12.1 &  4 & 2.2 & 56 \\
VV 114   & 0.020 & 5.2 & 11.7 & 15 & 6.9 & 21 \\
He 2--10 & 0.003 & 6.1 & 10.1 & \nodata & 6.5 & 0.43 \\
NGC 2623 & 0.018 & 4.5 & 11.5 & 13 & 5.7 & 14 \\
Mrk 266  & 0.028 & 1.7 & 11.5 & 9 & 2.0 & 12 \\
Arp 193  & 0.023 & 3.4 & 11.6 & 27 & 11 & 42 \\
NGC 1377 & 0.006 & 1.8 & 10.2 & 4 & 2.2 & 0.59\\
IC 694 (Arp 299) & 0.010 & 14.7 & 11.5 & 24 & 9.9 & 7.2 \\
NGC 3690 (Arp 299) &  0.010 & 9.8 & 11.4 & 9 & 3.0 & 2.2 \\
\enddata

\tablecomments{ 
Col.(1): Object name, observed with the pre-ALMA interferometric facility, 
Nobeyama Millimeter Array 
\citep{ima04,ima06b,in06,ima07b,ima09b}. 
Col.(2): Redshift. 
Col.(3): IRAS-measured infrared (8--1000 $\mu$m) flux, calculated with 
$f_{\rm IR}$ = 1.8 $\times$ 10$^{-14}$ $\times$ 
(13.48 $\times$ $f_{12}$ + 5.16 $\times$ $f_{25}$ + 
2.58 $\times$ $f_{60}$ + $f_{100}$) [W m$^{-2}$] \citep{sam96}, 
or equivalently 
$f_{\rm IR}$ = 4.5 $\times$ 10$^{4}$ $\times$ 
(13.48 $\times$ $f_{12}$ + 5.16 $\times$ $f_{25}$ + 
$2.58 \times f_{60} + f_{100}$) [10$^{6}$ L$_{\odot}$ Mpc$^{-2}$]. 
Col.(4): Decimal logarithm of IRAS-measured infrared (8$-$1000 $\mu$m)
luminosity in units of solar luminosity (L$_{\odot}$), calculated with 
$L_{\rm IR} = 2.1 \times 10^{39} \times$ D(Mpc)$^{2}$ 
$\times$ (13.48 $\times$ $f_{12}$ + 5.16 $\times$ $f_{25}$ + 
$2.58 \times f_{60} + f_{100}$) [ergs s$^{-1}$] \citep{sam96}, where 
we adopt H$_{0}$ $=$ 71 km s$^{-1}$ Mpc$^{-1}$, $\Omega_{\rm M}$ = 0.27, 
and $\Omega_{\rm \Lambda}$ = 0.73 \citep{kom09}, to estimate 
the luminosity distance D (Mpc) from the redshift. 
Col.(5): HCN J=1--0 emission peak in [mJy], based on Gaussian fit 
\citep{ima04,ima06b,in06,ima07b,ima09b}. 
Col.(6): HCN J=1--0 flux in [Jy km s$^{-1}$], based on Gaussian fit 
\citep{ima04,ima06b,in06,ima07b,ima09b}. 
Col.(7): HCN J=1--0 luminosity in [10$^{7}$ K km s$^{-1}$ pc$^{2}$], 
calculated using the equation (3) of \citet{sol05}. 
}

\end{deluxetable}

%%%%%%%%%% Table 8 %%%%%%%%%
\begin{deluxetable}{lcc}
%\tabletypesize{\small}
\tabletypesize{\scriptsize}
\tablecaption{Comparison of Expected and Observed HCN J=4--3 Emission 
Peaks \label{tbl-8}}  
\tablewidth{0pt}
\tablehead{
\colhead{Object} & \colhead{Expected} & \colhead{Observed} \\  
\colhead{} & \colhead{[mJy]} & \colhead{[mJy]} \\
\colhead{(1)} & \colhead{(2)} & \colhead{(3)} 
}
\startdata 
IRAS 08572$+$3915 & $\sim$70  & $\sim$10 \\
IRAS 12127$-$1412 & $\sim$15  & $\sim$2  \\ 
IRAS 00183$-$7111 & $\sim$10  & $\sim$2  \\
IRAS 20551$-$4250 & $\sim$110 & $\sim$50 \\
IRAS 22491$-$1808 & $\sim$40  & $\sim$20 \\
NGC 1614          & $\sim$340 & $\sim$20 \\
\enddata

\tablecomments{ 
Col.(1): Object name. 
Col.(2): Expected HCN J=4--3 peak flux in [mJy], in the case that 
(a) HCN J=1--0 emission peak follows the dashed line in Figure 11 
({\it Left}), and (b) HCN is thermalized at up to J=4--3, where the peak 
flux is 16 times higher at J=4--3 than at J=1--0. 
Col.(3): Observed HCN J=4--3 peak flux in [mJy].
}

\end{deluxetable}

\clearpage

%--- Figure 1 ---%
\begin{figure}
\begin{center}
\includegraphics[angle=0,scale=.42]{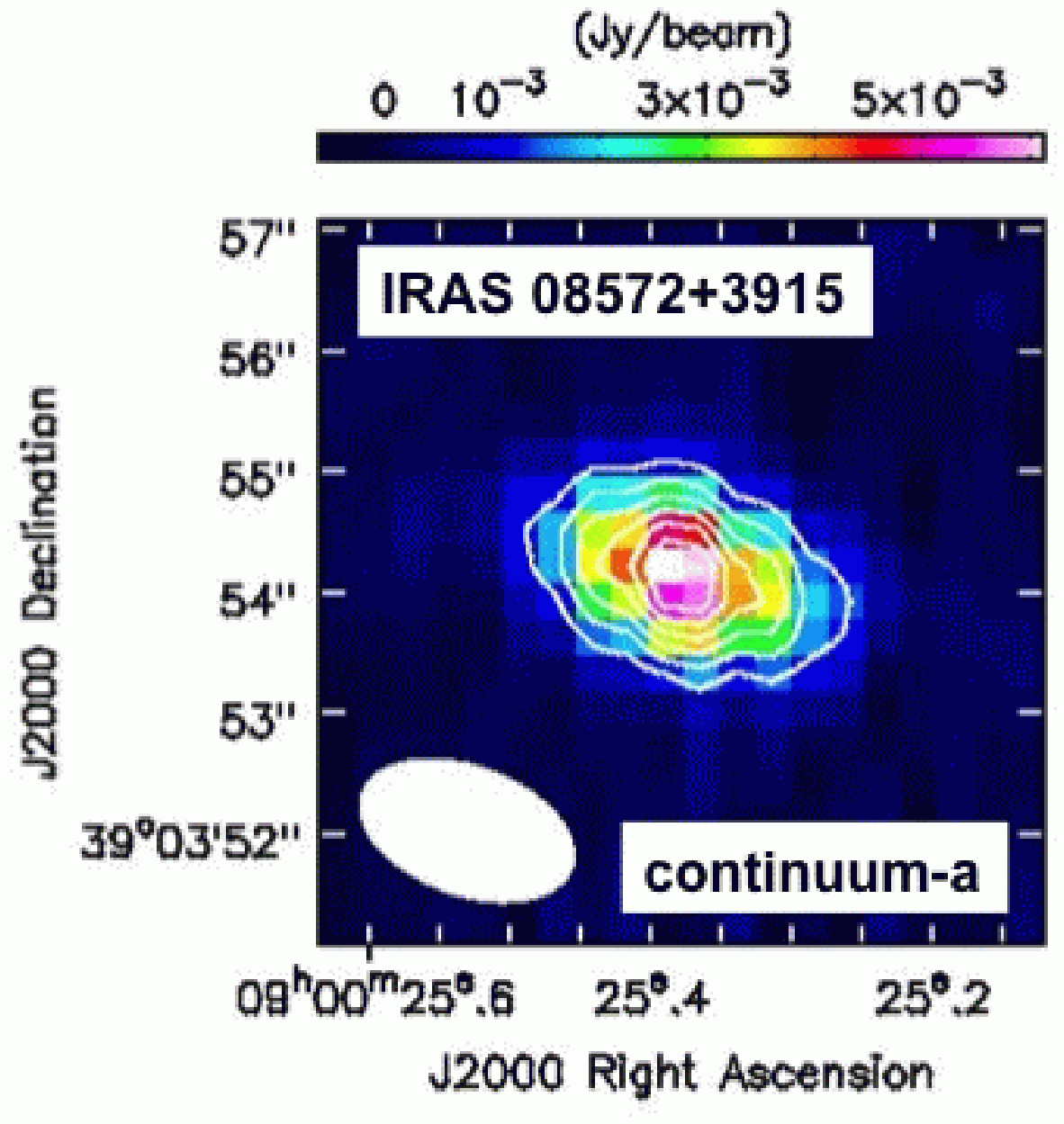} \\
\includegraphics[angle=0,scale=.42]{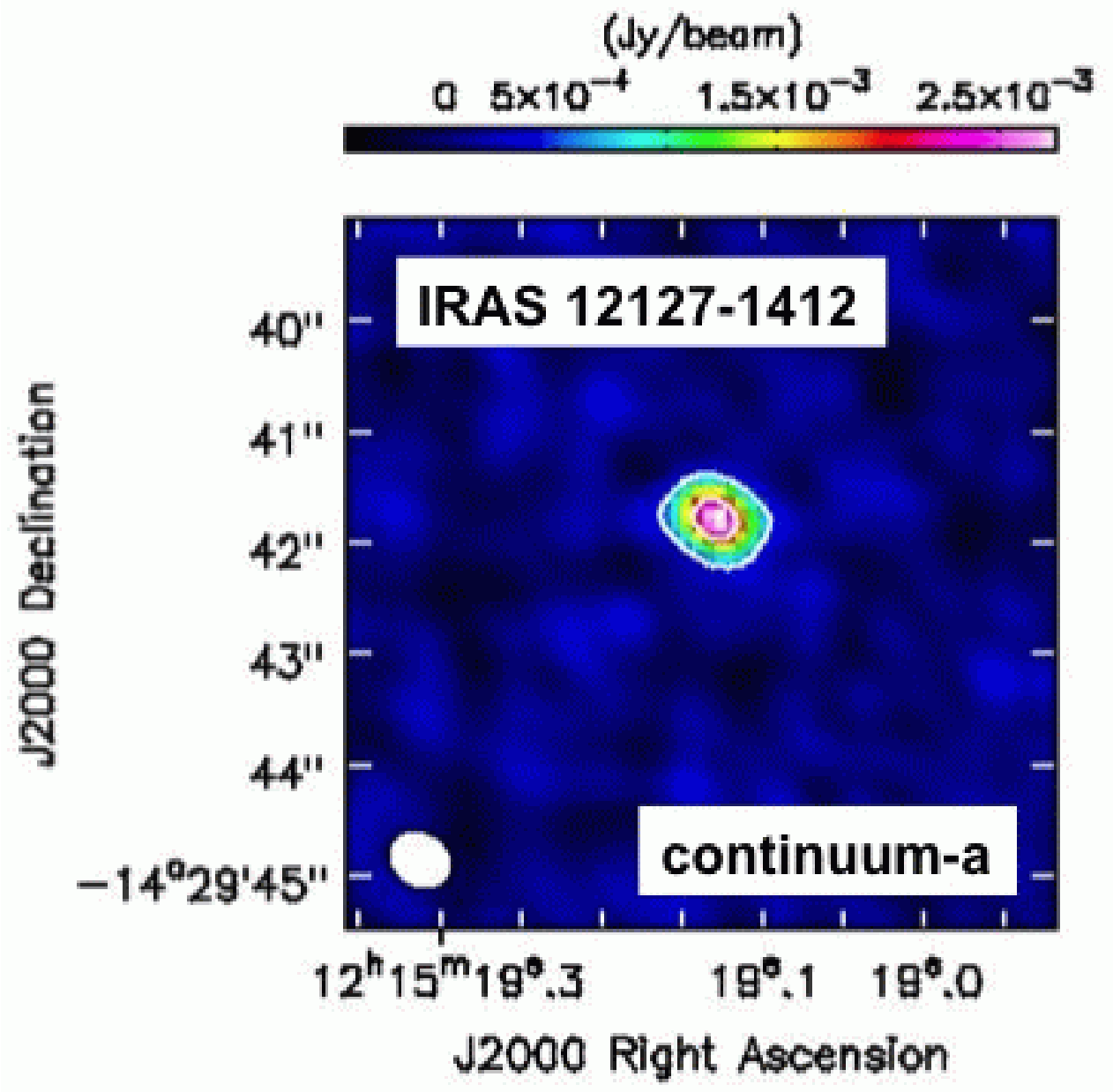} 
\includegraphics[angle=0,scale=.42]{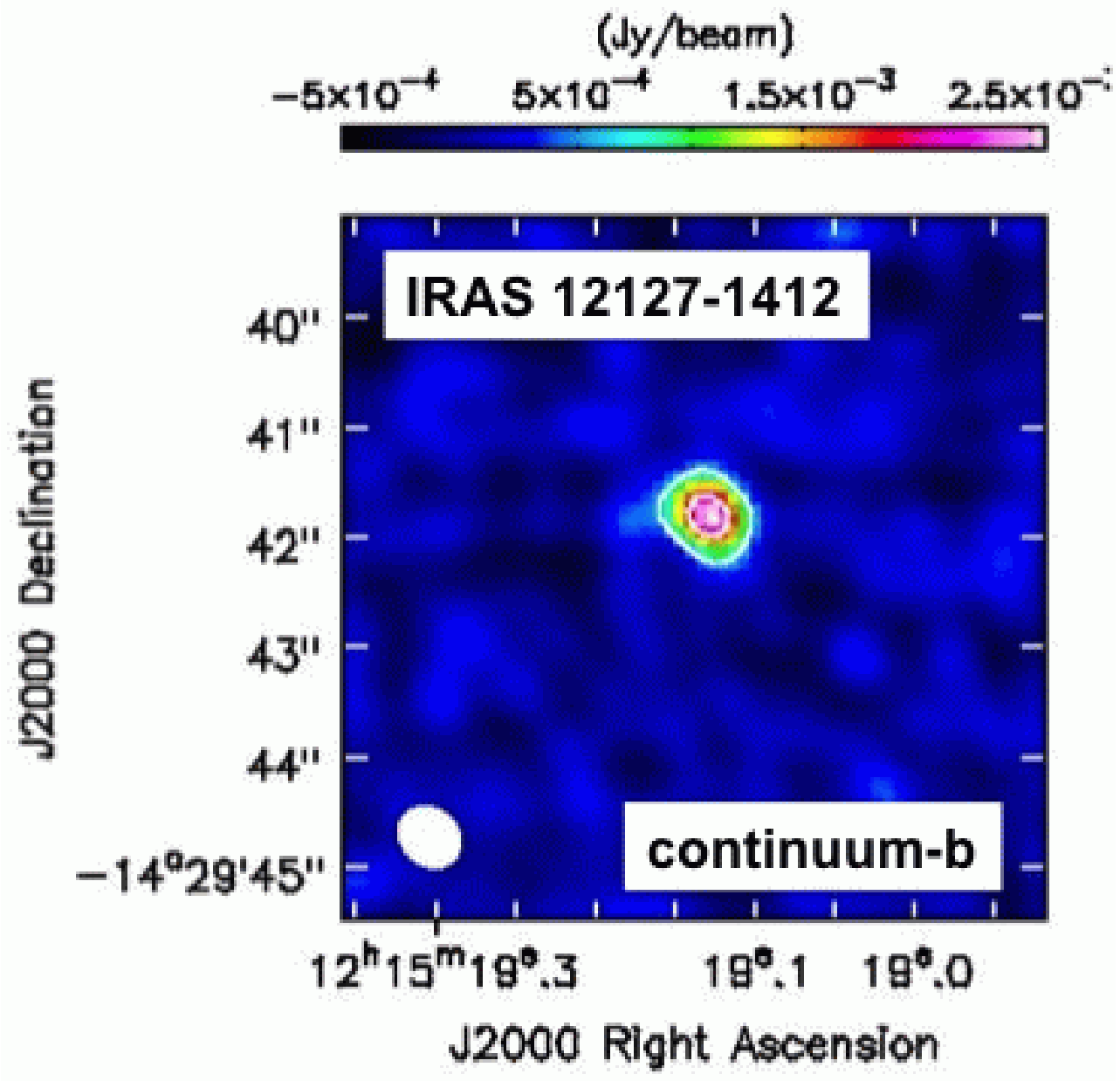} \\
\includegraphics[angle=0,scale=.42]{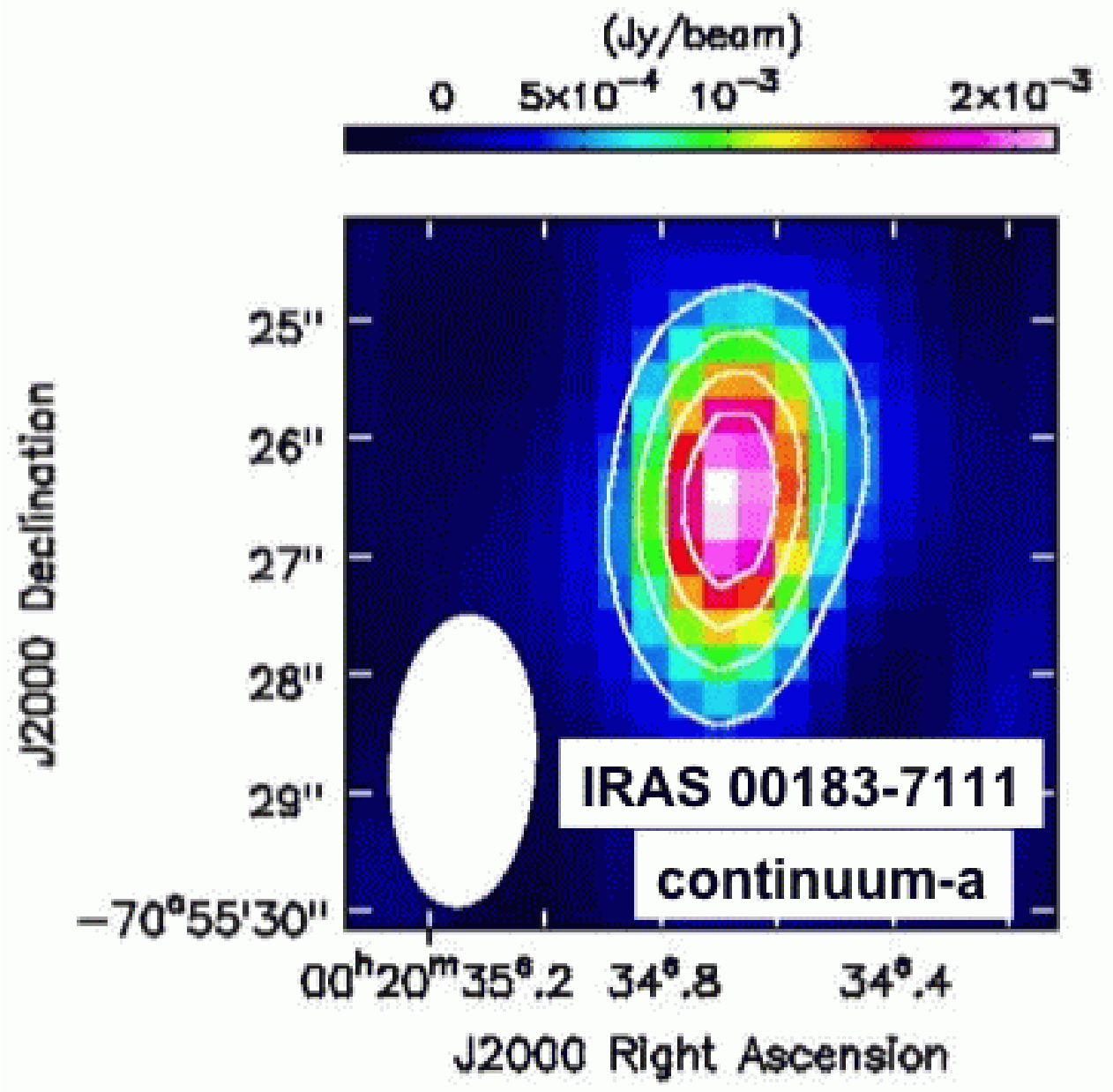} 
\includegraphics[angle=0,scale=.42]{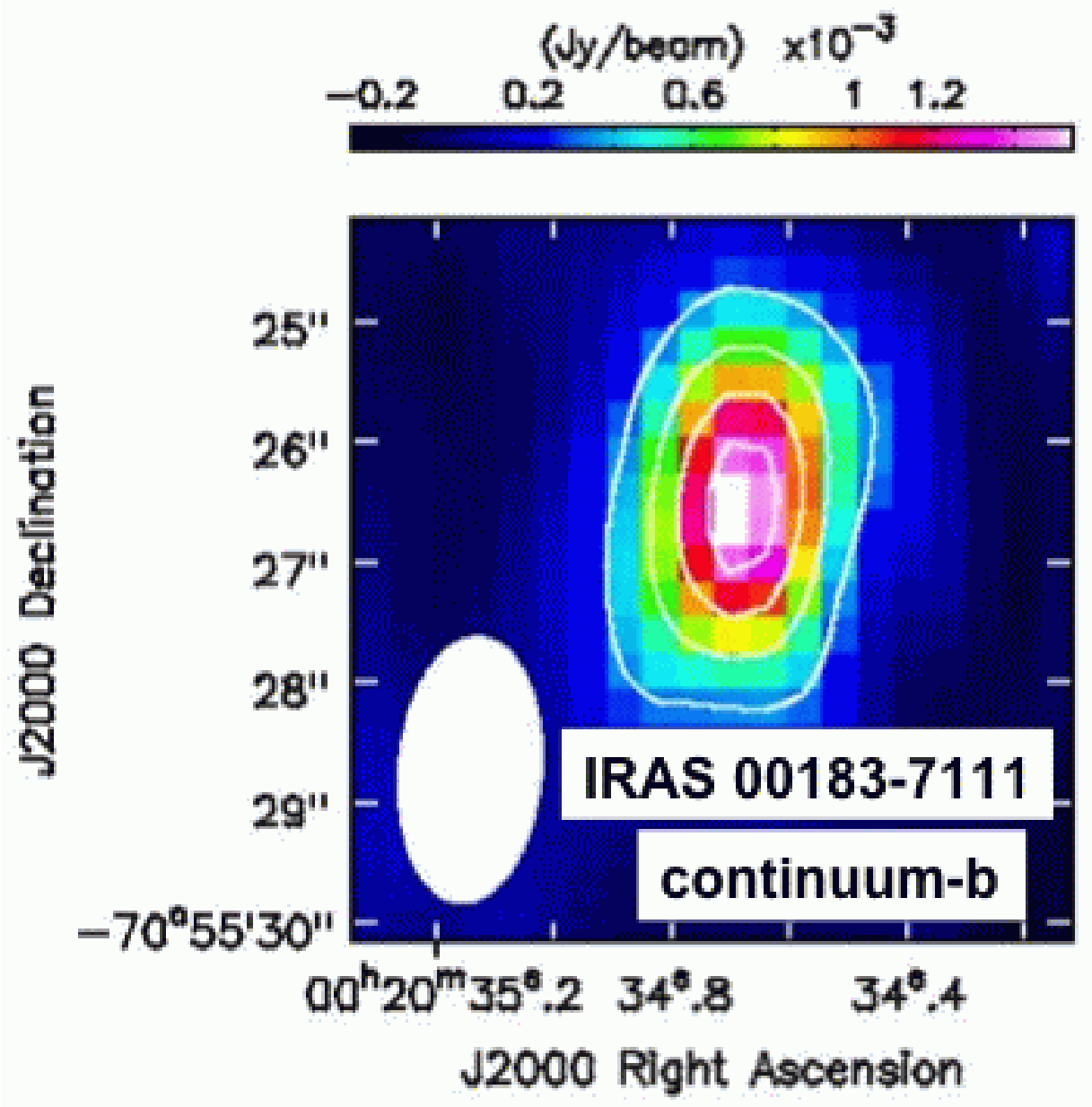} \\
\end{center}
\end{figure}

\clearpage

\begin{figure}
\begin{center}
\includegraphics[angle=0,scale=.42]{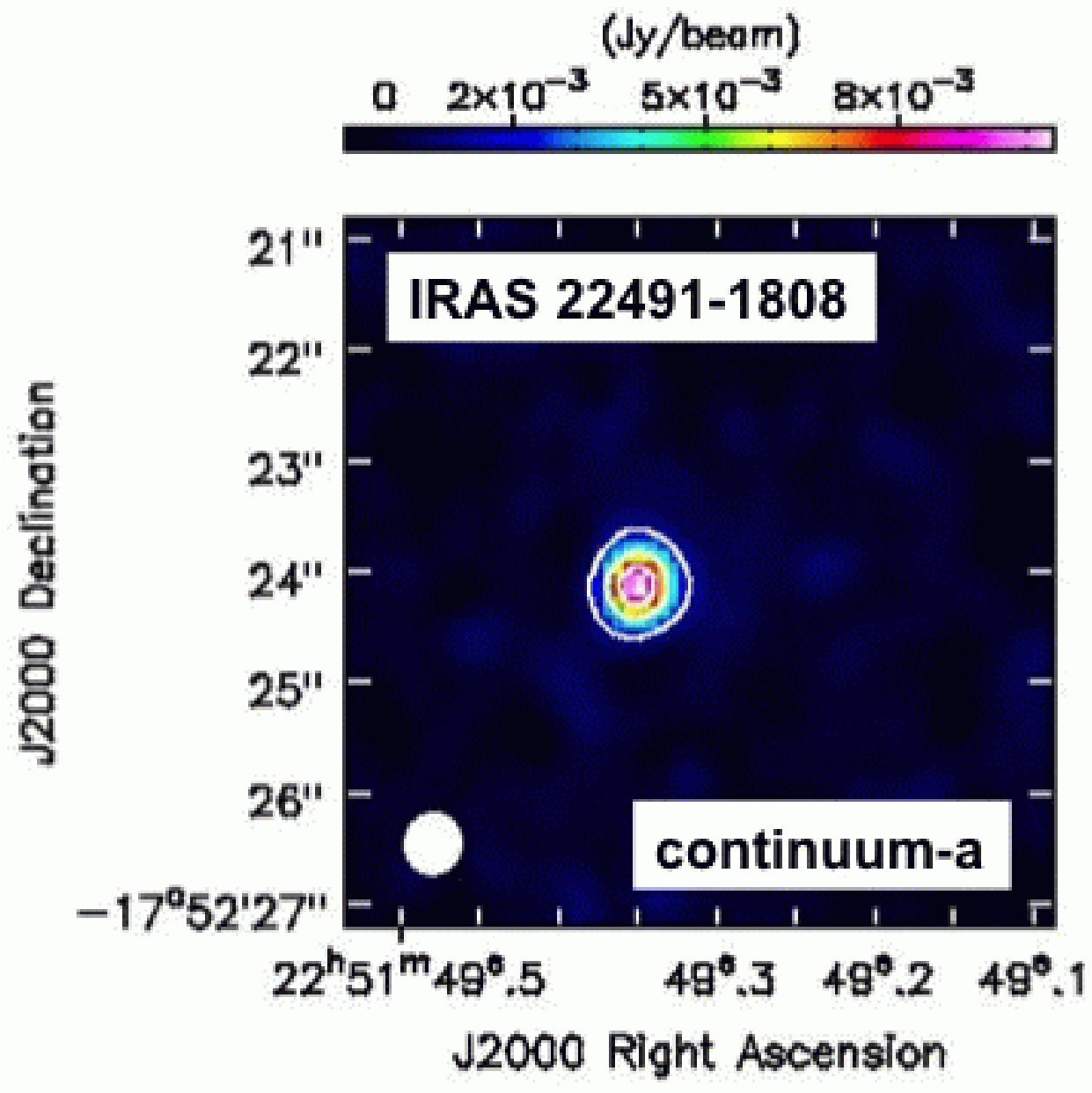} 
\includegraphics[angle=0,scale=.42]{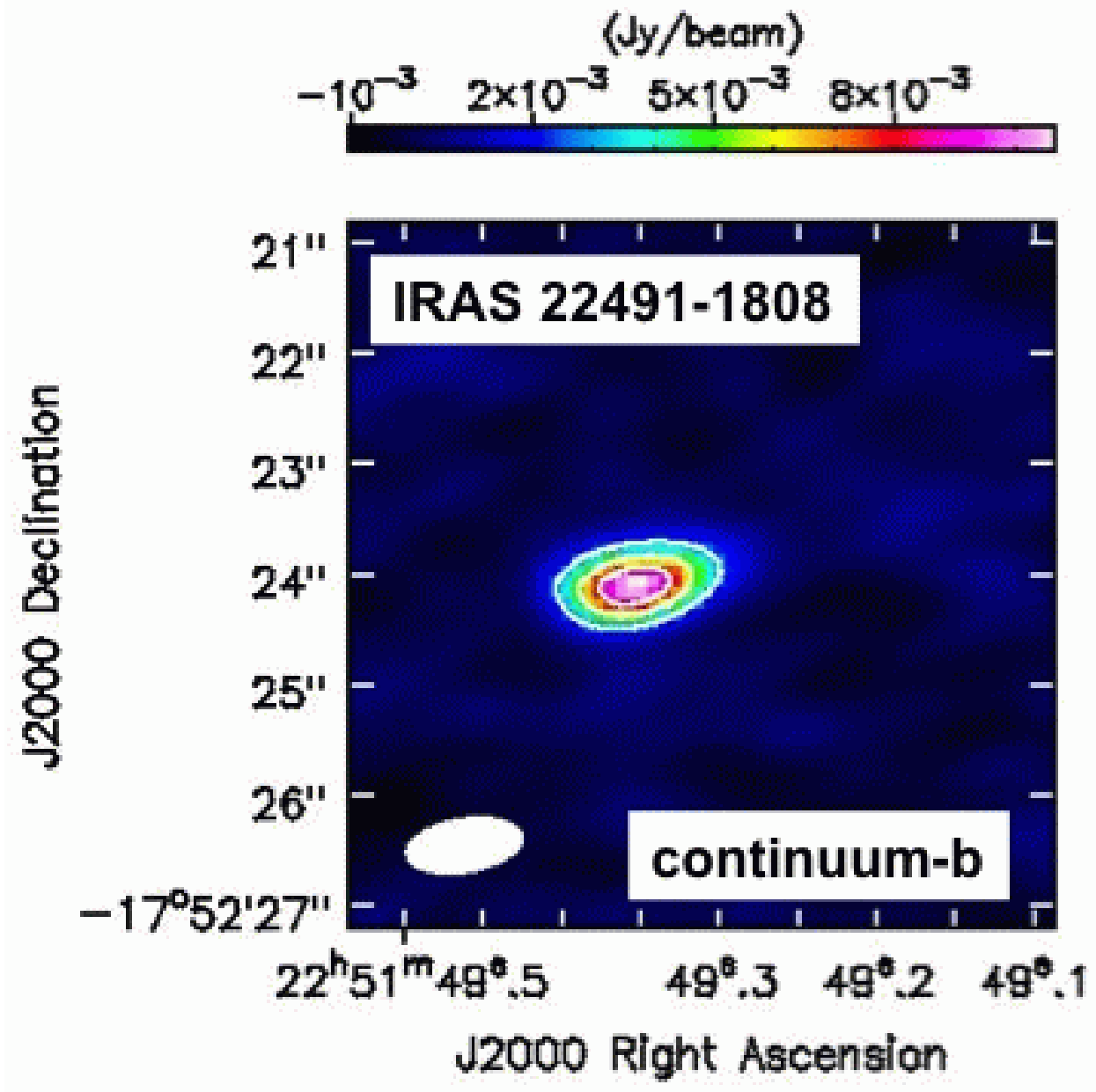} \\
\end{center}
\caption{
Continuum emission map. 
Continuum-a and -b data were taken simultaneously with 
HCN/HCO$^{+}$ and HNC observations, respectively. 
For IRAS 08572+3915, contours are 8$\sigma$, 13$\sigma$, 18$\sigma$, 
23$\sigma$, 28$\sigma$, 33$\sigma$, 38$\sigma$ for continuum-a. 
For IRAS 12127$-$1412, contours are 5$\sigma$, 20$\sigma$ for 
continuum-a, and 5$\sigma$, 15$\sigma$ for continuum-b. 
For IRAS 00183$-$7111, contours are 5$\sigma$, 10$\sigma$, 15$\sigma$, 
20$\sigma$ for continuum-a, and 5$\sigma$, 10$\sigma$, 15$\sigma$, 
20$\sigma$ for continuum-b. 
For IRAS 22491$-$1808, contours are 10$\sigma$, 30$\sigma$, 50$\sigma$ 
for continuum-a, and 10$\sigma$, 20$\sigma$, 30$\sigma$ for continuum-b. 
The rms noise level is different in each image, and is summarized in 
Table 4.
}
\end{figure}

\clearpage

%--- Figure 2 ---%
\begin{figure}
\includegraphics[angle=0,scale=.43]{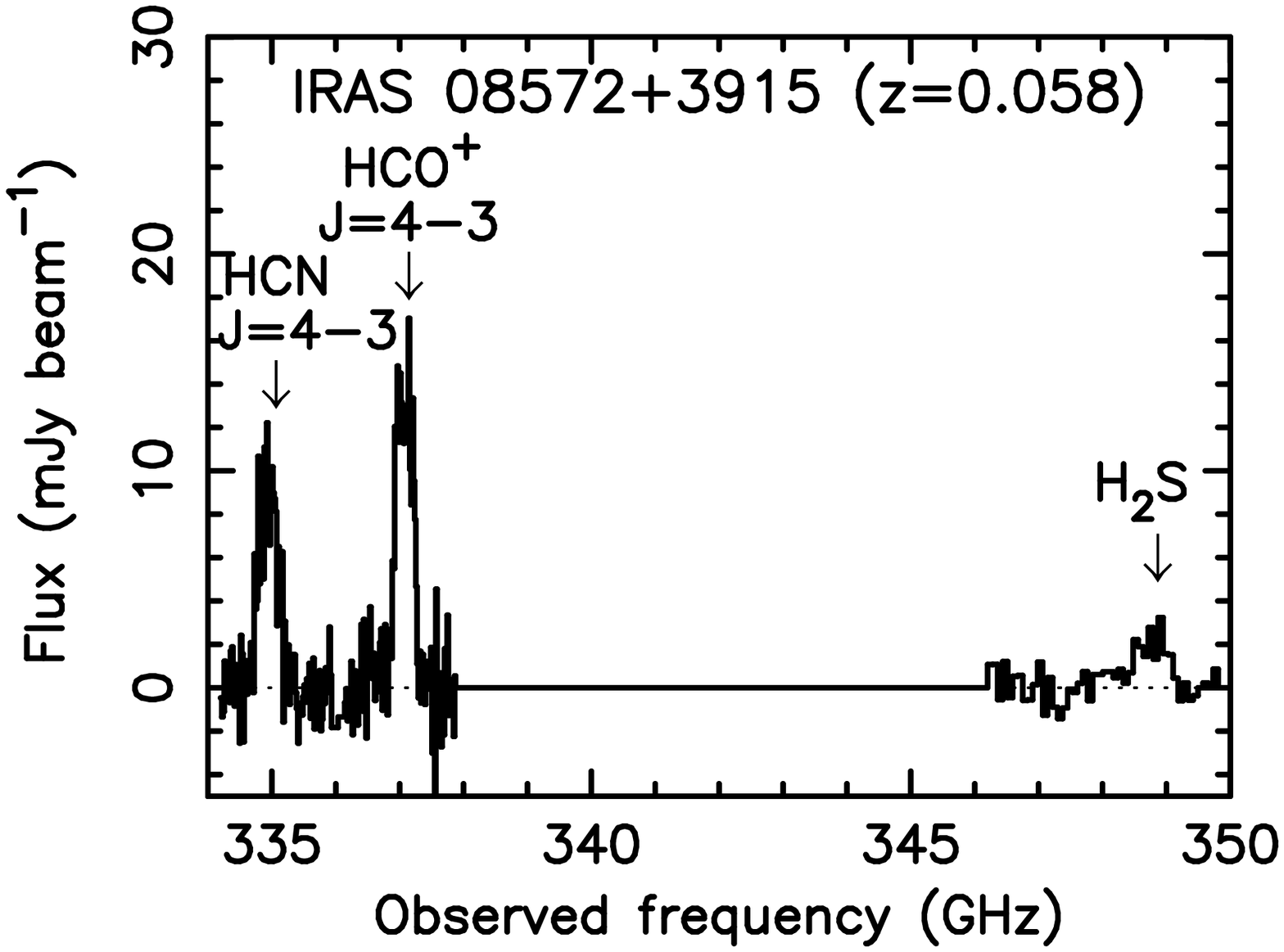} 
\includegraphics[angle=0,scale=.43]{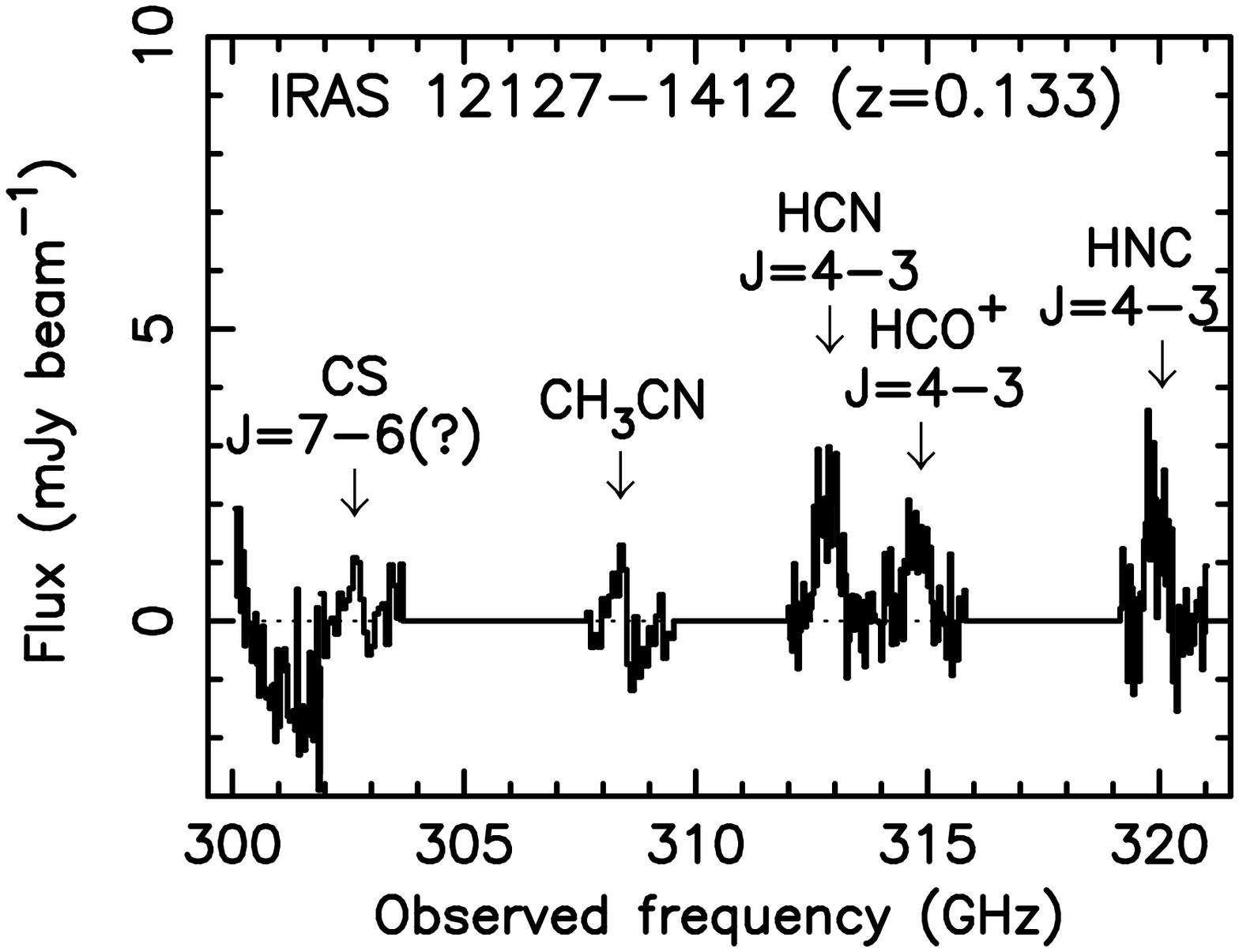} \\
\includegraphics[angle=0,scale=.43]{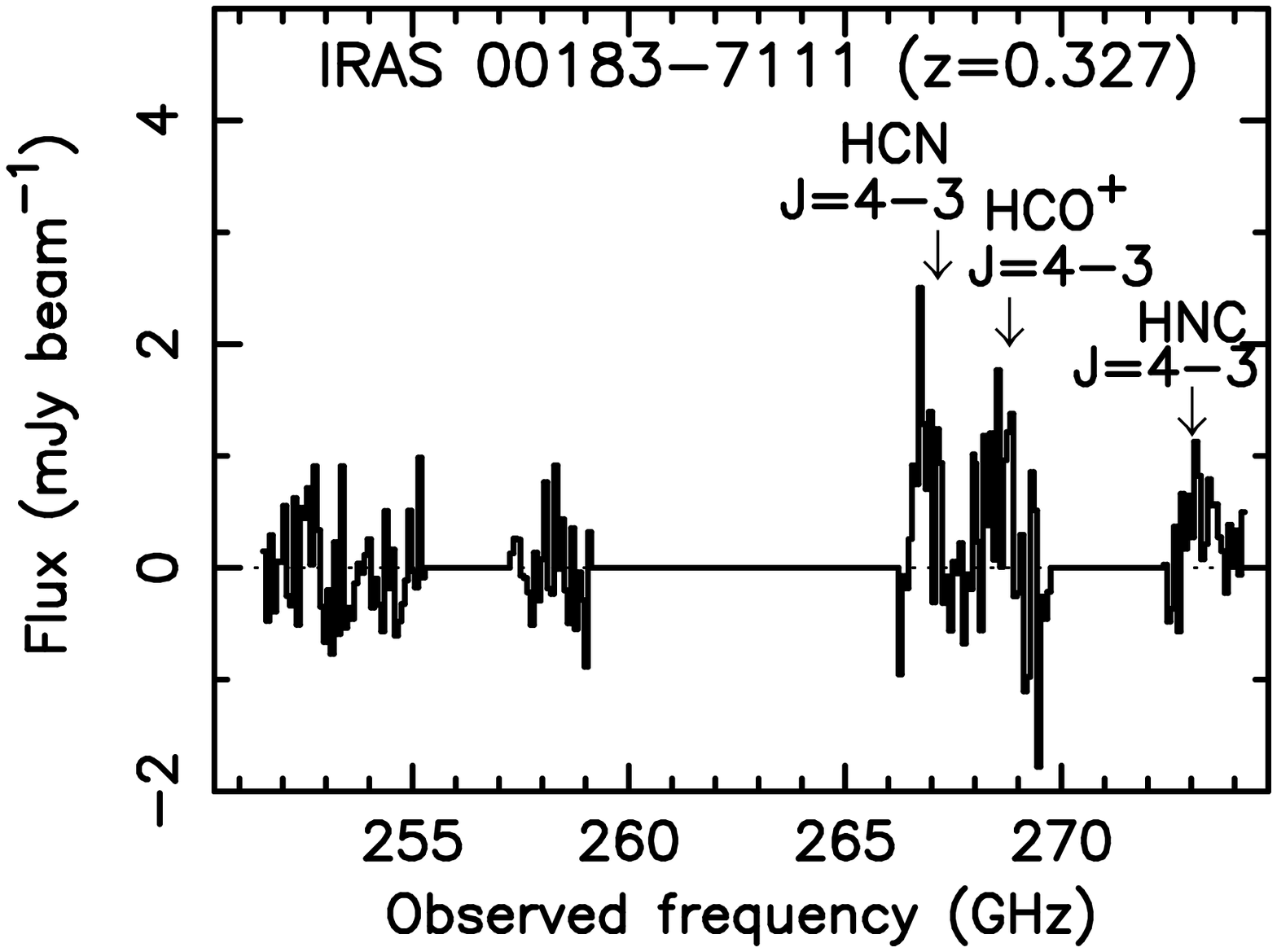} 
\includegraphics[angle=0,scale=.43]{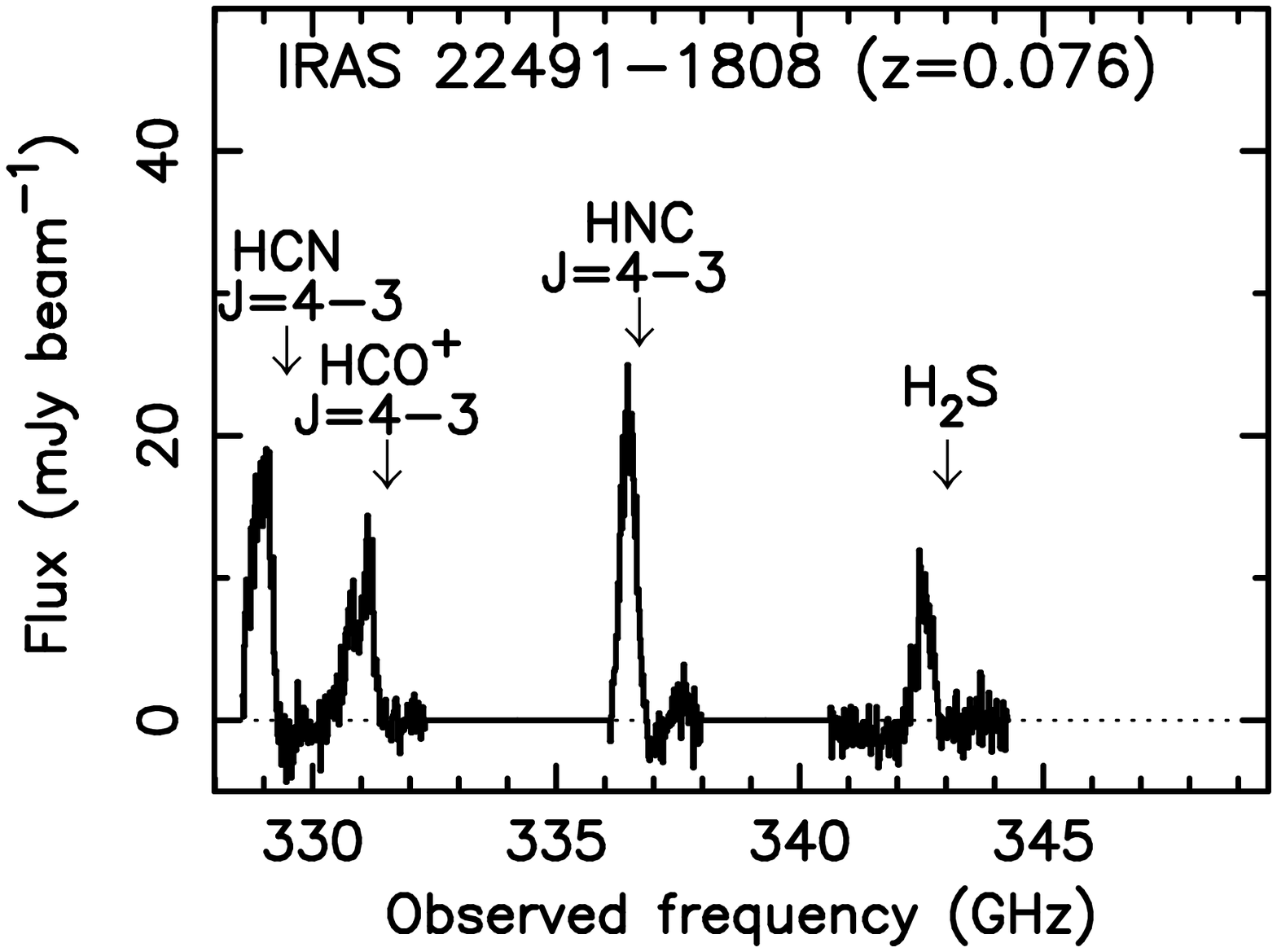} 
\caption{
Full-frequency coverage spectra at the continuum peak position within 
the beam size. 
The abscissa shows the observed frequency in [GHz], and the ordinate shows flux 
in [mJy beam$^{-1}$]. 
For HCN, HCO$^{+}$, H$_{2}$S, and CS lines, 
spectra are extracted at the continuum-a peak positions. 
For HNC and CH$_{3}$CN lines, spectra are extracted at the 
continuum-b peak positions. 
Since each spectrum was taken from two separate observations
(Tables 2 and 3), actually achieved noise level is not uniform
throughout the observed frequency.
The expected frequencies of individual emission lines from optical 
redshifts are shown as downward arrows. 
For IRAS 22491$-$1808, the molecular gas emission peaks are 
significantly shifted to lower frequency than those expected from the 
adopted optical redshift of $z$=0.076 \citep{kim98}.
}
\end{figure}

%--- Figure 3 ---%
\begin{figure}
%\vspace*{-1.9cm}
\begin{center}
\includegraphics[angle=0,scale=.42]{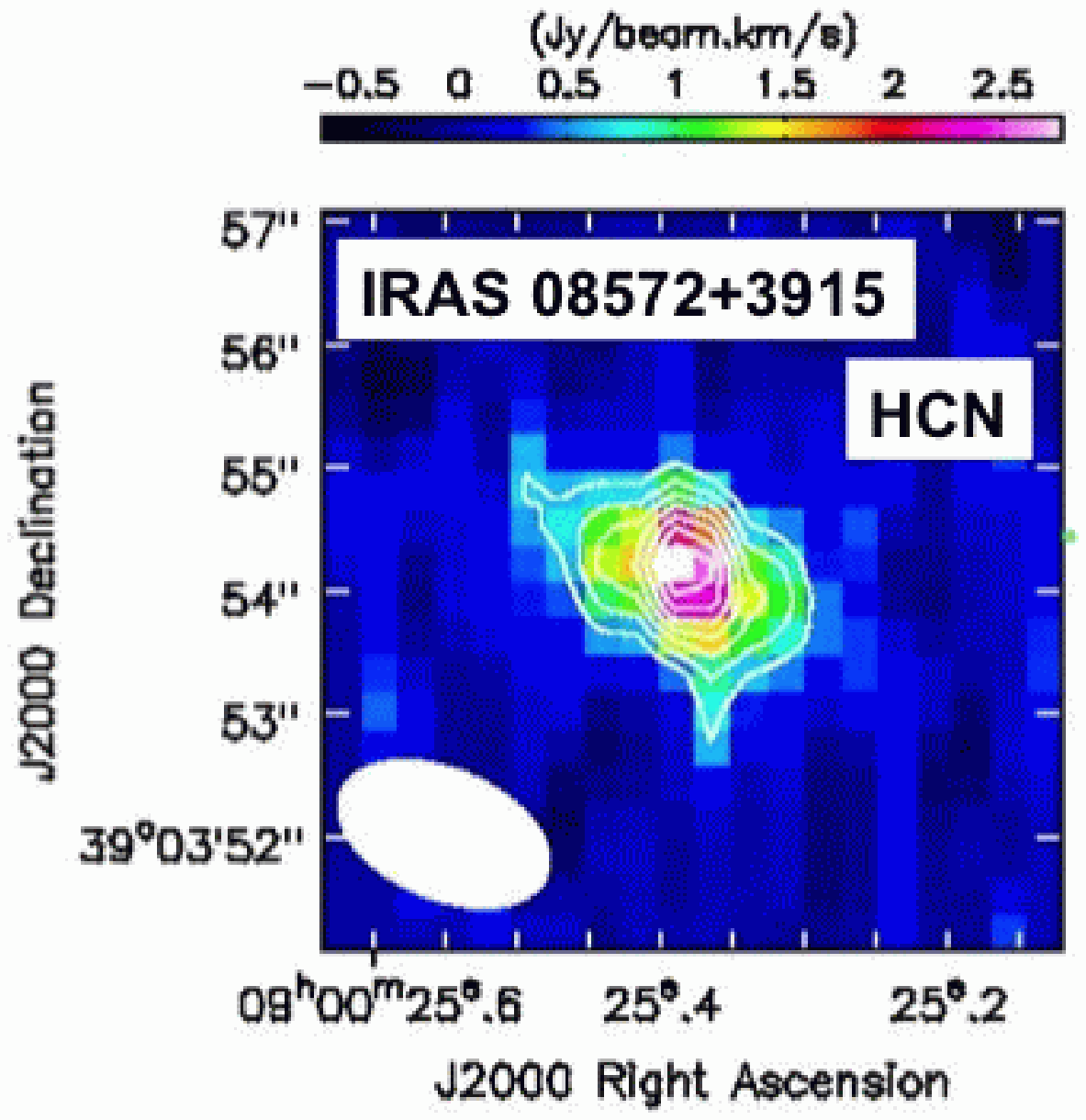} 
\includegraphics[angle=0,scale=.42]{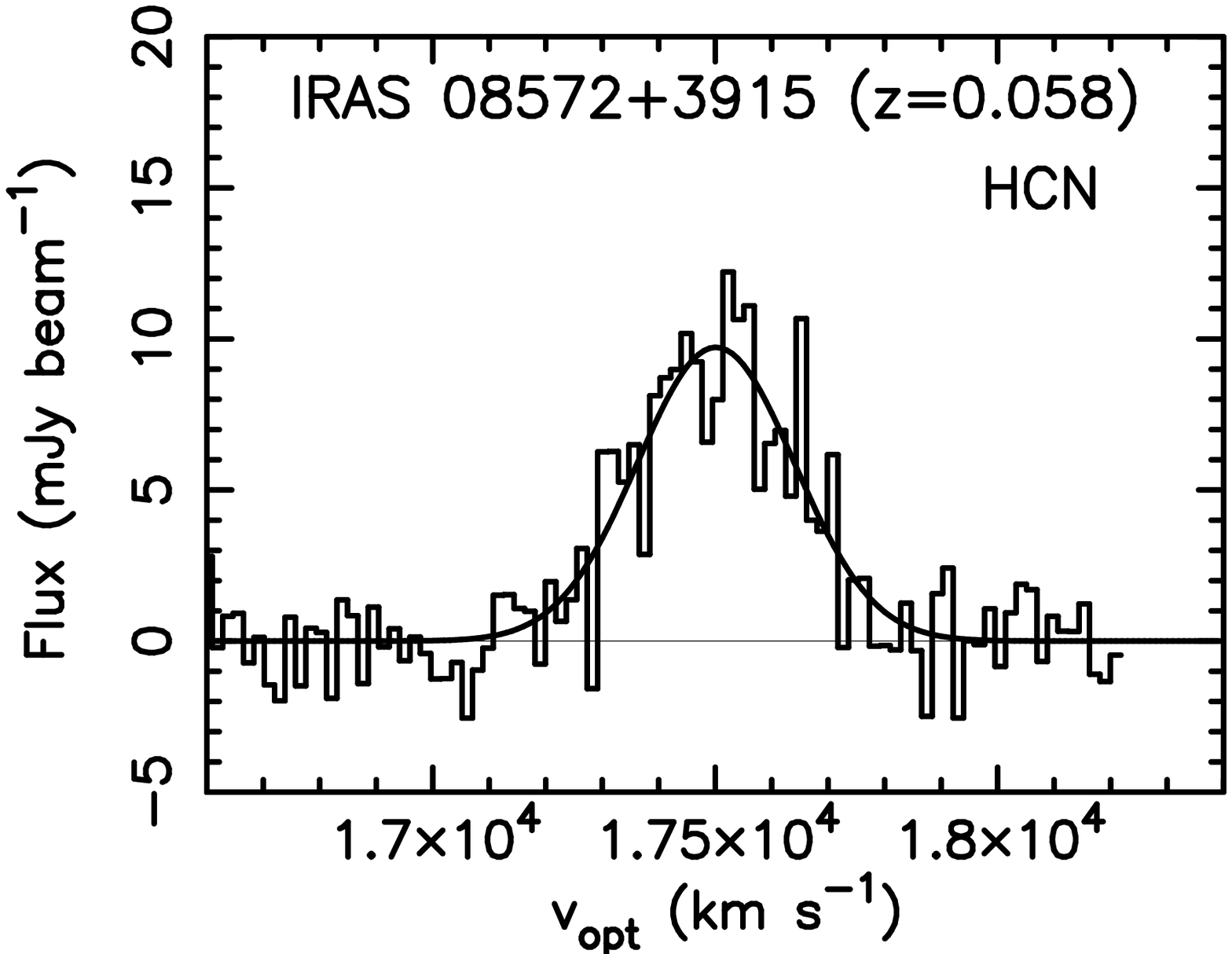}\\ 
\includegraphics[angle=0,scale=.42]{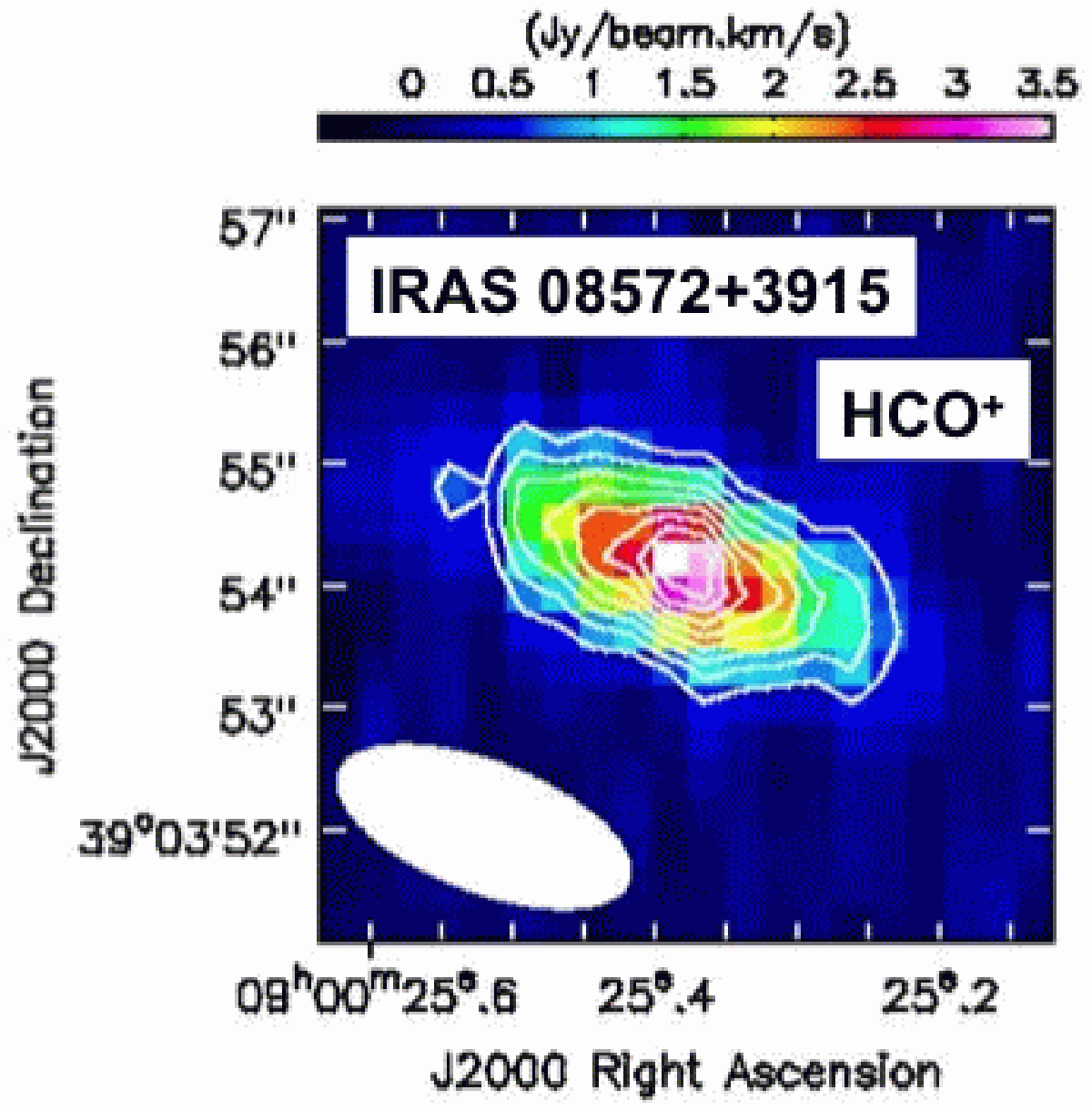} 
\includegraphics[angle=0,scale=.42]{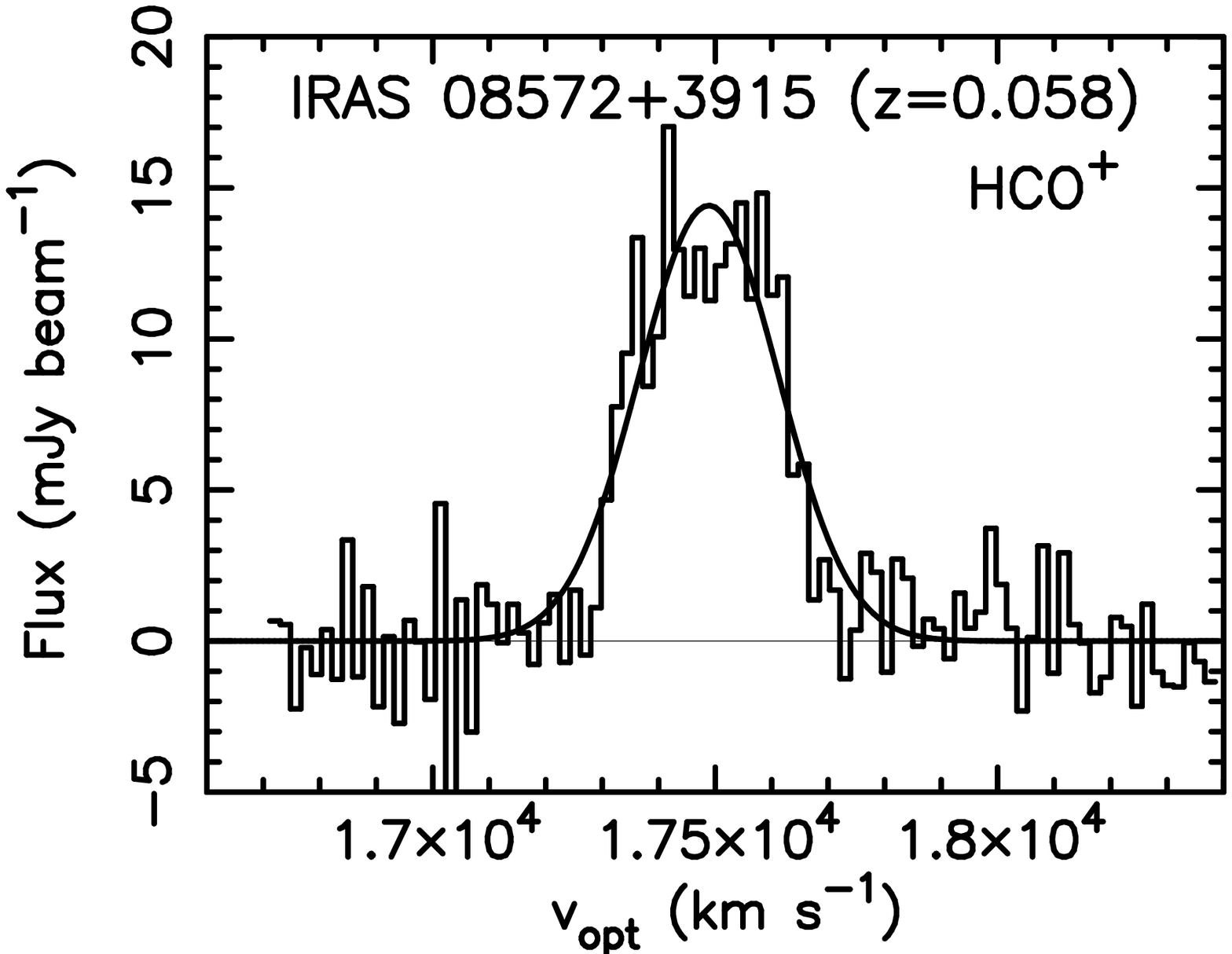}\\ 
\includegraphics[angle=0,scale=.42]{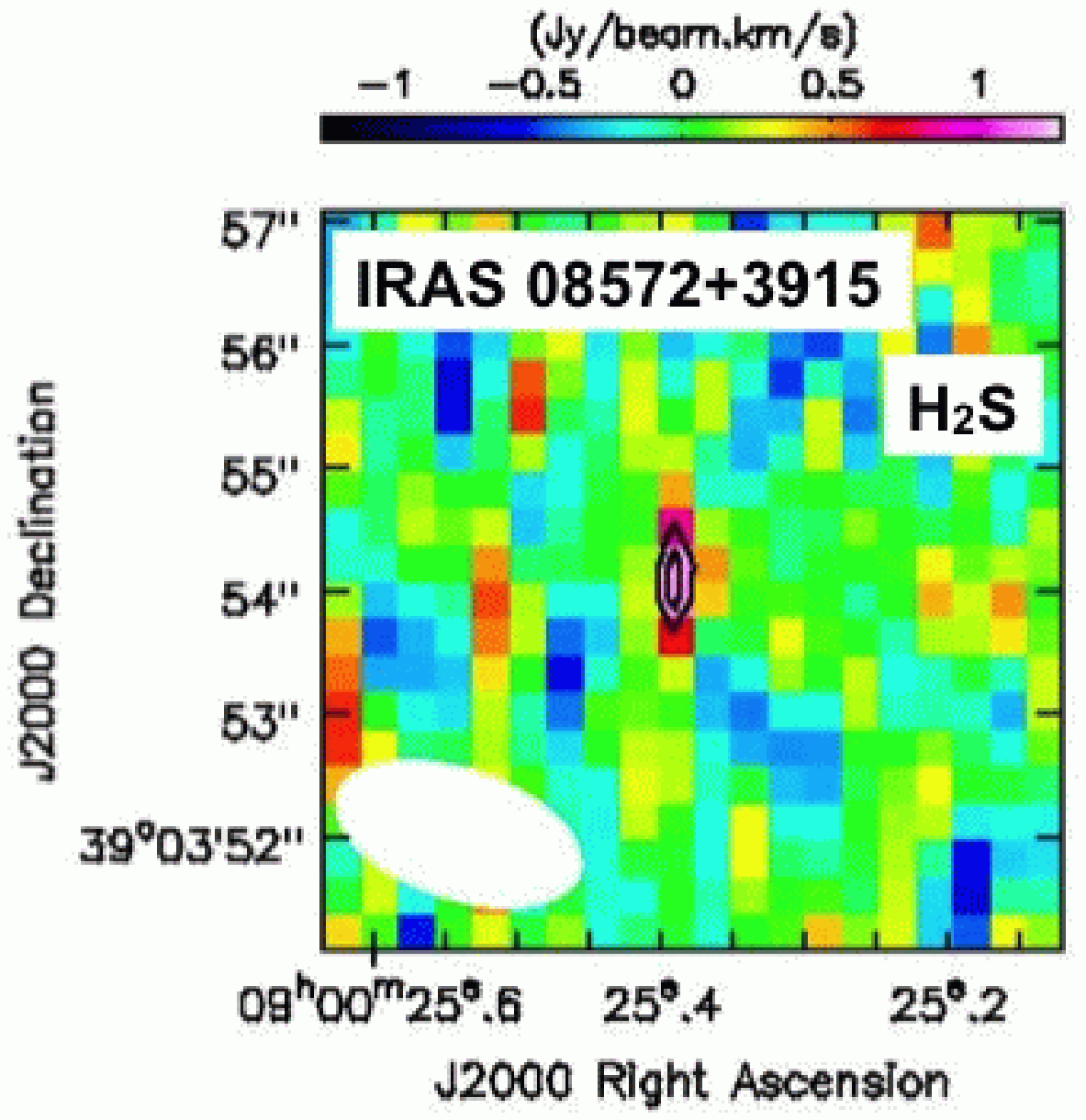} 
\includegraphics[angle=0,scale=.42]{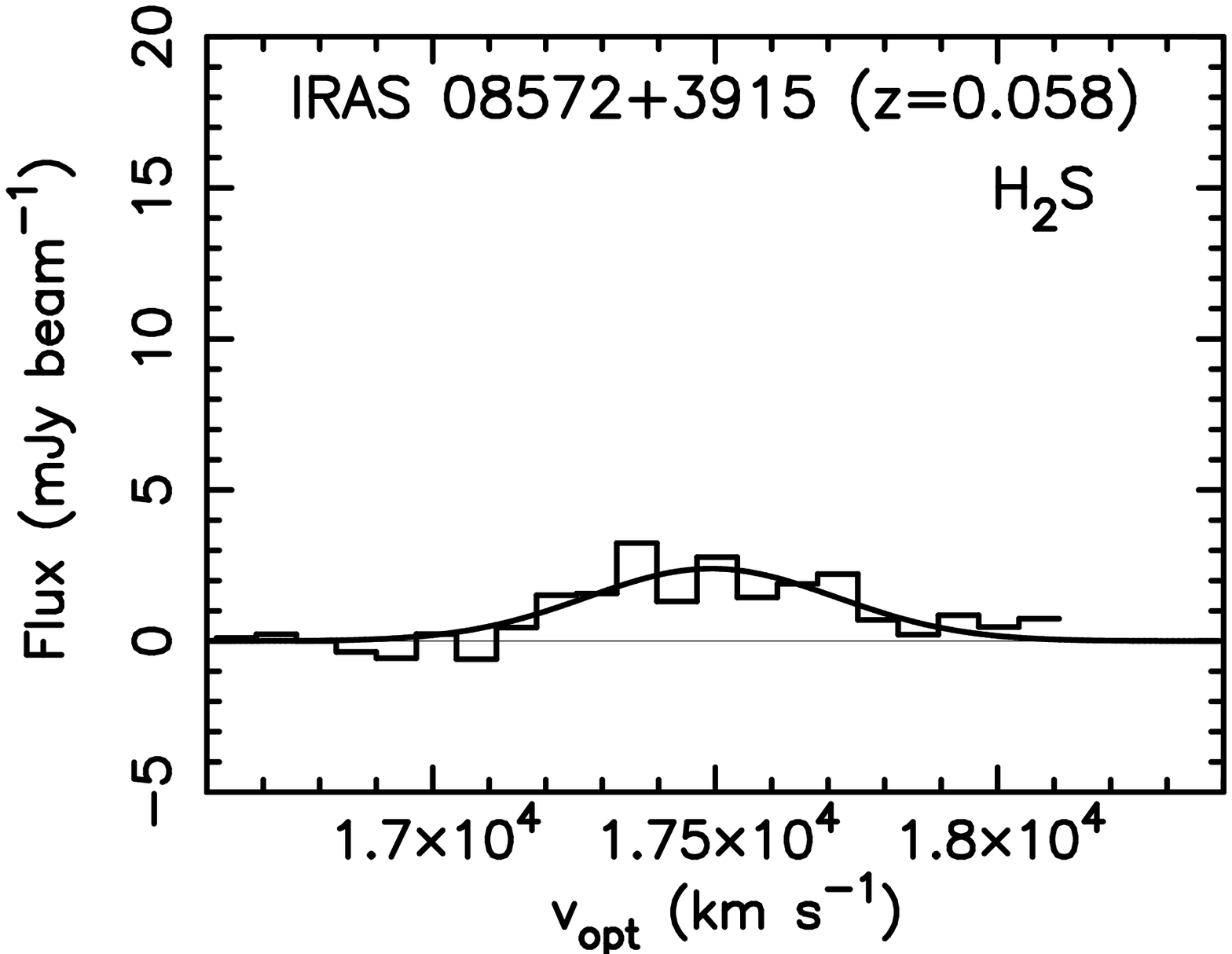}\\ 
\end{center}
%\vspace{-0.9cm}
\caption{
Integrated intensity (moment 0) maps (left) and spectra at the continuum 
peak position, within the beam size (right), of HCN, HCO$^{+}$, and 
H$_{2}$S for IRAS 08572$+$3915. 
For moment 0 maps, contours are 4$\sigma$, 6$\sigma$, 8$\sigma$, 
10$\sigma$, 12$\sigma$, 14$\sigma$, 16$\sigma$ for HCN, 
5$\sigma$, 8$\sigma$, 11$\sigma$, 14$\sigma$, 17$\sigma$, 20$\sigma$, 
23$\sigma$, 26$\sigma$ for HCO$^{+}$, 
3$\sigma$, 4$\sigma$ for H$_{2}$S. 
Detection of the H$_{2}$S emission is marginal and confirmation is required, 
because the signal spatial profile differs significantly from the beam
pattern.  
For the spectra, the abscissa shows v$_{\rm opt}$ $\equiv$ c 
($\lambda-\lambda_{\rm 0}$)/$\lambda_{\rm 0}$ in [km s$^{-1}$], and the 
ordinate shows flux in [mJy beam$^{-1}$].
The best Gaussian fits (Table 5) are overplotted as solid curved lines.
The spectrum of H$_{2}$S is shown with four spectral elements binning.
}
\end{figure}

%--- Figure 4 ---%
\begin{figure}
\begin{center}
\includegraphics[angle=0,scale=.42]{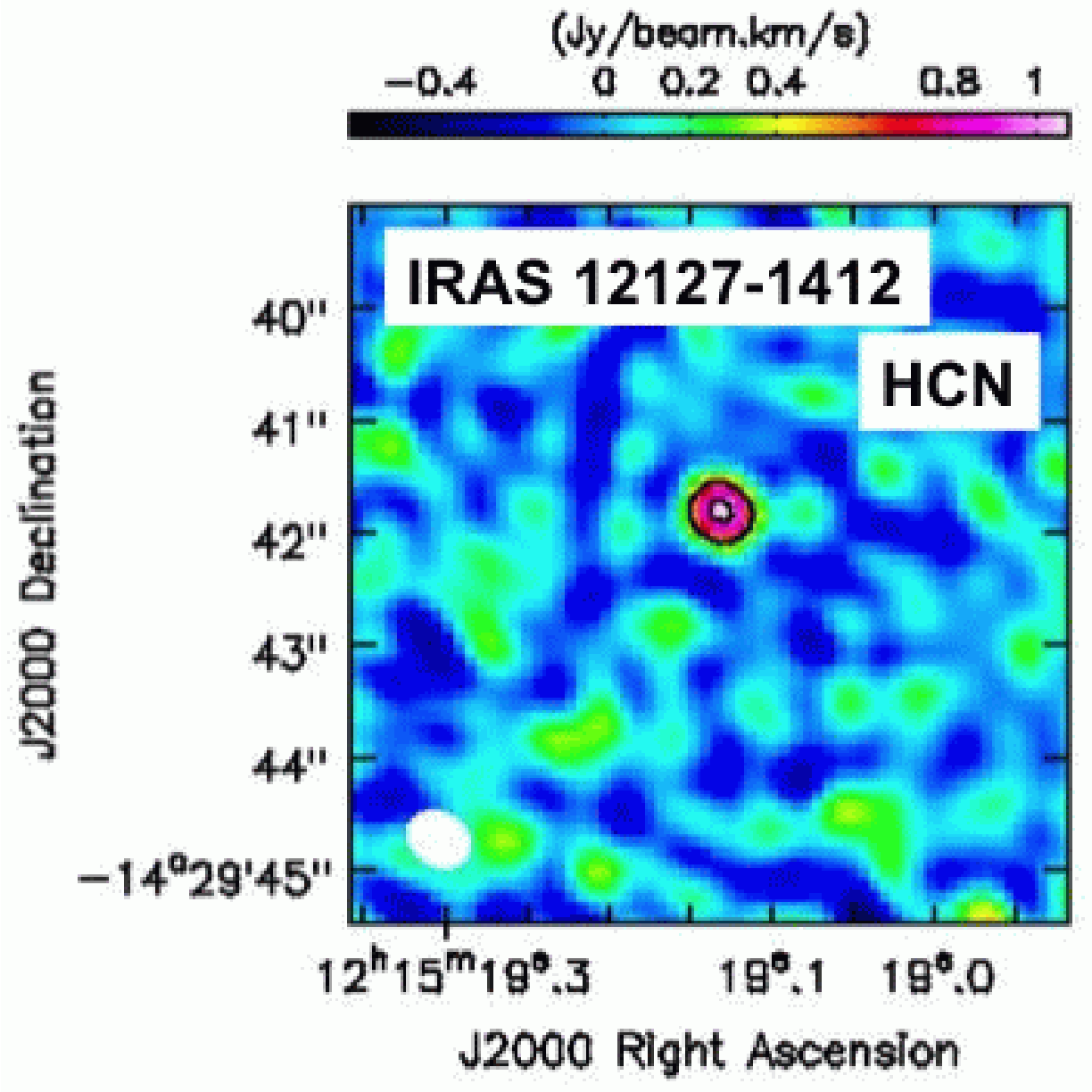} 
\includegraphics[angle=0,scale=.42]{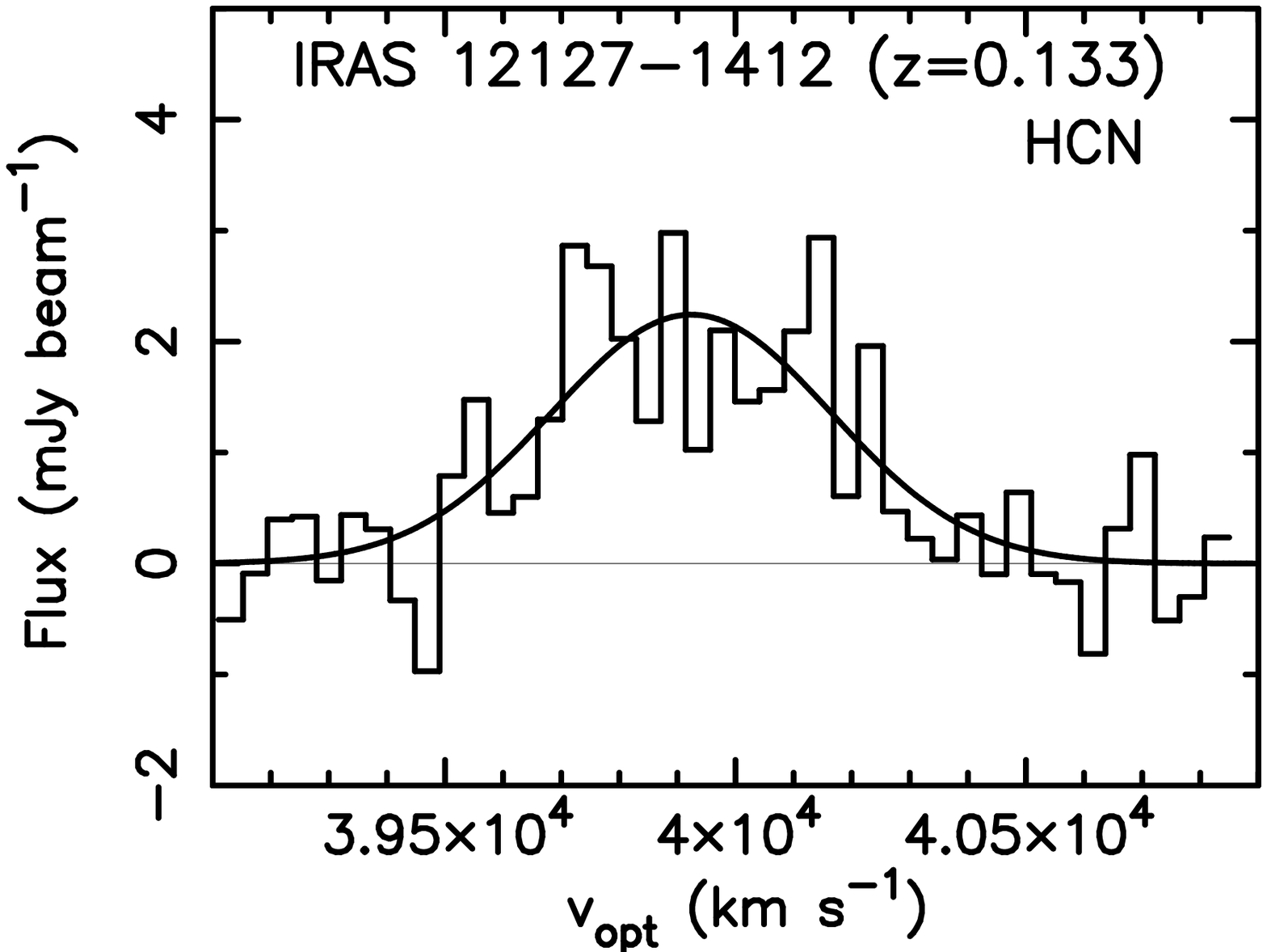}\\ 
\includegraphics[angle=0,scale=.42]{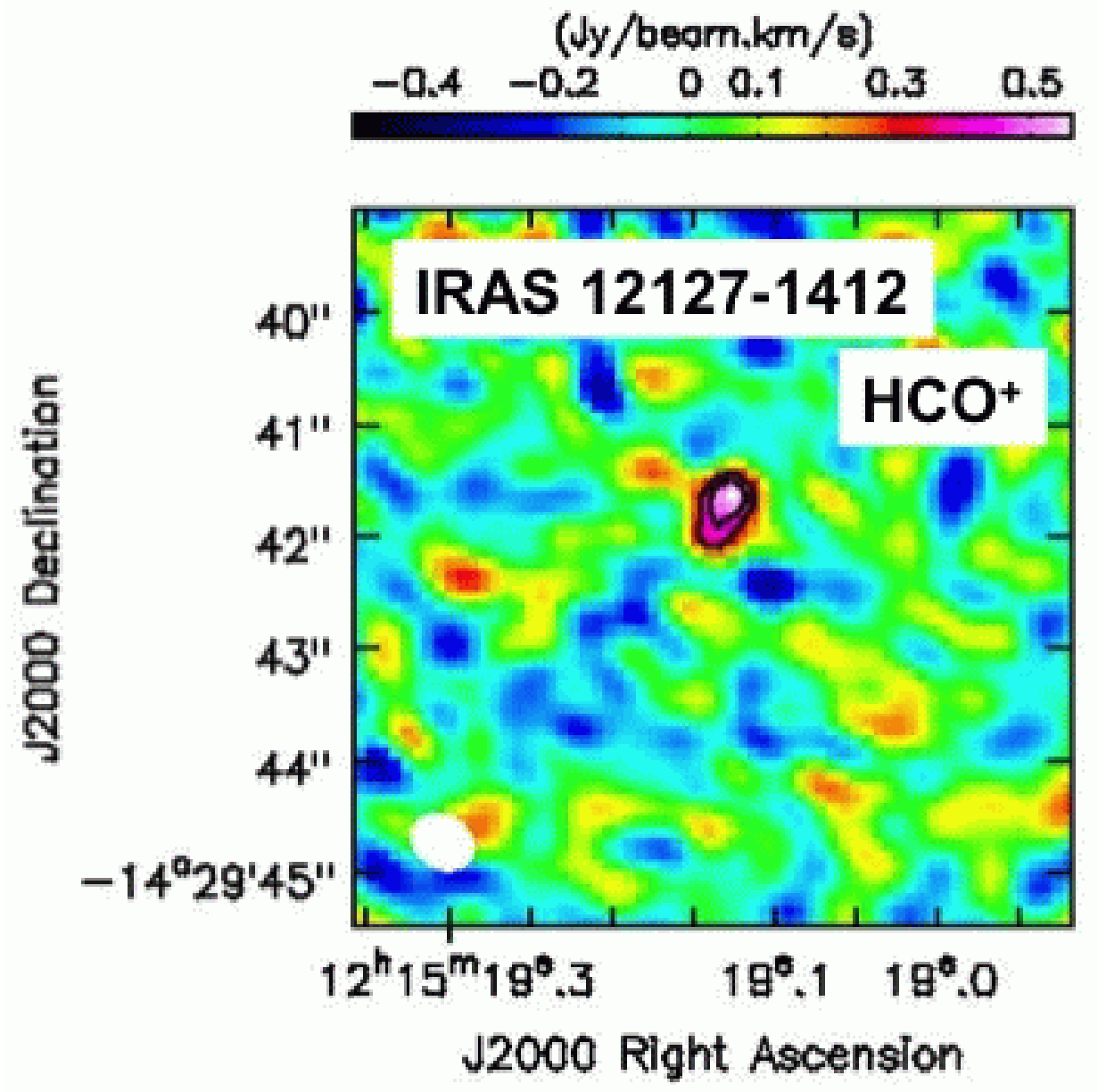} 
\includegraphics[angle=0,scale=.42]{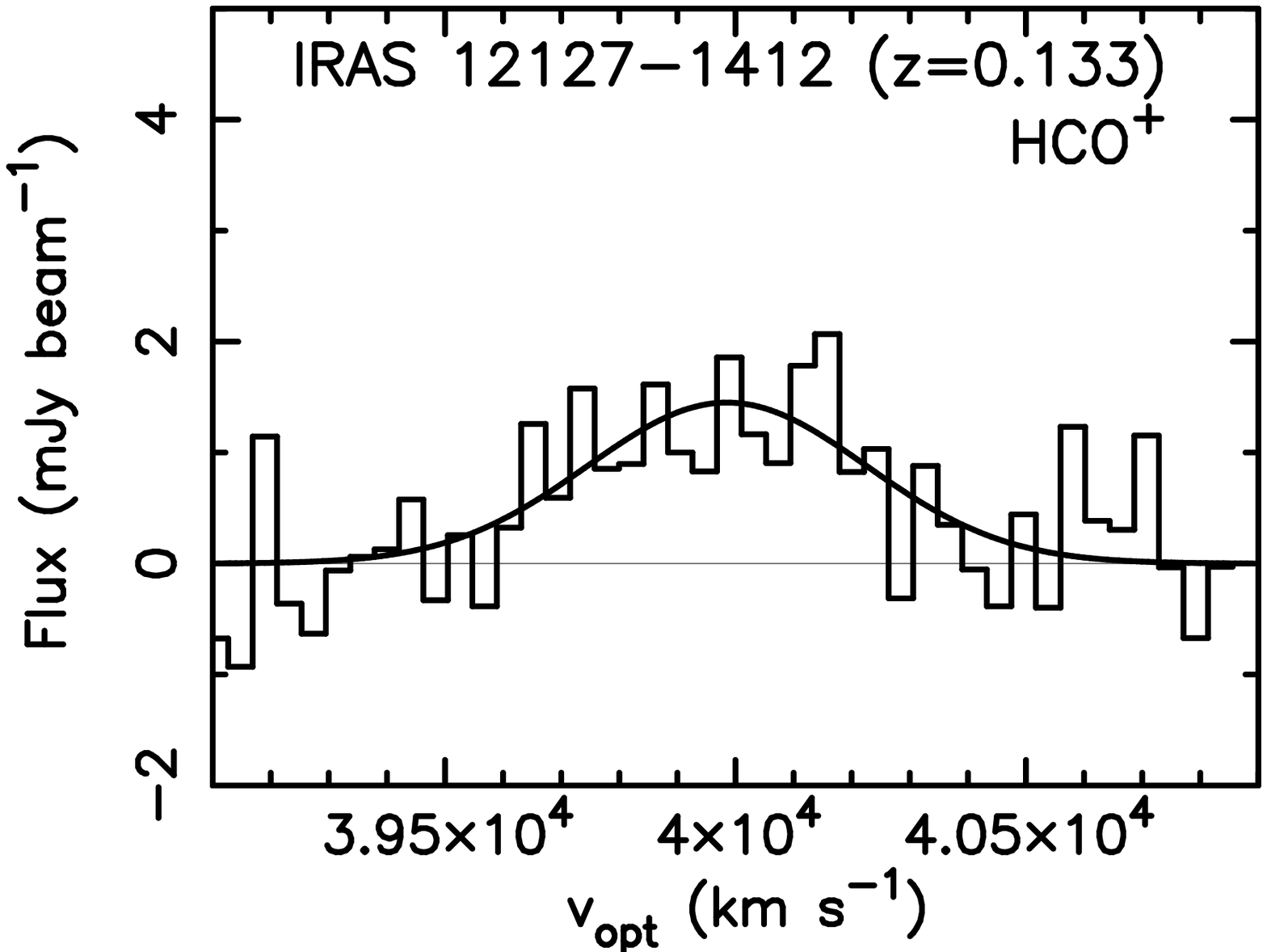}\\ 
\includegraphics[angle=0,scale=.42]{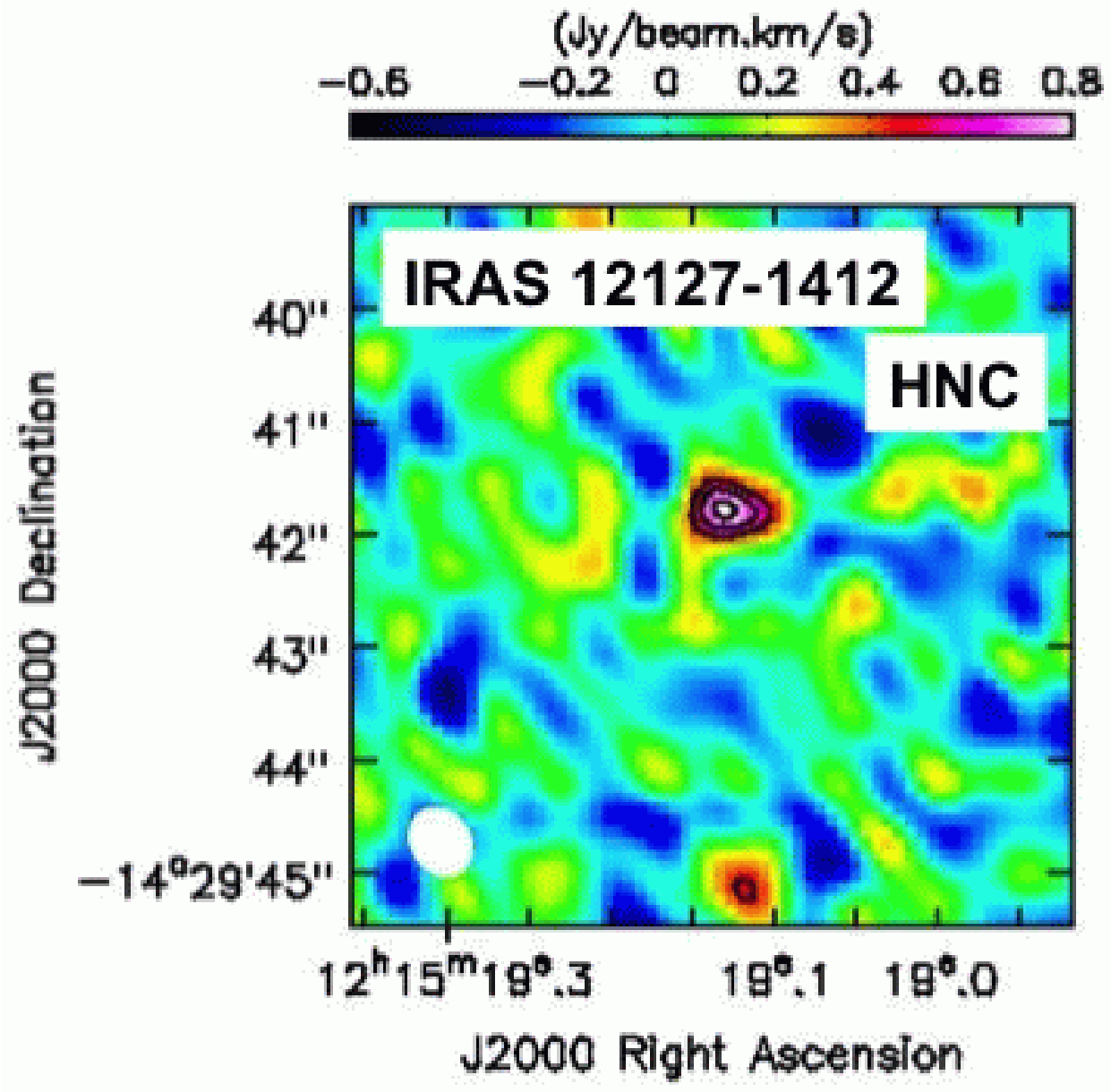} 
\includegraphics[angle=0,scale=.42]{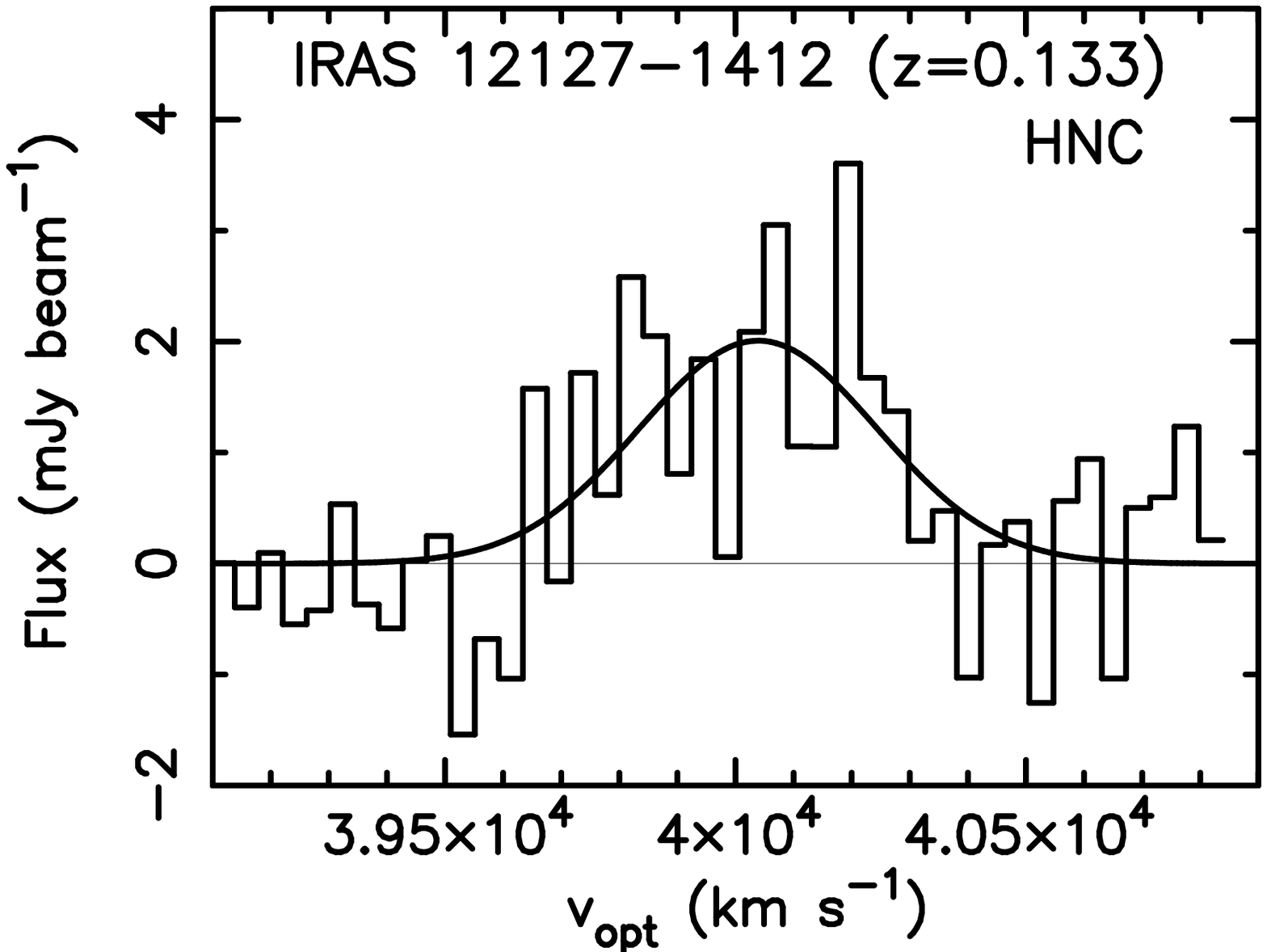}\\ 
\end{center}
\end{figure}

\begin{figure}
\begin{center}
\includegraphics[angle=0,scale=.42]{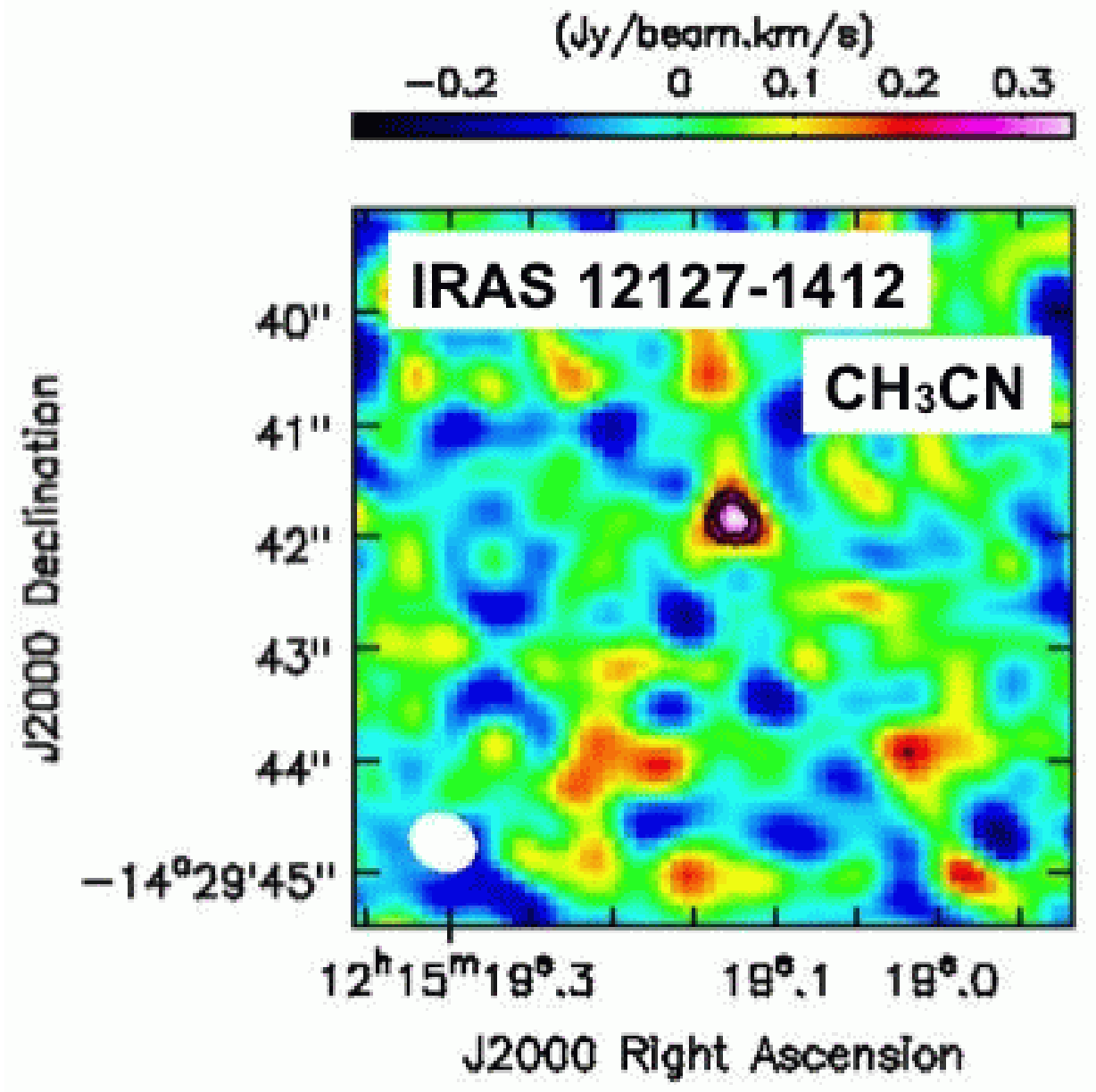} 
\includegraphics[angle=0,scale=.45]{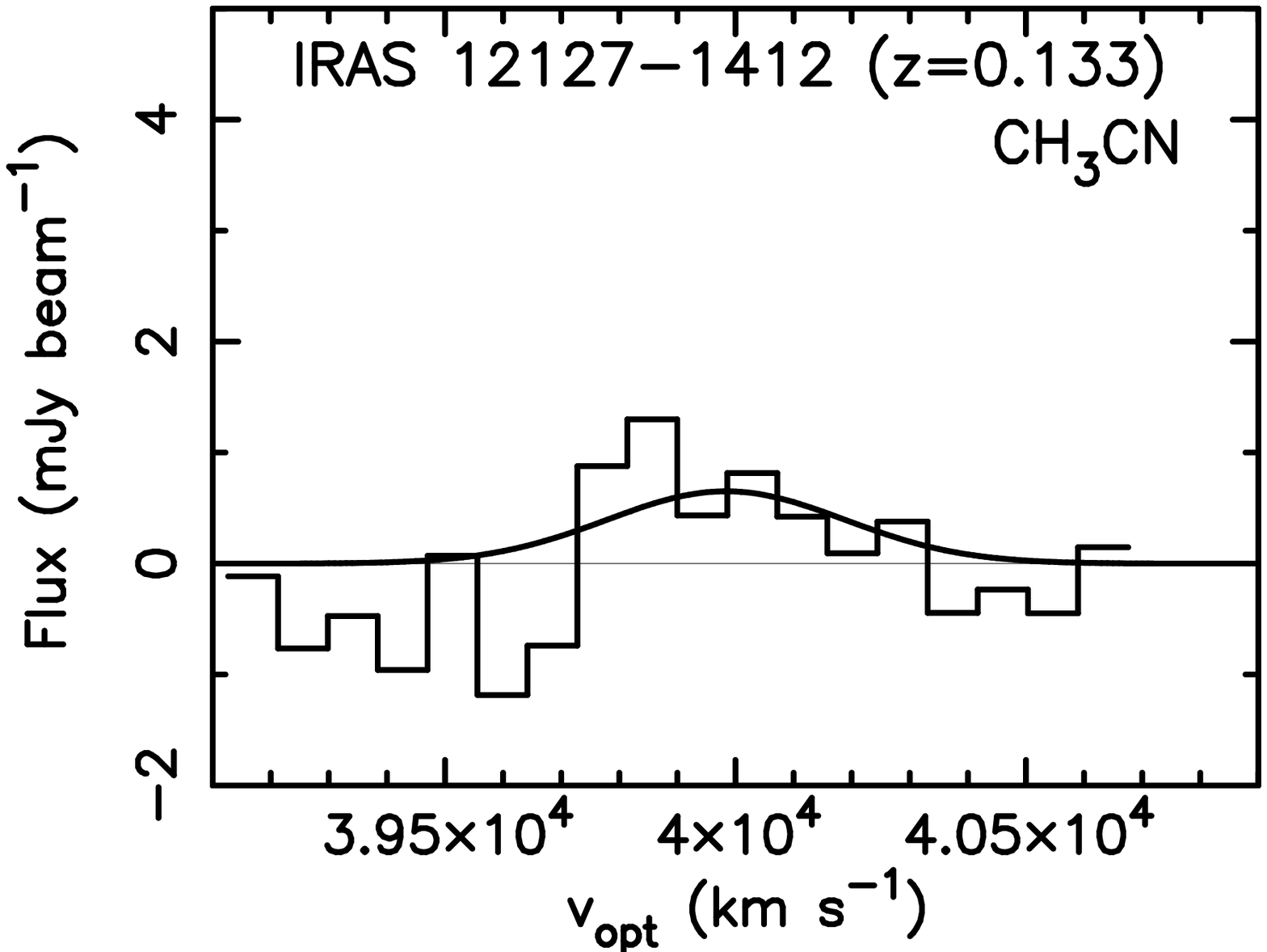}\\ 
\includegraphics[angle=0,scale=.42]{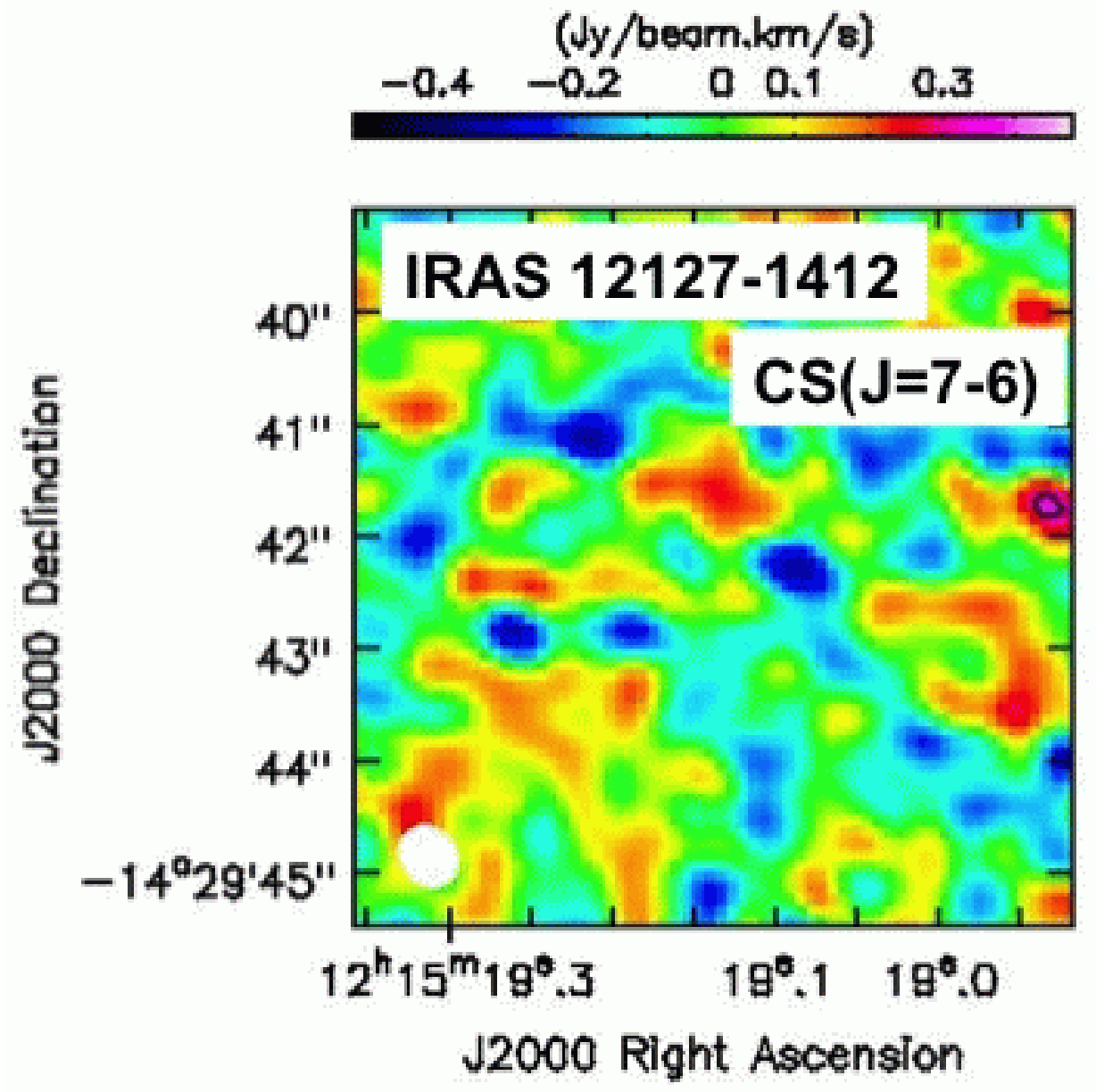} 
\includegraphics[angle=0,scale=.45]{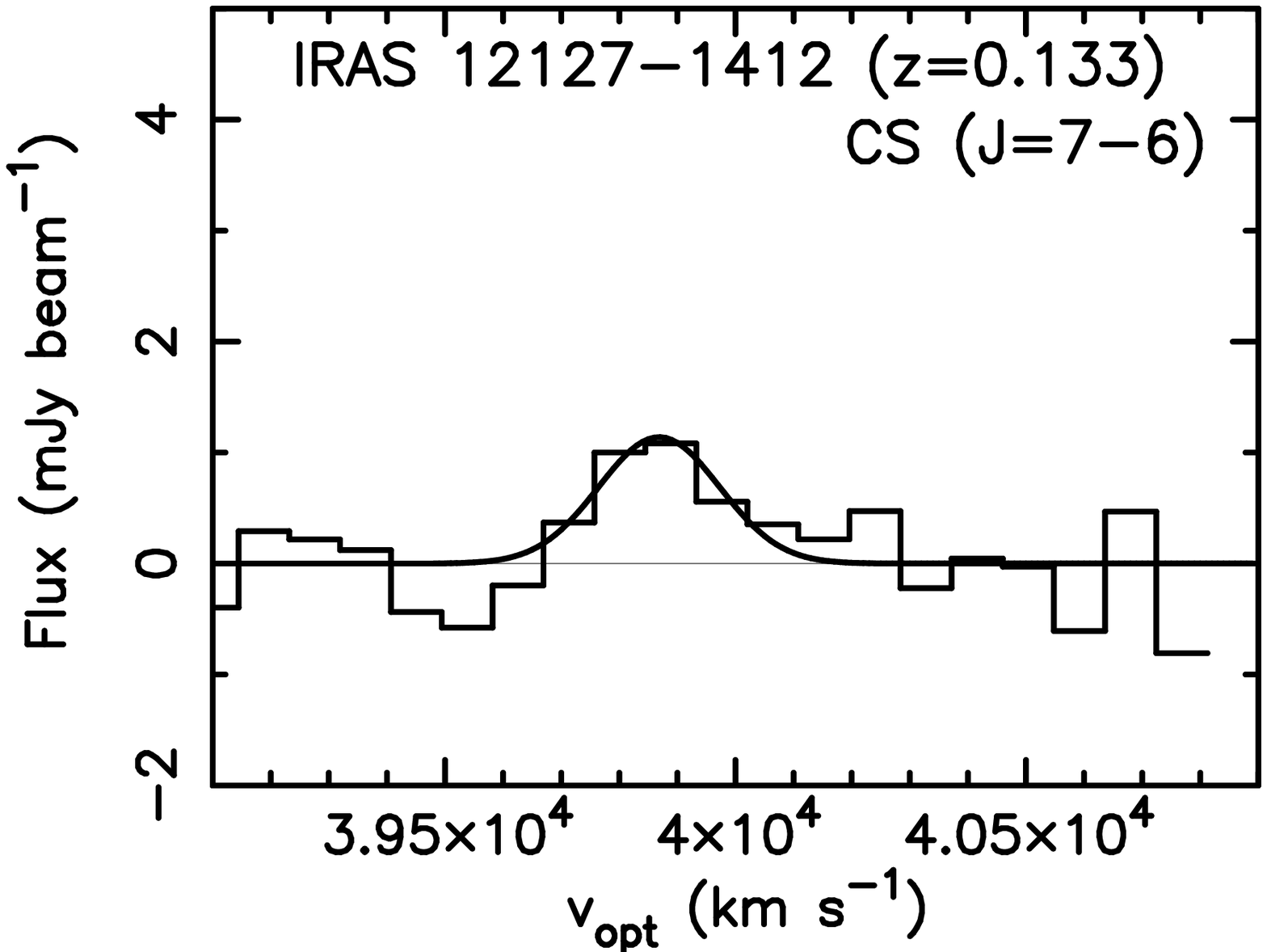}\\ 
\end{center}
\caption{
Integrated intensity (moment 0) maps (left) and spectra at the continuum 
peak position, within the beam size (right), of HCN, HCO$^{+}$, HNC, 
CH$_{3}$CN, and CS for IRAS 12127$-$1412. 
Contours are 
4$\sigma$, 8$\sigma$ for HCN, 
3$\sigma$, 4$\sigma$ for HCO$^{+}$, 
3$\sigma$, 4$\sigma$, 5$\sigma$ for HNC, 
3$\sigma$, 4$\sigma$ for CH$_{3}$CN. 
For CS J=7--6, no emission feature with $\gtrsim$3$\sigma$ is seen. 
For the spectra, the abscissa shows v$_{\rm opt}$ $\equiv$ c 
($\lambda-\lambda_{\rm 0}$)/$\lambda_{\rm 0}$ in [km s$^{-1}$], and the 
ordinate shows flux in [mJy beam$^{-1}$].
The best Gaussian fits (Table 5) are overplotted as solid curved lines.
The spectra of HCN, HCO$^{+}$, and HNC are shown with 
two spectral elements binning.
Those of CH$_{3}$CN and CS are shown with four spectral 
elements binning.
}
\end{figure}

%--- Figure 5 ---%
\begin{figure}
%\vspace*{-0.7cm}
\begin{center}
\includegraphics[angle=0,scale=.42]{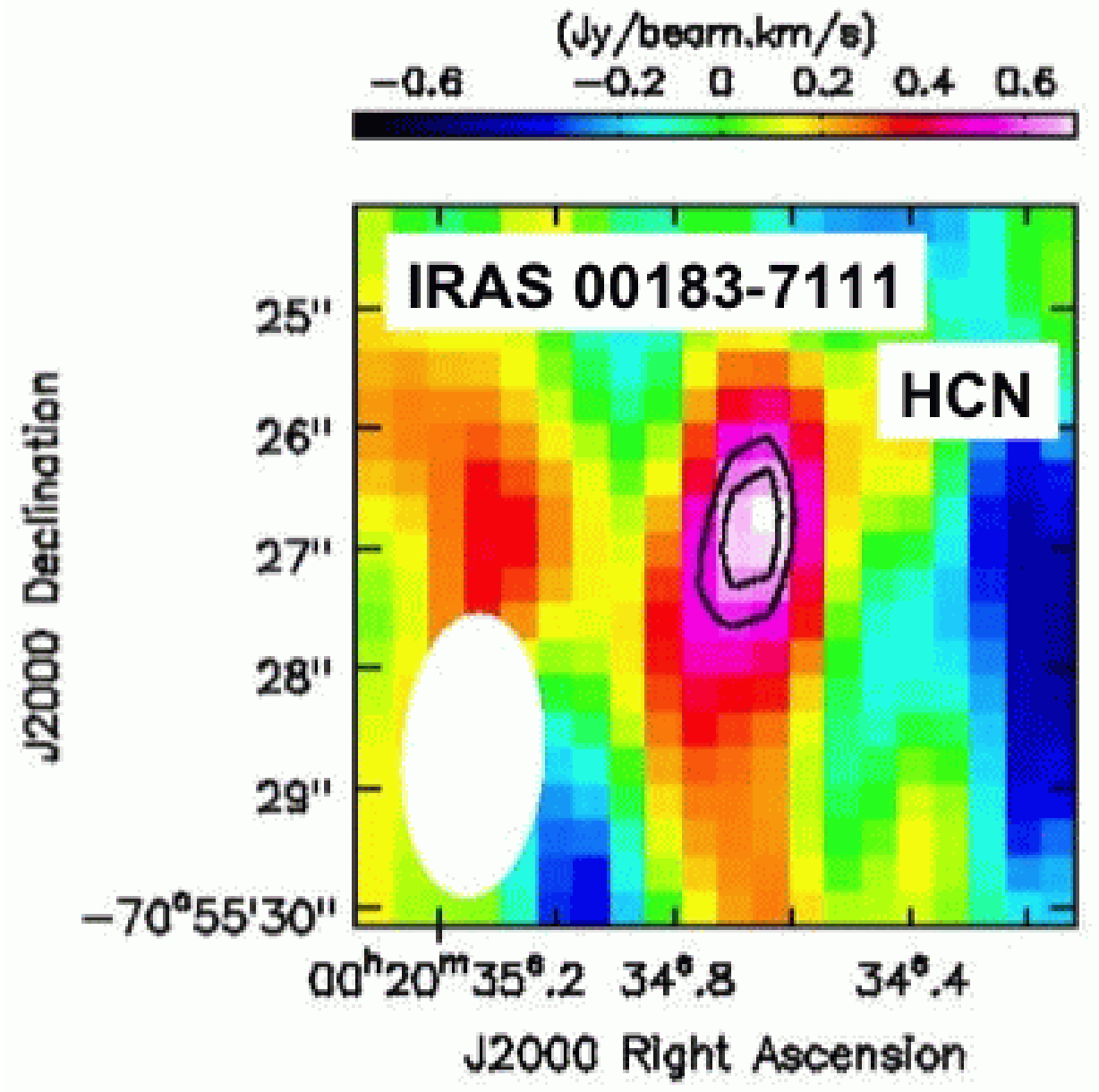} 
\includegraphics[angle=0,scale=.42]{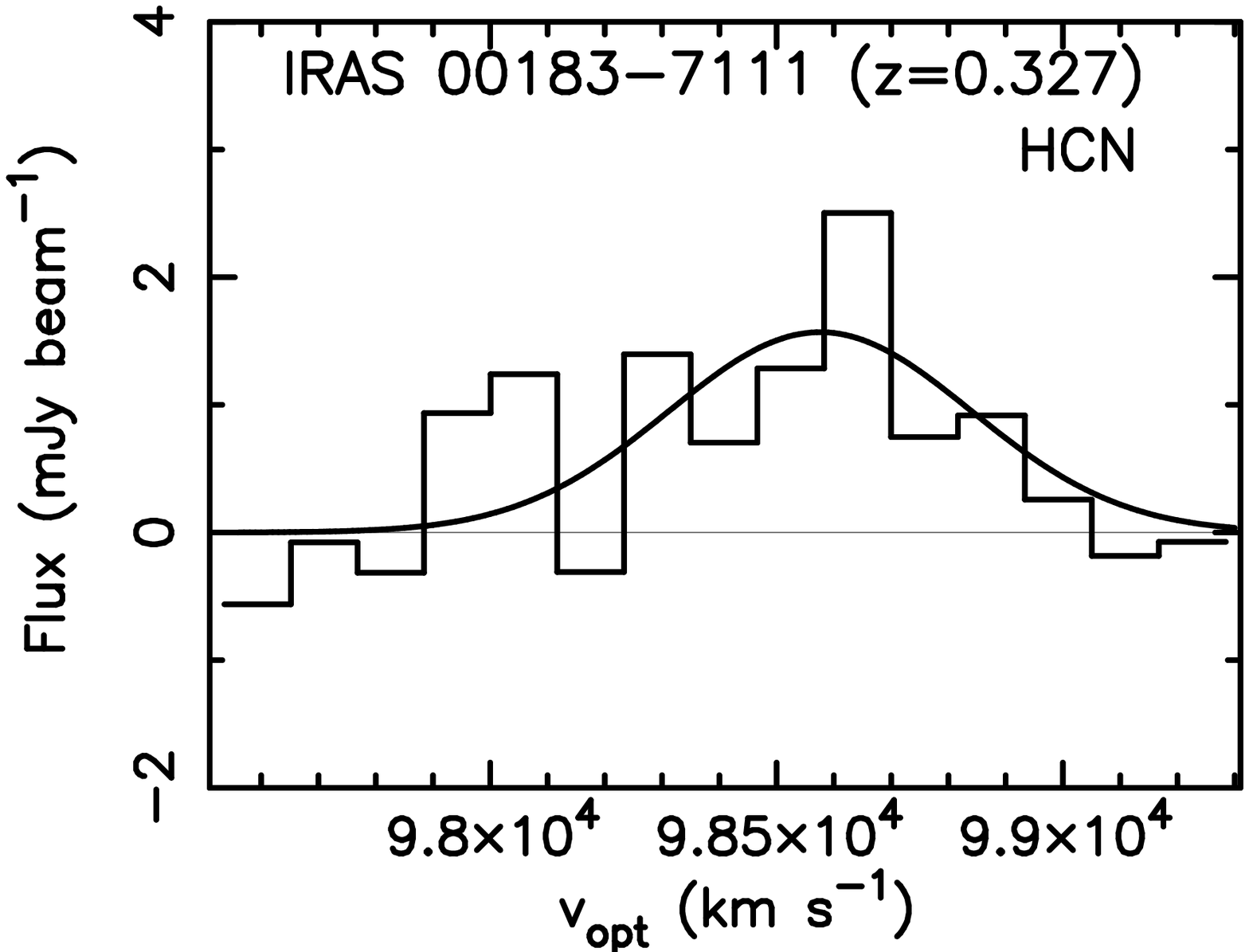}\\ 
\includegraphics[angle=0,scale=.42]{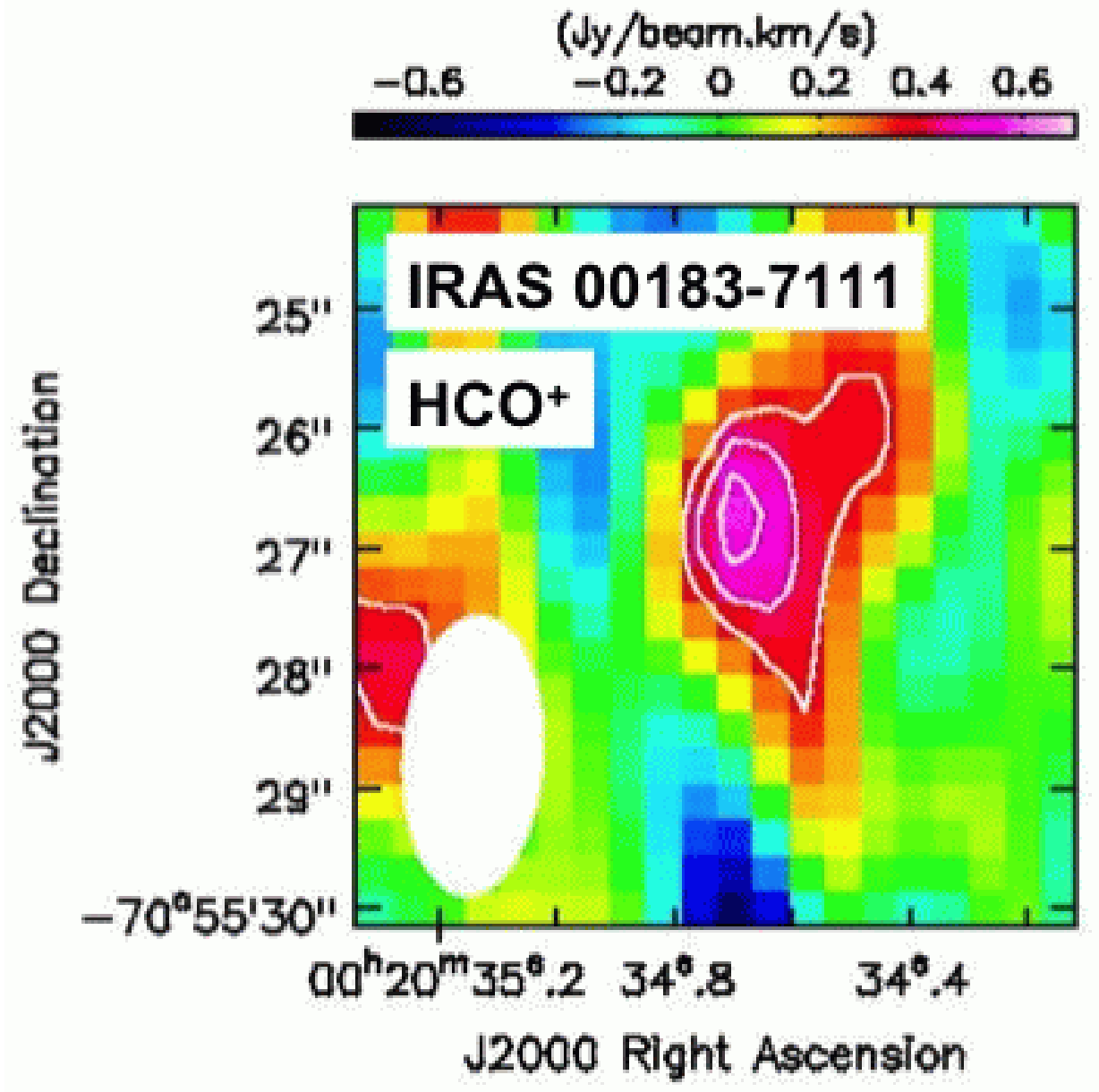} 
\includegraphics[angle=0,scale=.42]{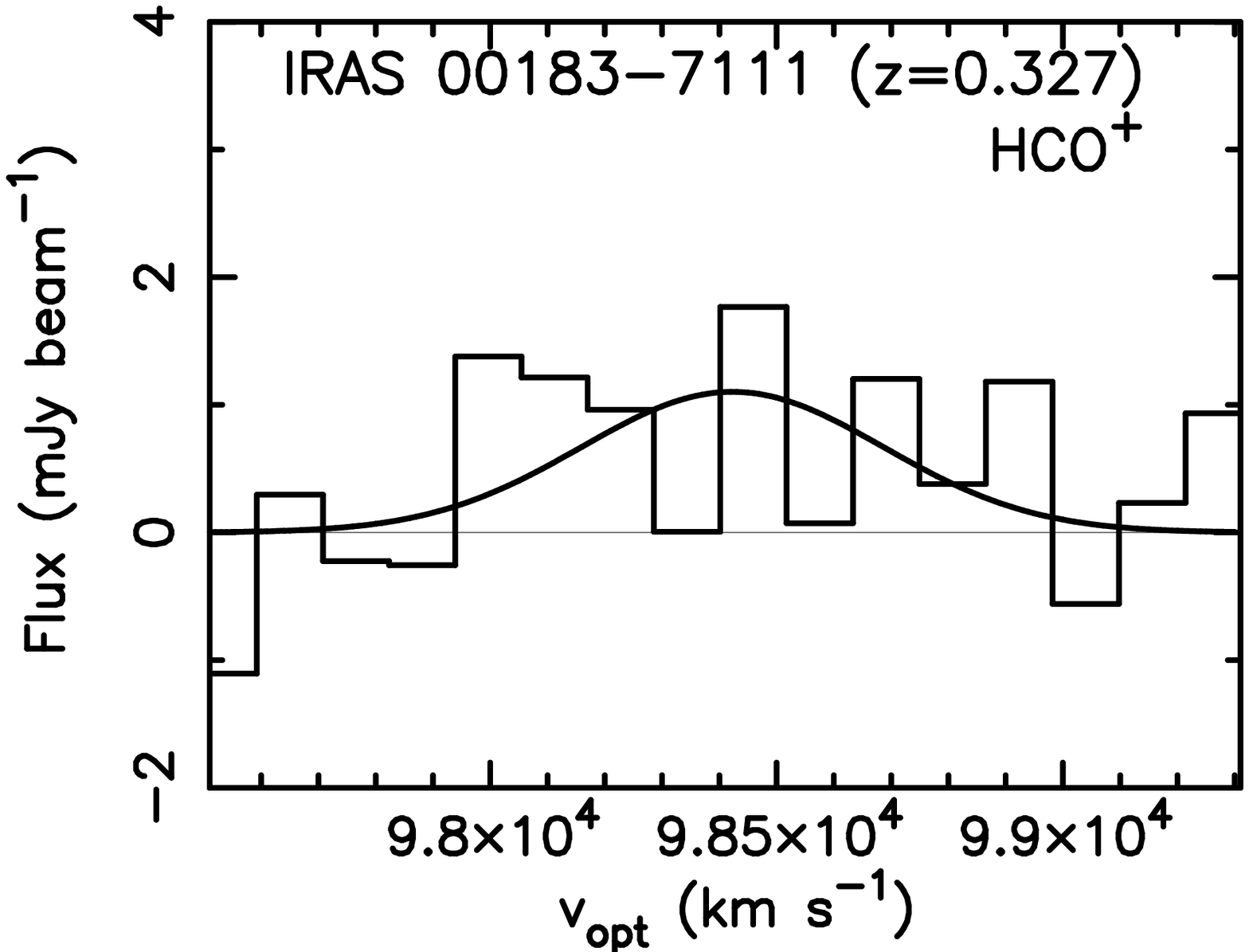}\\ 
\includegraphics[angle=0,scale=.42]{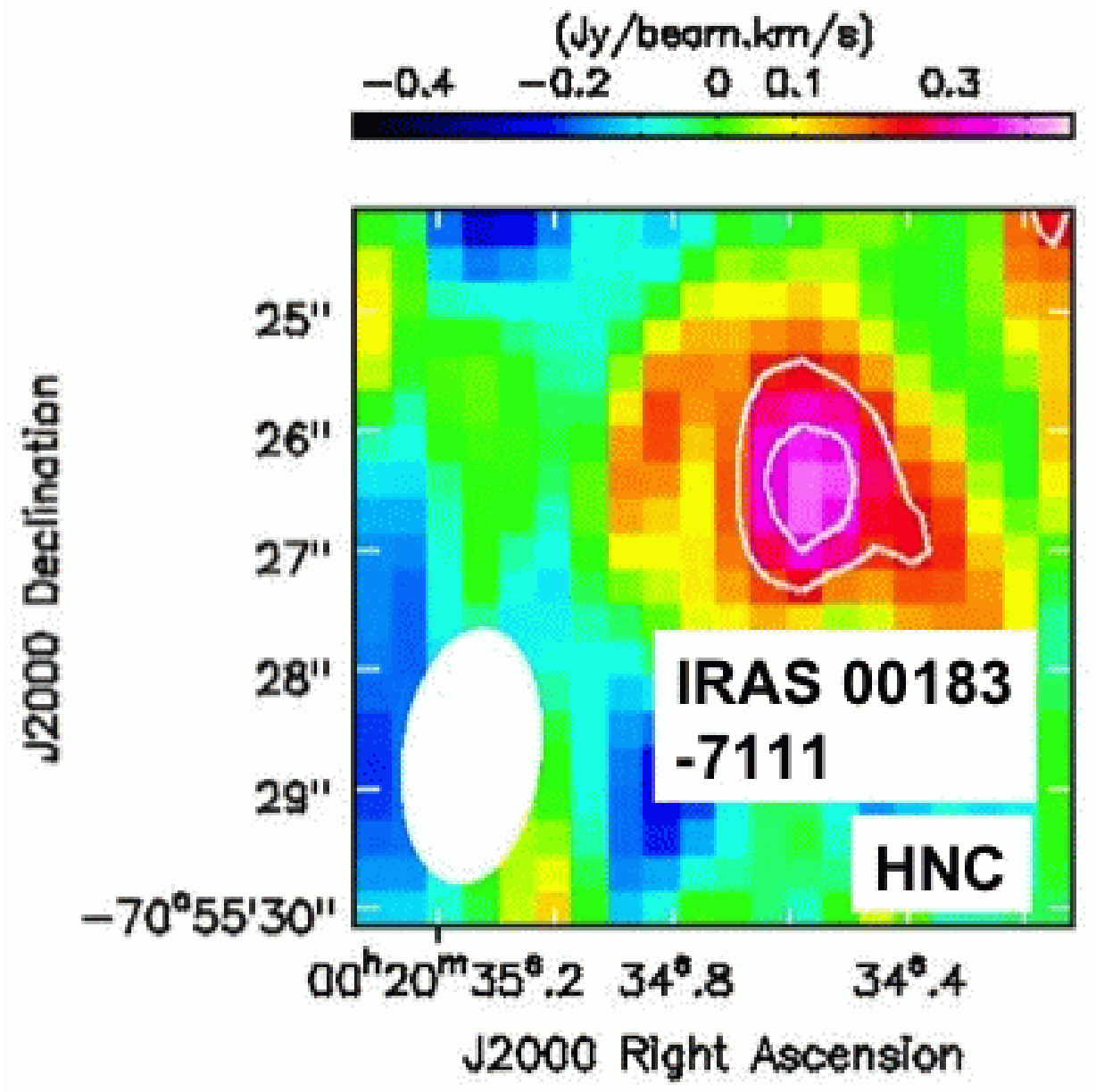} 
\includegraphics[angle=0,scale=.42]{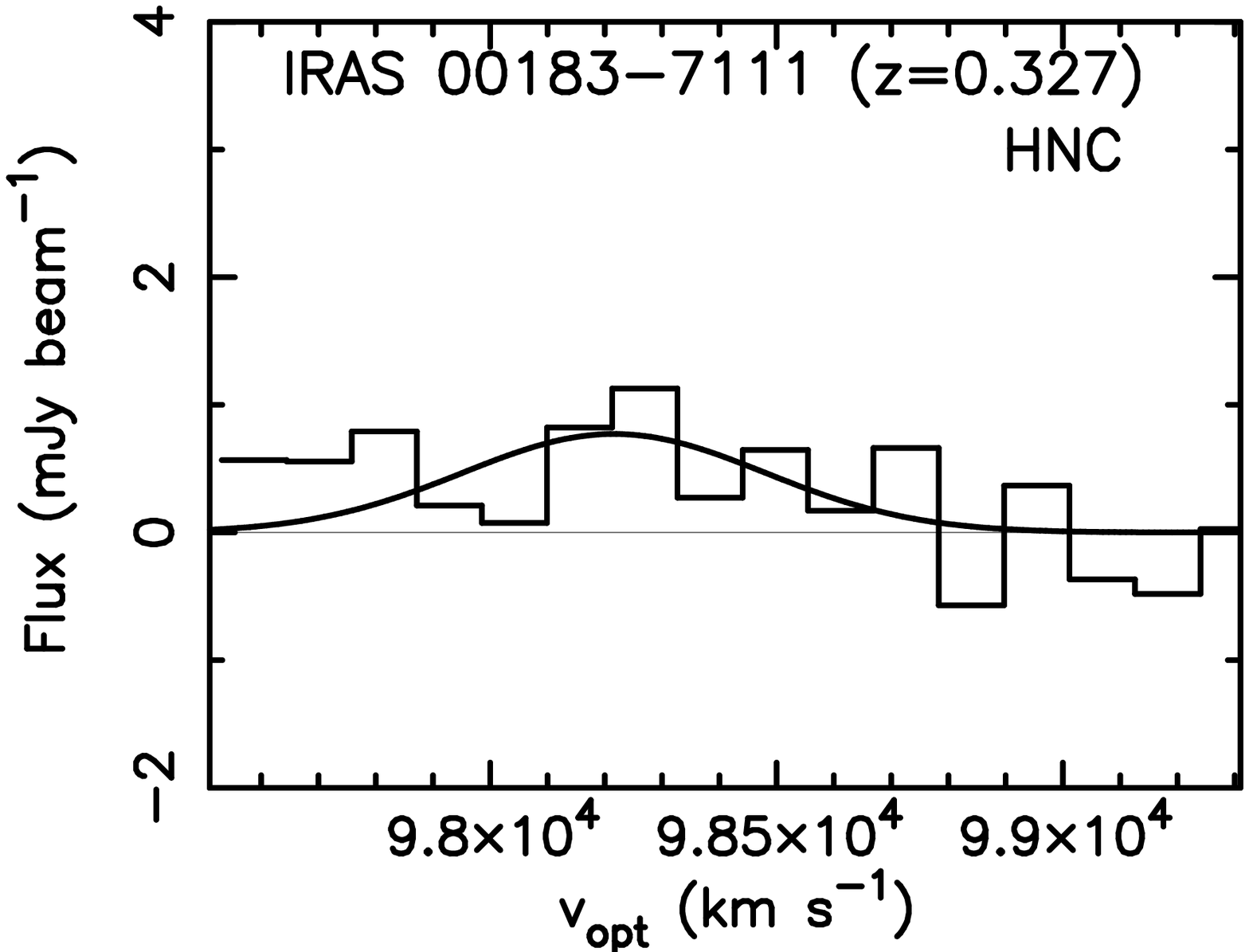}\\ 
\end{center}
%\vspace{-0.5cm}
\caption{
Integrated intensity (moment 0) maps (left) and spectra at the continuum 
peak position, within the beam size (right), of HCN, HCO$^{+}$, and HNC 
for IRAS 00183$-$7111. 
Contours are 
3$\sigma$, 3.5$\sigma$ for HCN, 
2$\sigma$, 2.5$\sigma$, 3$\sigma$ for HCO$^{+}$, 
2$\sigma$, 3$\sigma$ for HNC. 
For the spectra, the abscissa shows v$_{\rm opt}$ $\equiv$ c 
($\lambda-\lambda_{\rm 0}$)/$\lambda_{\rm 0}$ in [km s$^{-1}$], and the 
ordinate shows flux in [mJy beam$^{-1}$].
The best Gaussian fits (Table 5) are overplotted as solid curved lines.
Spectra are shown with four spectral elements binning.
}
\end{figure}

%--- Figure 6 ---%
\begin{figure}
\begin{center}
\includegraphics[angle=0,scale=.42]{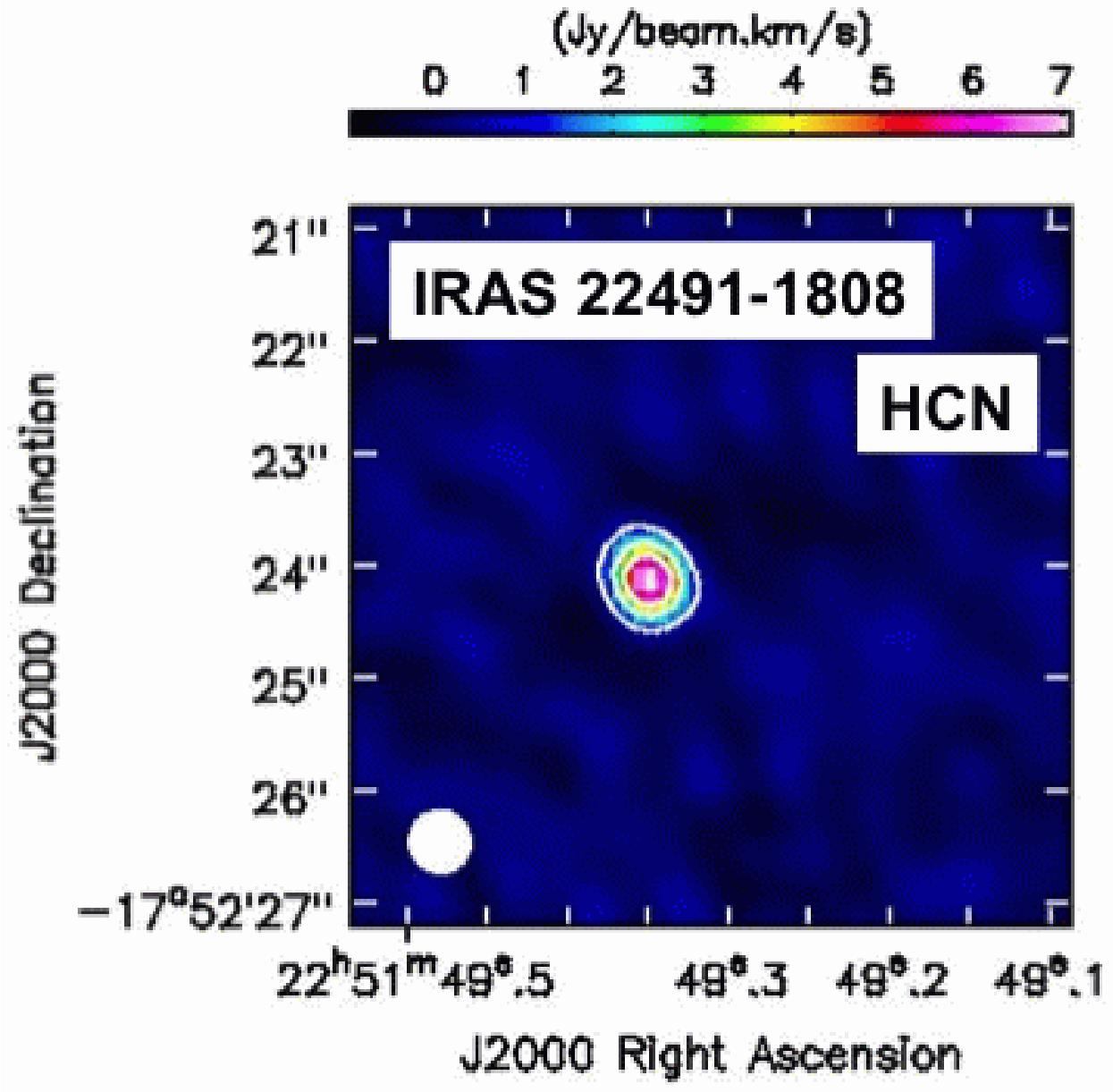} 
\includegraphics[angle=0,scale=.45]{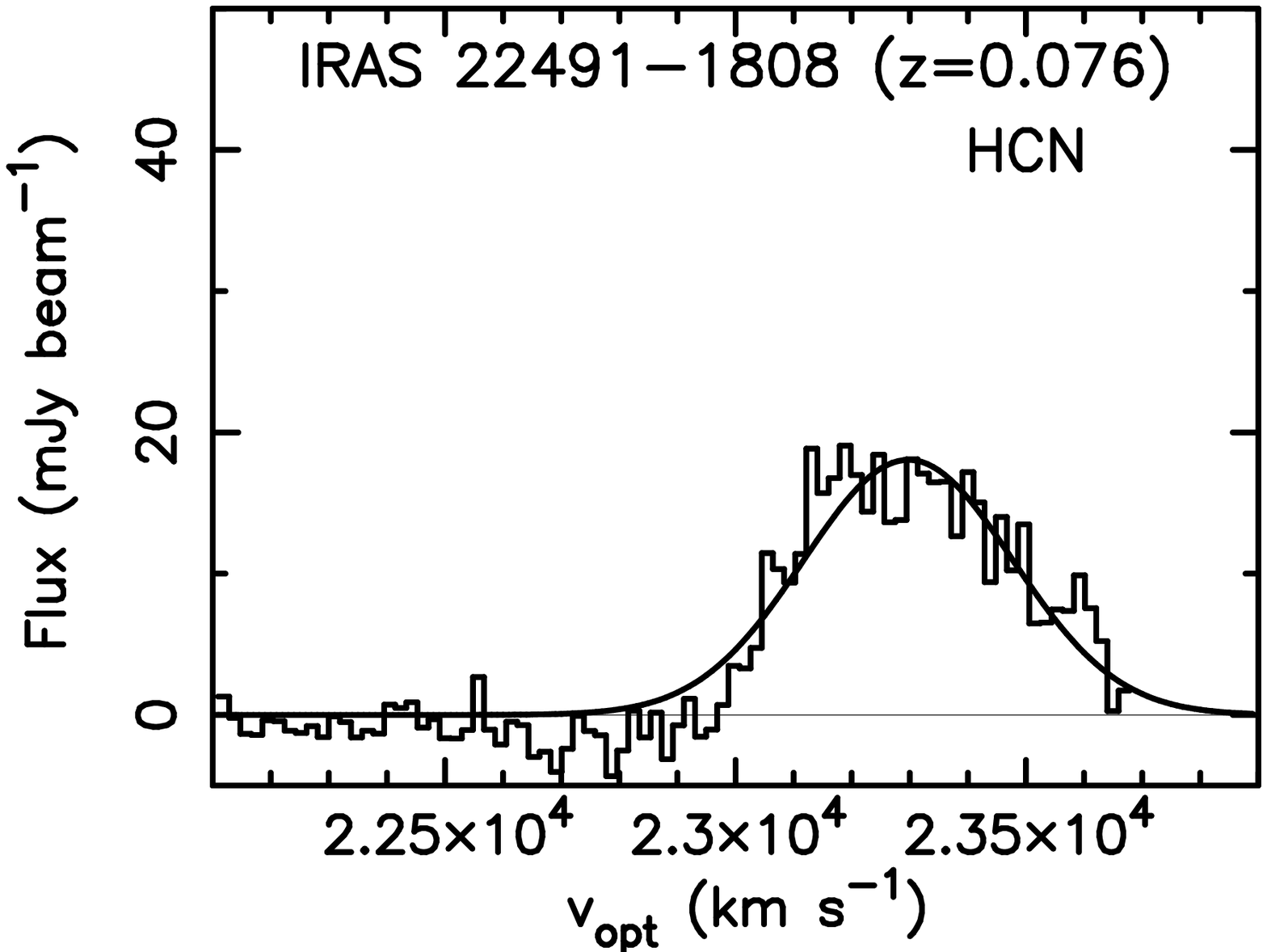}\\ 
\includegraphics[angle=0,scale=.42]{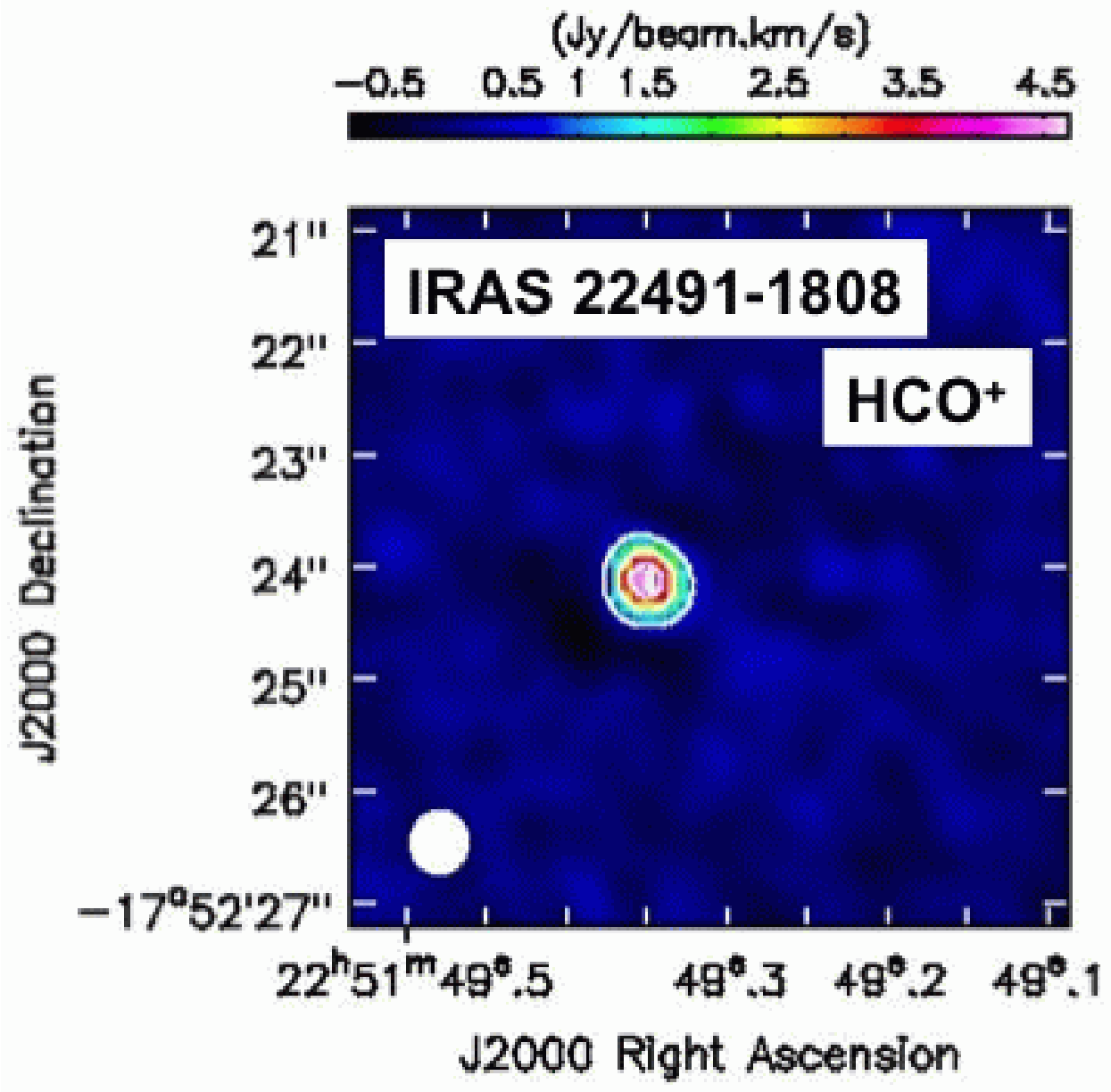} 
\includegraphics[angle=0,scale=.45]{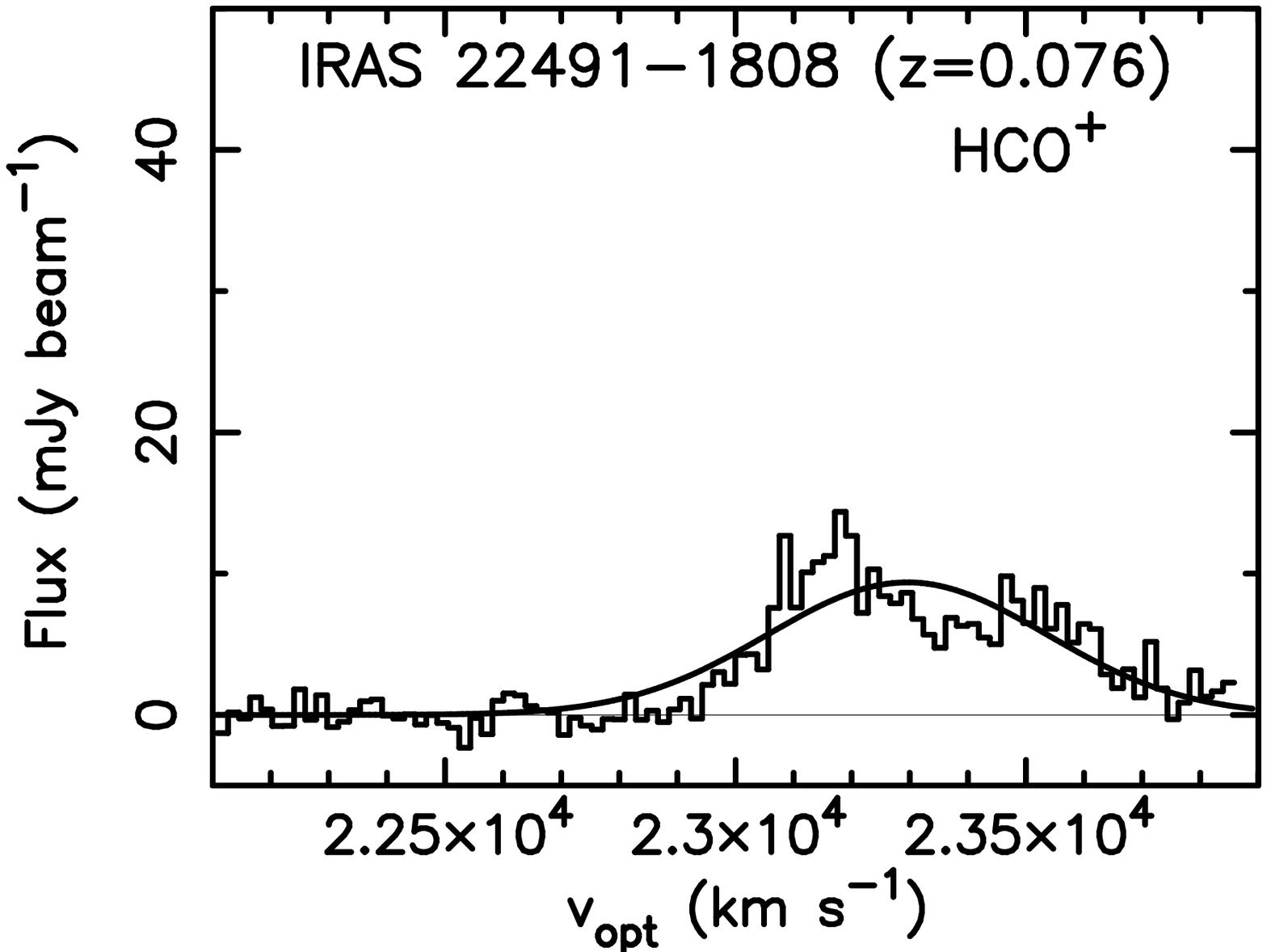}\\ 
\includegraphics[angle=0,scale=.42]{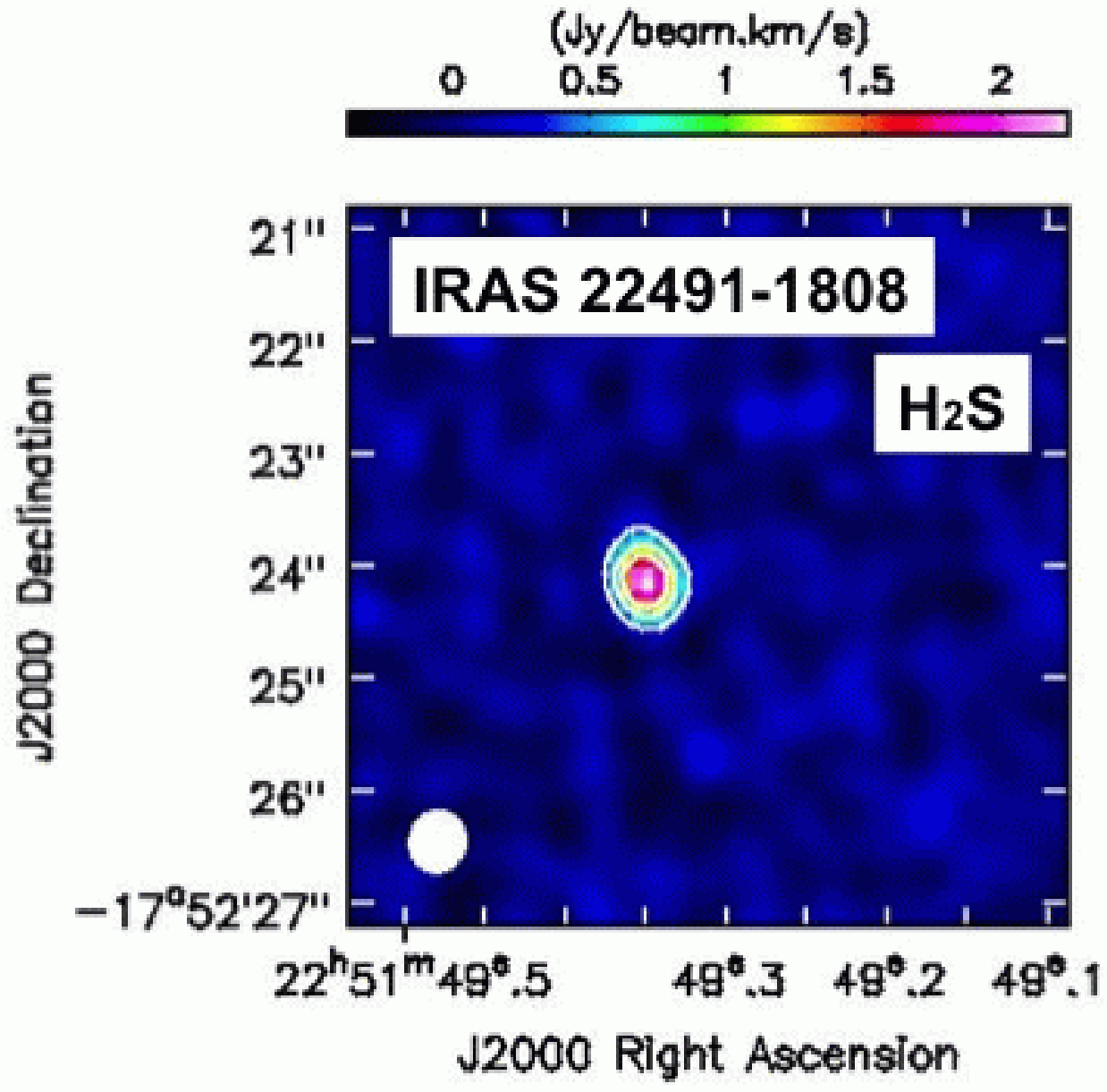} 
\includegraphics[angle=0,scale=.45]{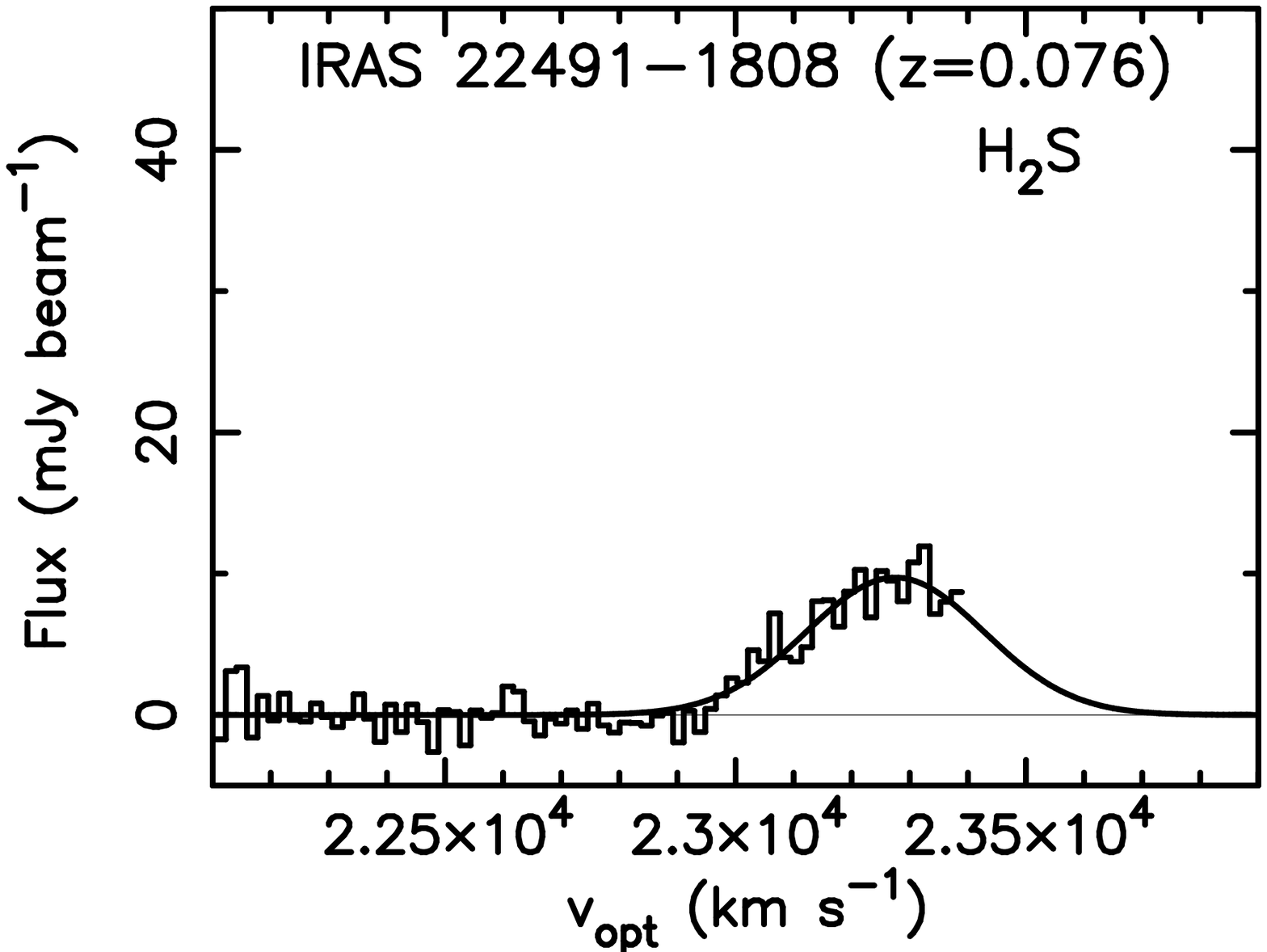}\\ 
\end{center}
\end{figure}

\begin{figure}
\begin{center}
\includegraphics[angle=0,scale=.42]{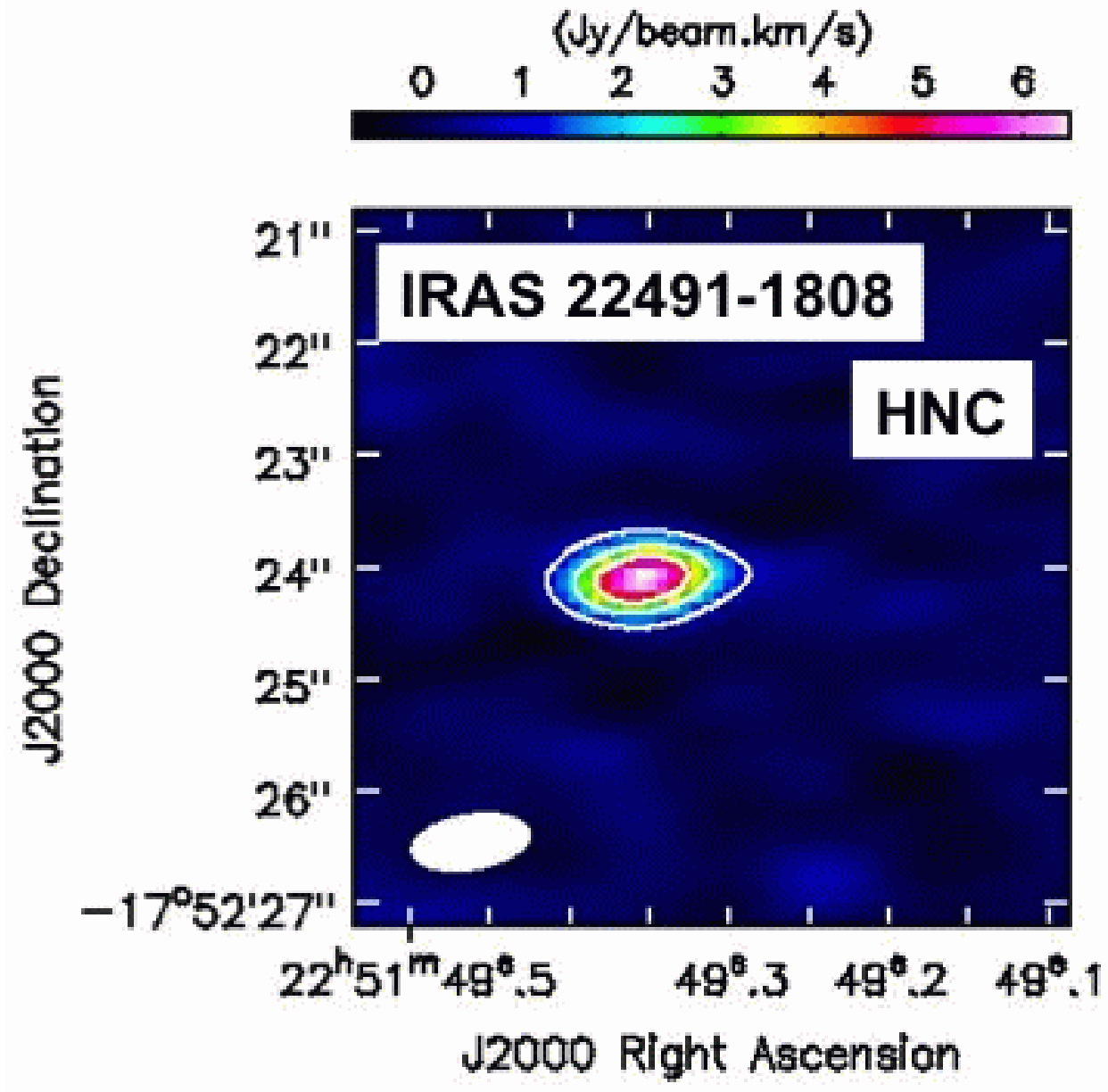} 
\includegraphics[angle=0,scale=.45]{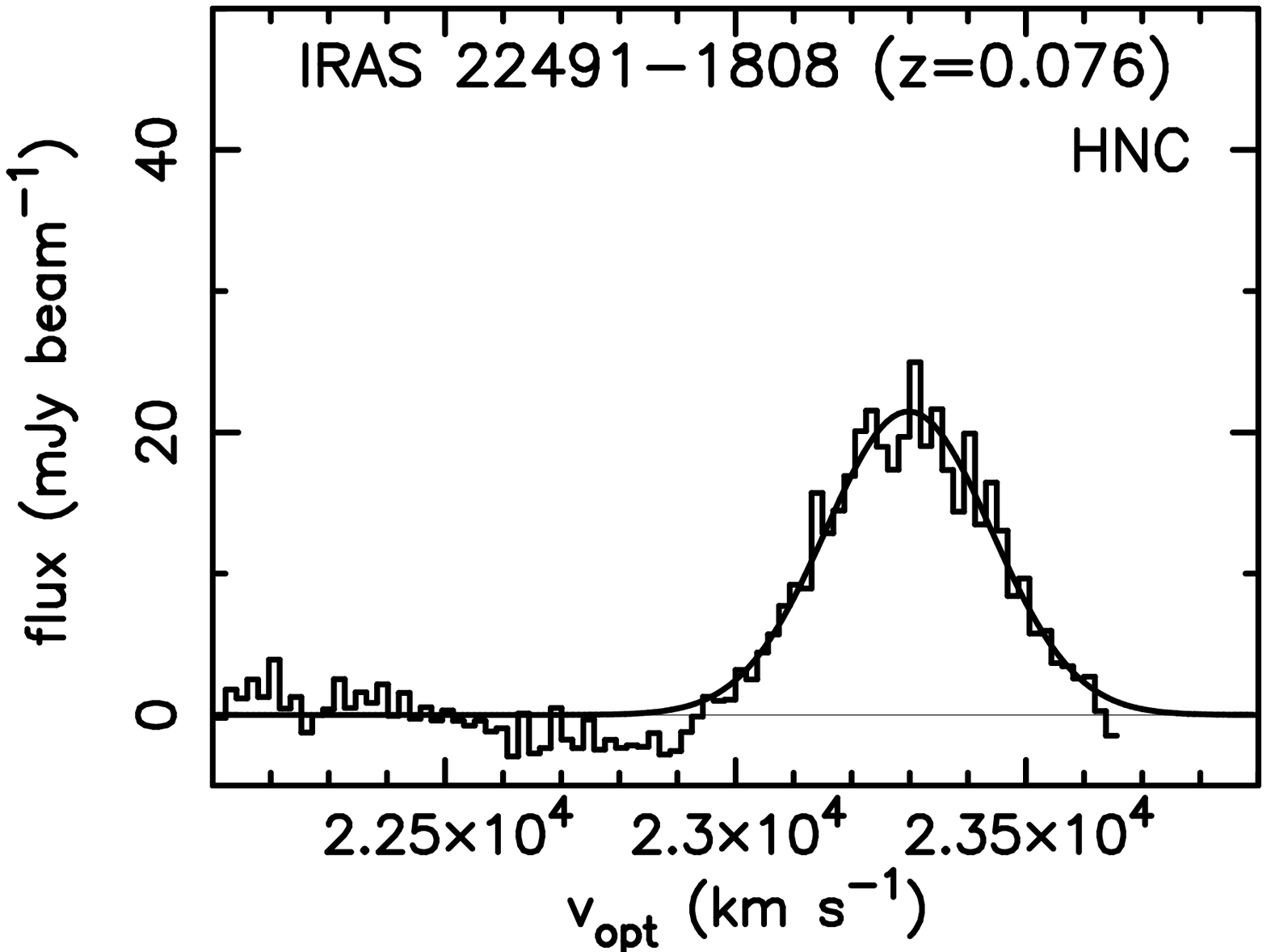}\\ 
\end{center}
\caption{
Integrated intensity (moment 0) maps (left) and spectra at the continuum 
peak position, within the beam size (right), of HCN, HCO$^{+}$, H$_{2}$S, 
and HNC for IRAS 22491$-$1808. 
Contours are 5$\sigma$, 15$\sigma$, 25$\sigma$ for HCN, 
5$\sigma$, 15$\sigma$, 25$\sigma$ for HCO$^{+}$, 
4$\sigma$, 10$\sigma$, 16$\sigma$ for H$_{2}$S, 
5$\sigma$, 15$\sigma$, 25$\sigma$ for HNC. 
For the spectra, the abscissa shows v$_{\rm opt}$ $\equiv$ c 
($\lambda-\lambda_{\rm 0}$)/$\lambda_{\rm 0}$ in [km s$^{-1}$], and the 
ordinate shows flux in [mJy beam$^{-1}$]. 
The best Gaussian fits (Table 5) are overplotted as solid curved lines.
From the optical redshift of $z=$ 0.076 \citep{kim98}, the expected peak 
velocity is v$_{\rm opt}$ $\sim$ 22,800 km s$^{-1}$. 
The observed peak velocities are $\sim$500 km s$^{-1}$ higher.
}
\end{figure}

%--- Figure 7 ---%
\begin{figure}
\begin{center}
\includegraphics[angle=0,scale=.45]{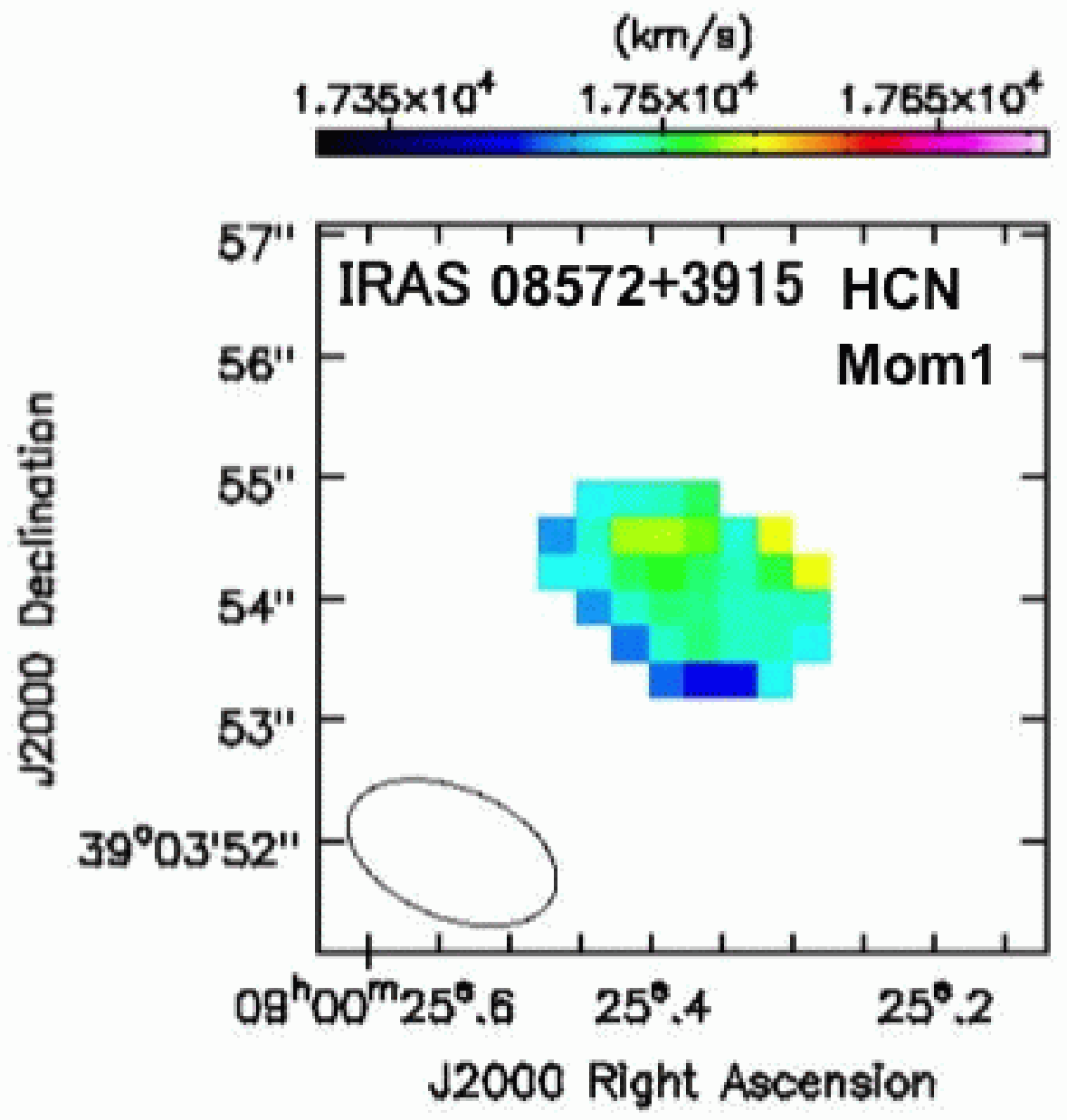} 
\includegraphics[angle=0,scale=.45]{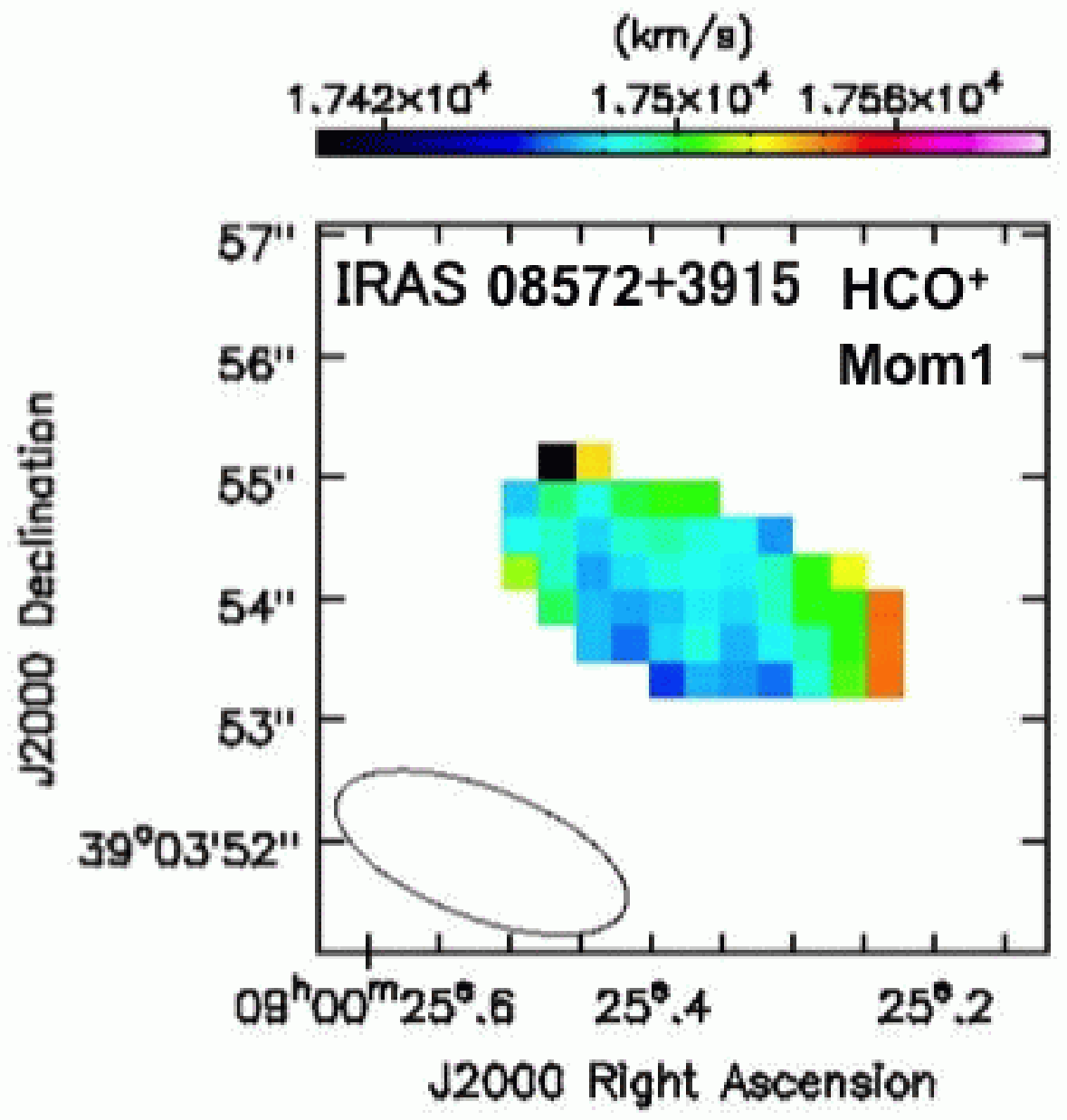} \\
\includegraphics[angle=0,scale=.45]{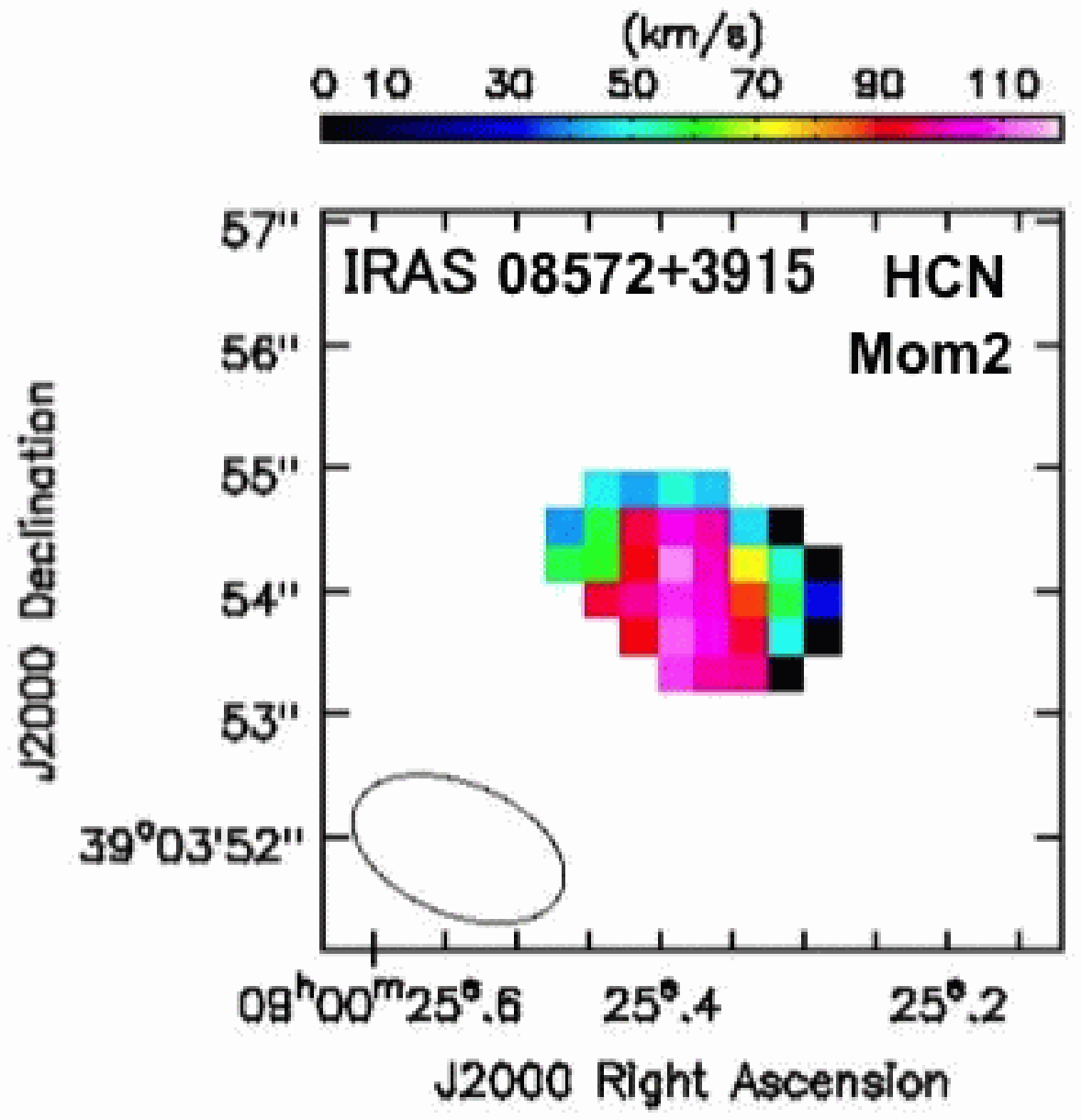} 
\includegraphics[angle=0,scale=.45]{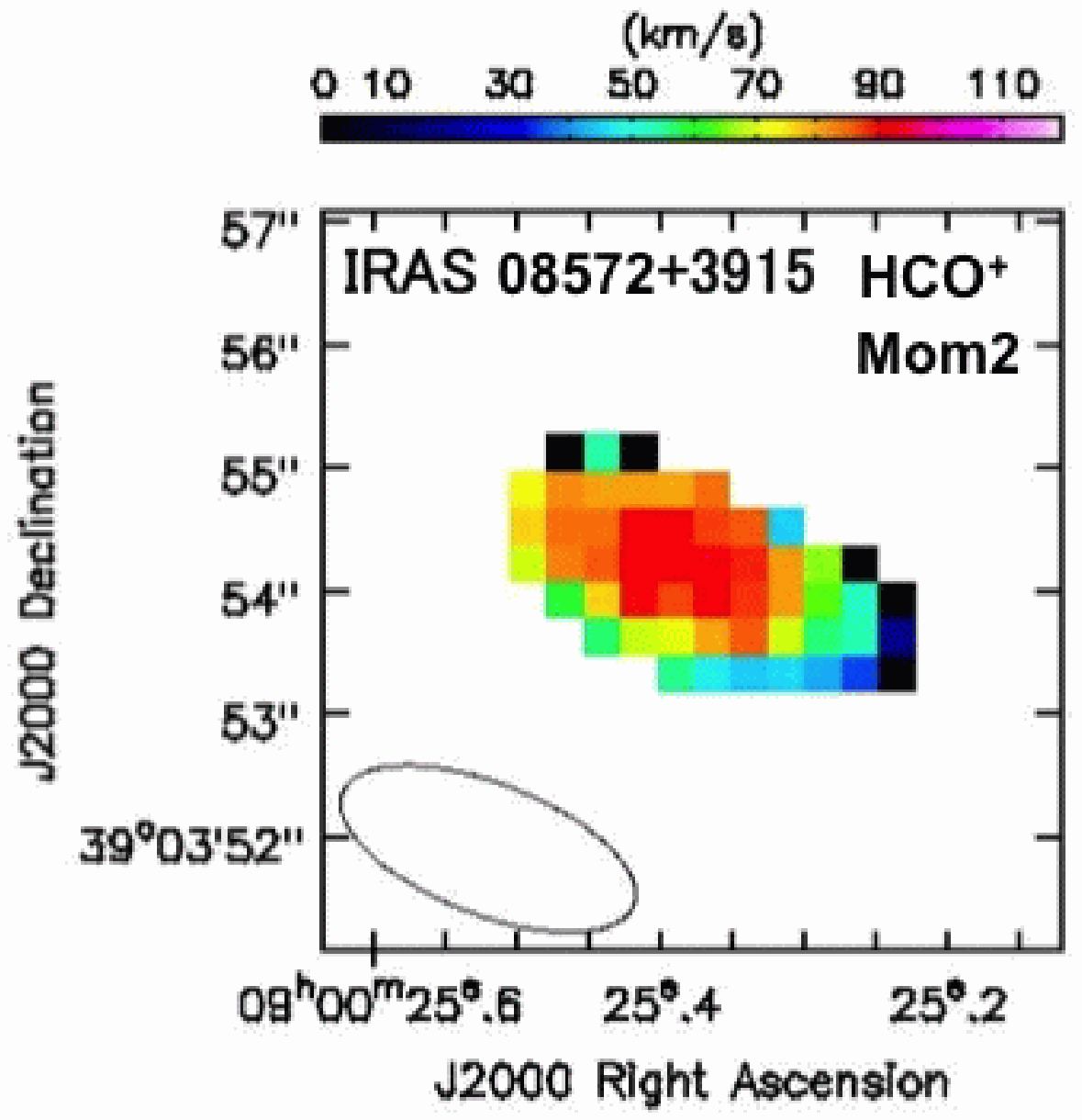} \\
\end{center}
\caption{
Intensity-weighted mean velocity (moment 1) and 
intensity-weighted velocity dispersion (moment 2) maps 
of HCN J=4--3 and HCO$^{+}$ J=4--3 emission lines for IRAS 08572$-$3915. 
For moment 1 maps, the velocity is in v$_{\rm opt}$$\equiv$ c 
($\lambda-\lambda_{\rm 0}$)/$\lambda_{\rm 0}$ in [km s$^{-1}$]. 
For moment 2 maps, the velocity is in [km s$^{-1}$].
}
\end{figure}

\clearpage

%--- Figure 8 ---%
\begin{figure}
\begin{center}
\includegraphics[angle=0,scale=.45]{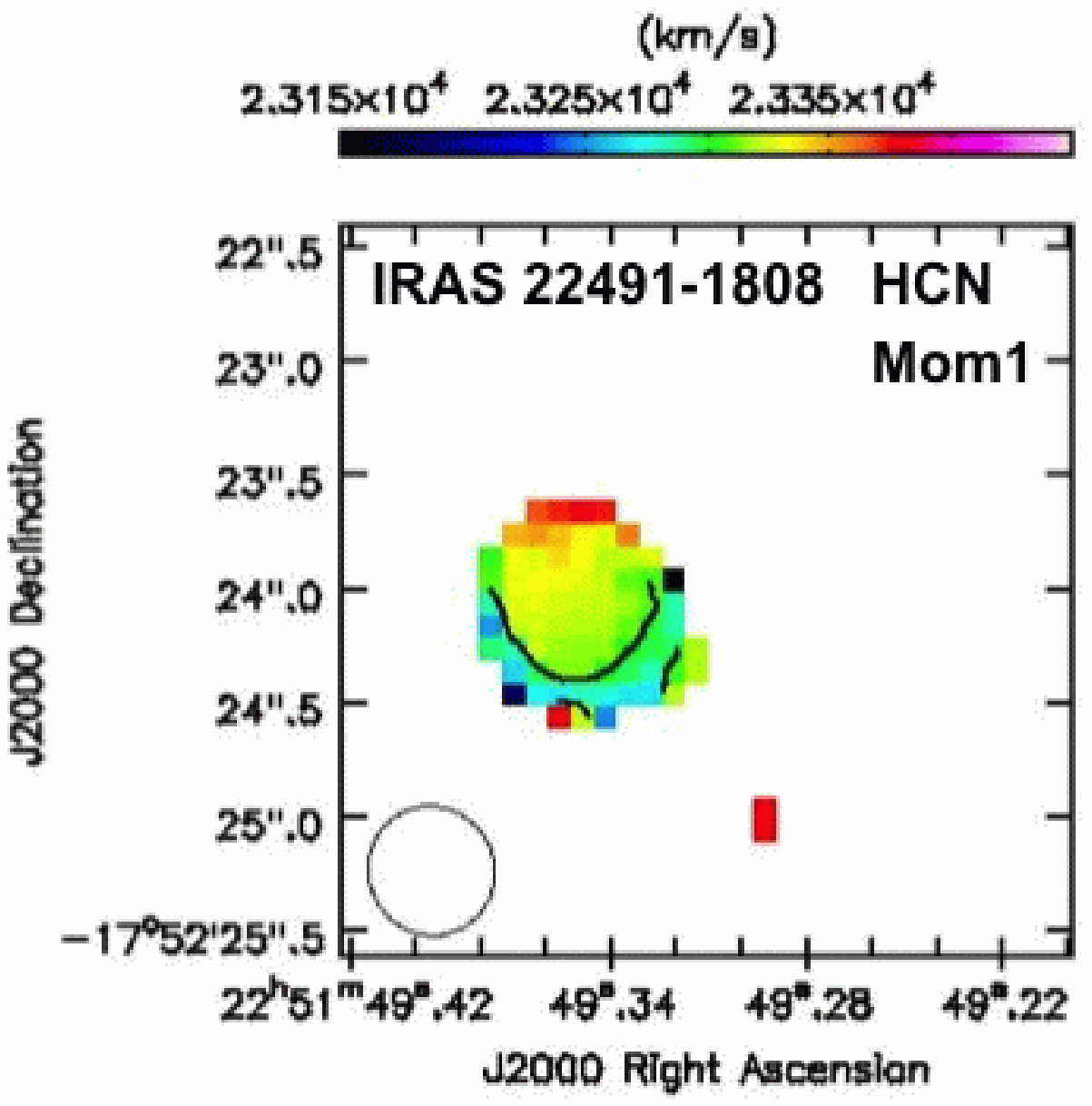} 
\includegraphics[angle=0,scale=.45]{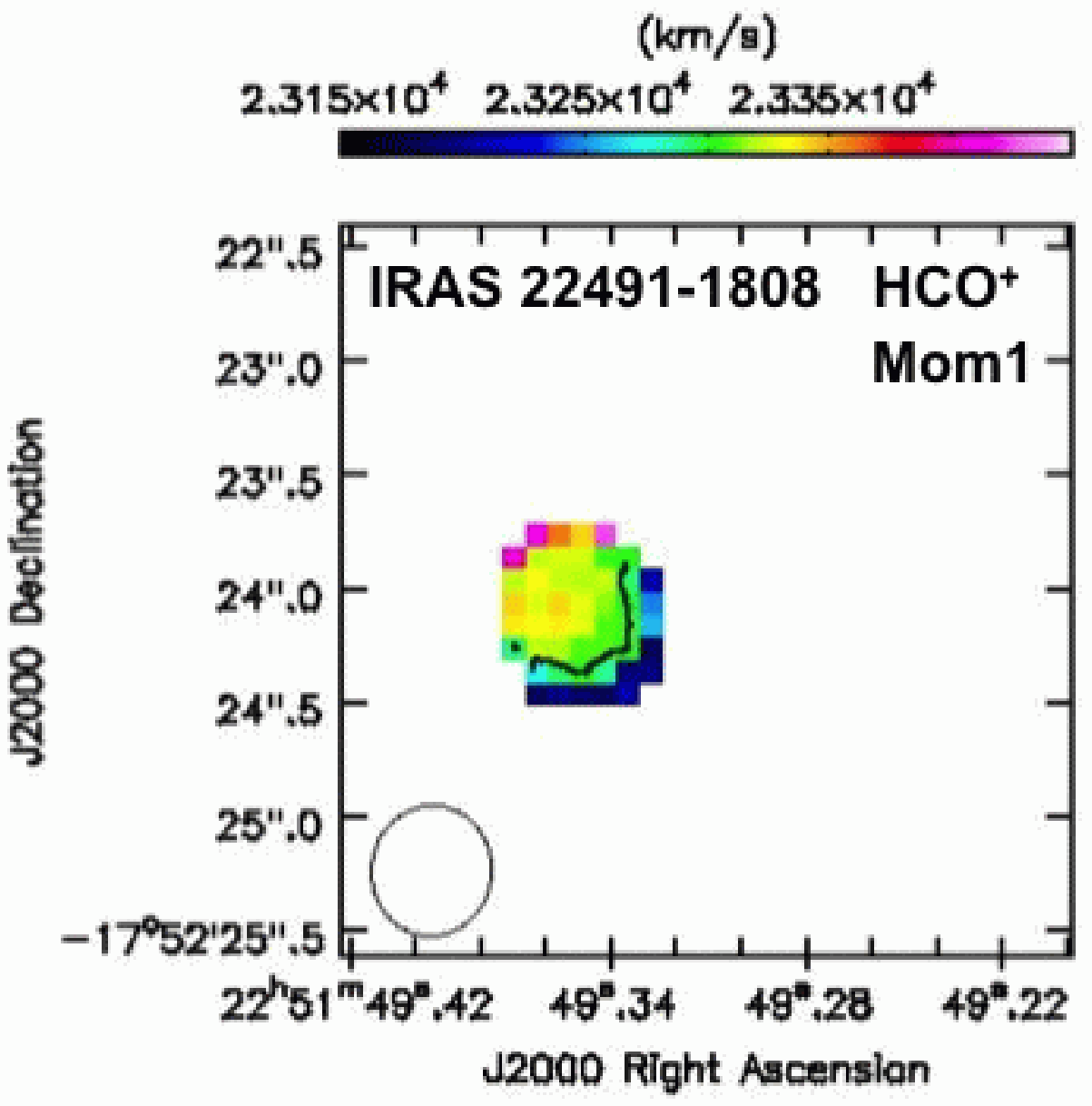} \\
\includegraphics[angle=0,scale=.45]{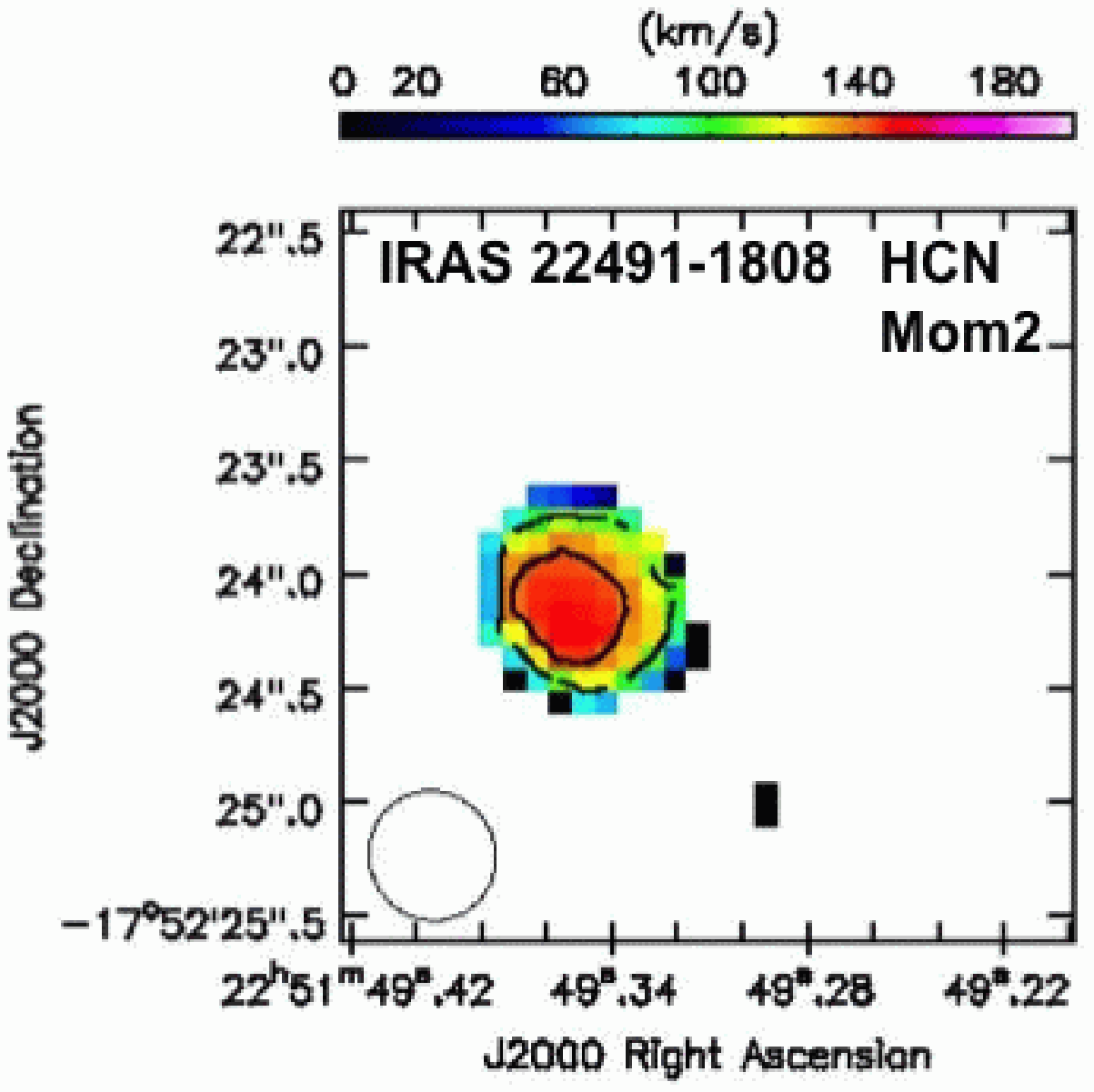} 
\includegraphics[angle=0,scale=.45]{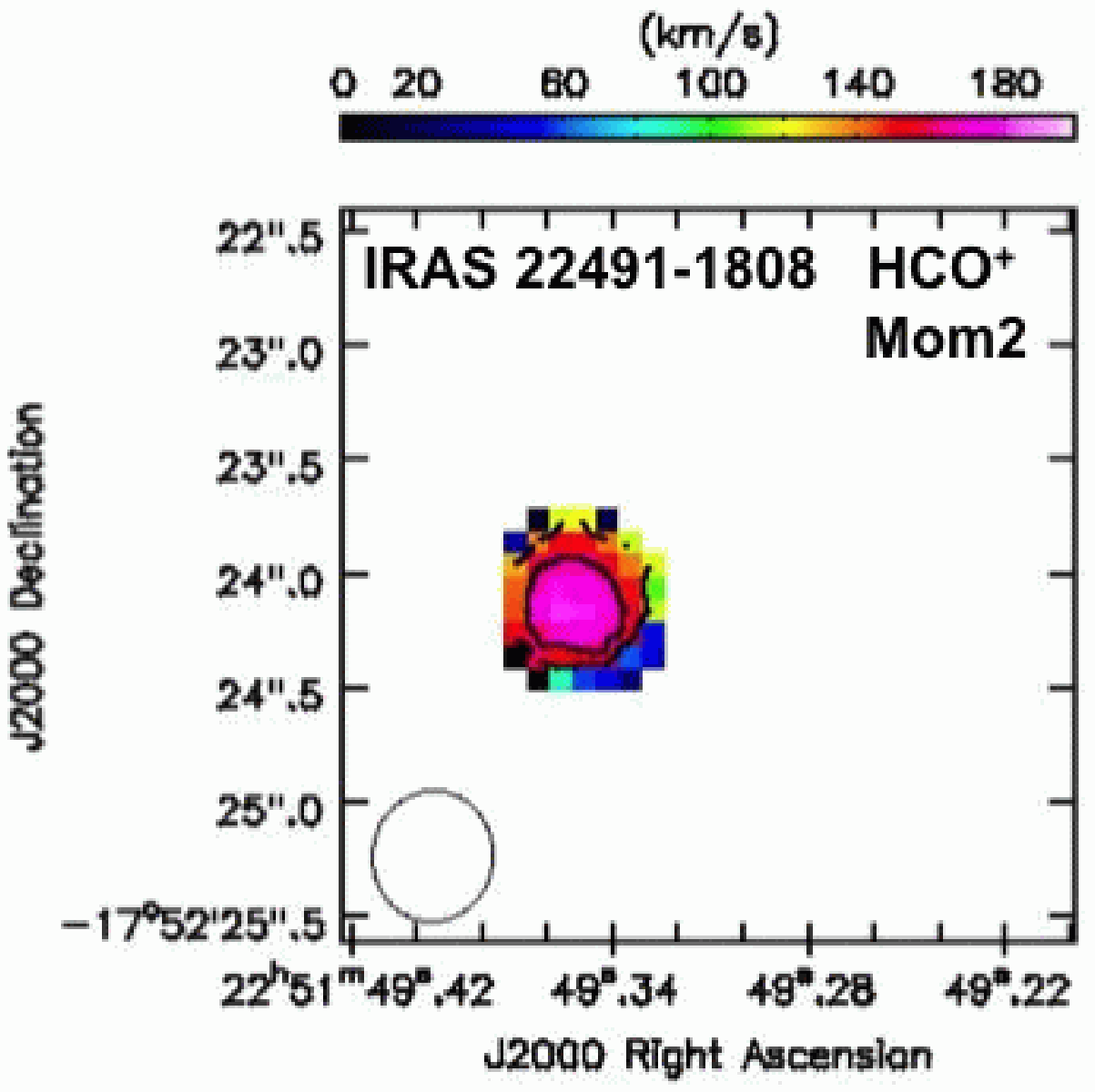} \\
\end{center}
\caption{
Intensity-weighted mean velocity (moment 1) and intensity-weighted 
velocity dispersion (moment 2) maps 
of HCN J=4--3 and HCO$^{+}$ J=4--3 emission lines for IRAS 22491$-$1808. 
For moment 1 maps, the velocity is in v$_{\rm opt}$$\equiv$ c 
($\lambda-\lambda_{\rm 0}$)/$\lambda_{\rm 0}$ in [km s$^{-1}$]. 
For moment 2 maps, the velocity is in [km s$^{-1}$]. 
The contours in moment 1 maps are 23300 km s$^{-1}$ for both HCN and 
HCO$^{+}$ J=4--3. 
The contours in moment 2 maps are 100 and 140 km s$^{-1}$ for HCN J=4--3, 
and 120 and 160 km s$^{-1}$ for HCO$^{+}$ J=4--3.
}
\end{figure}

\clearpage

%--- Figure 9 ---%
\begin{figure}
\includegraphics[angle=0,scale=.45]{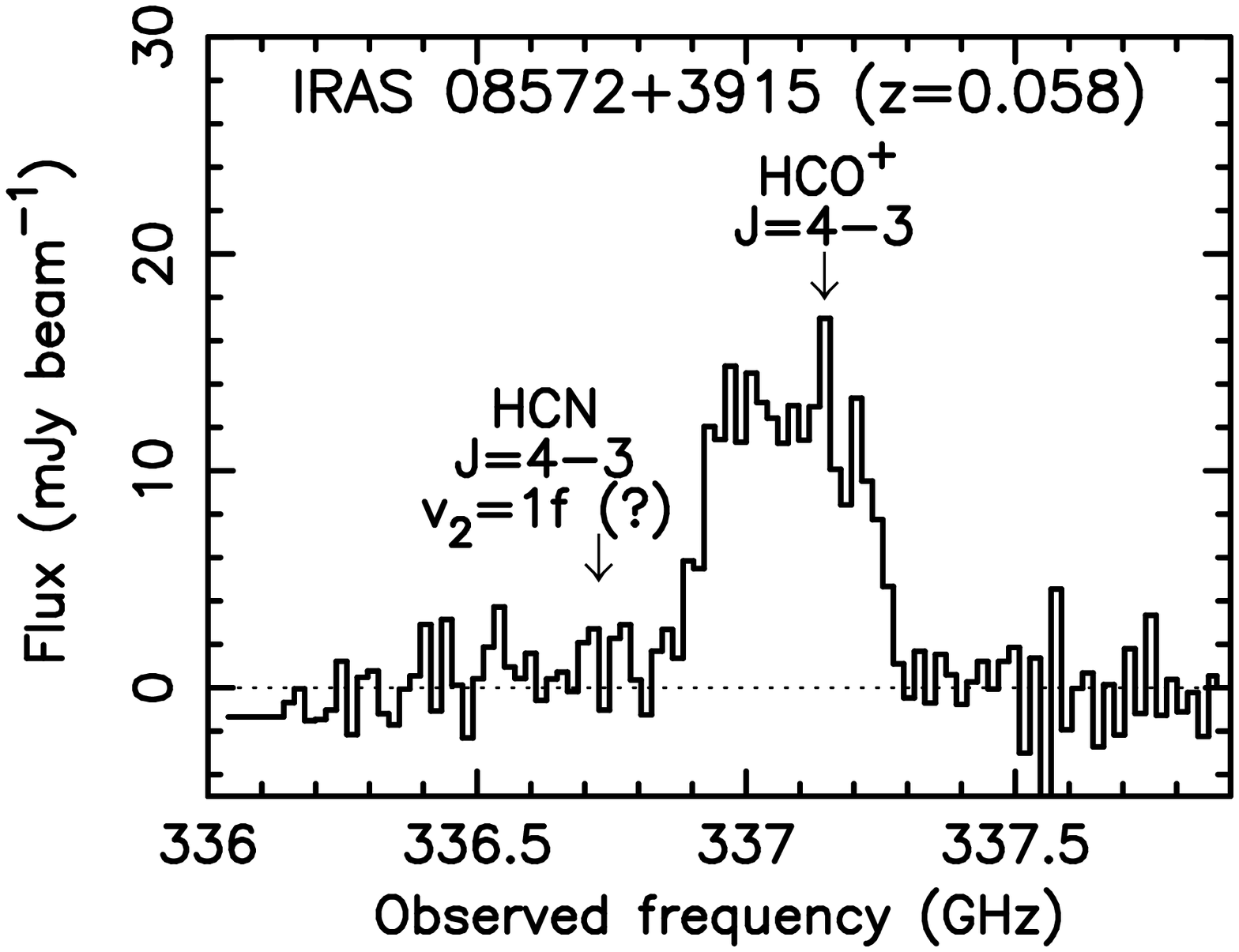}
\includegraphics[angle=0,scale=.45]{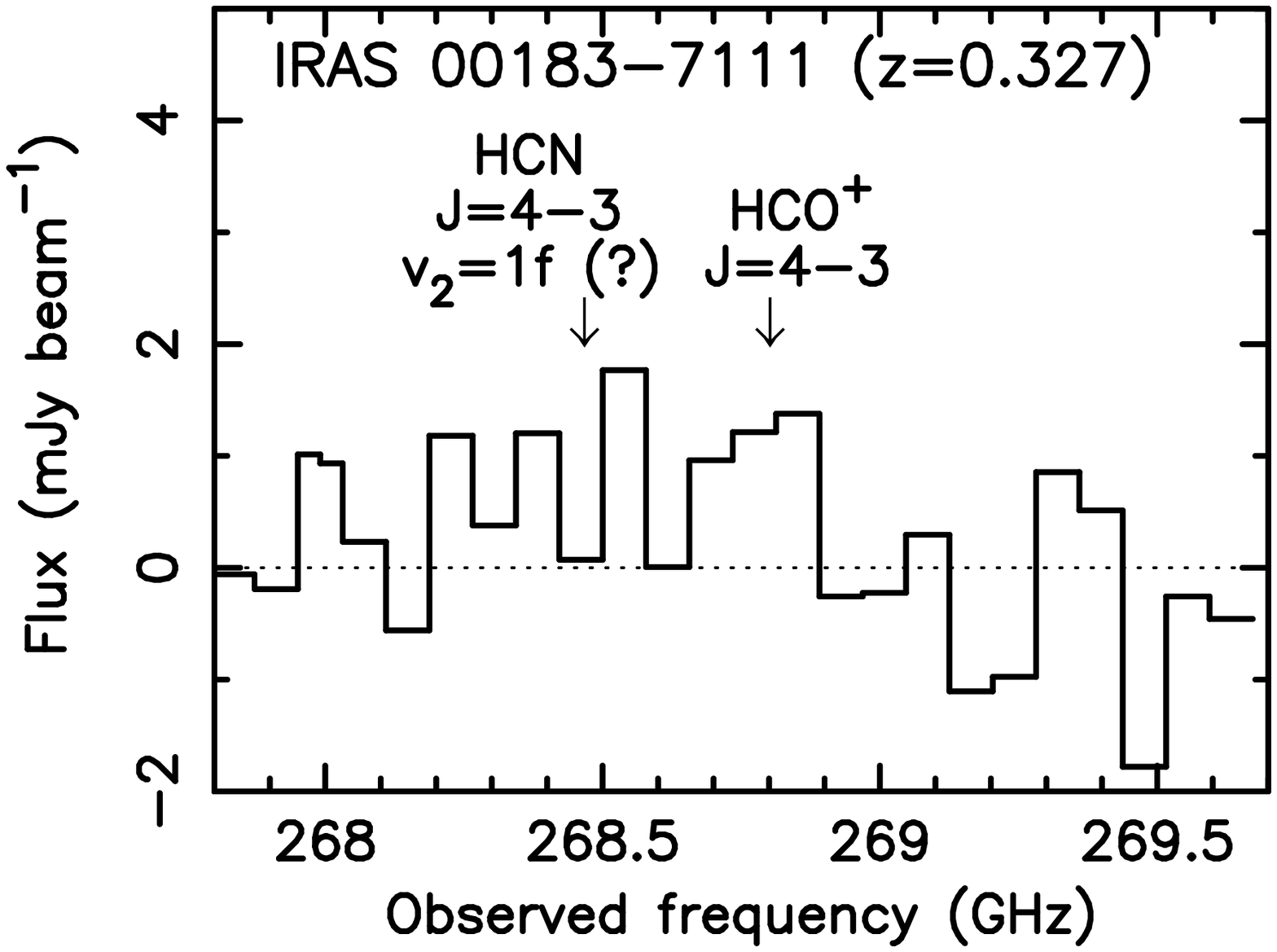}  
\includegraphics[angle=0,scale=.45]{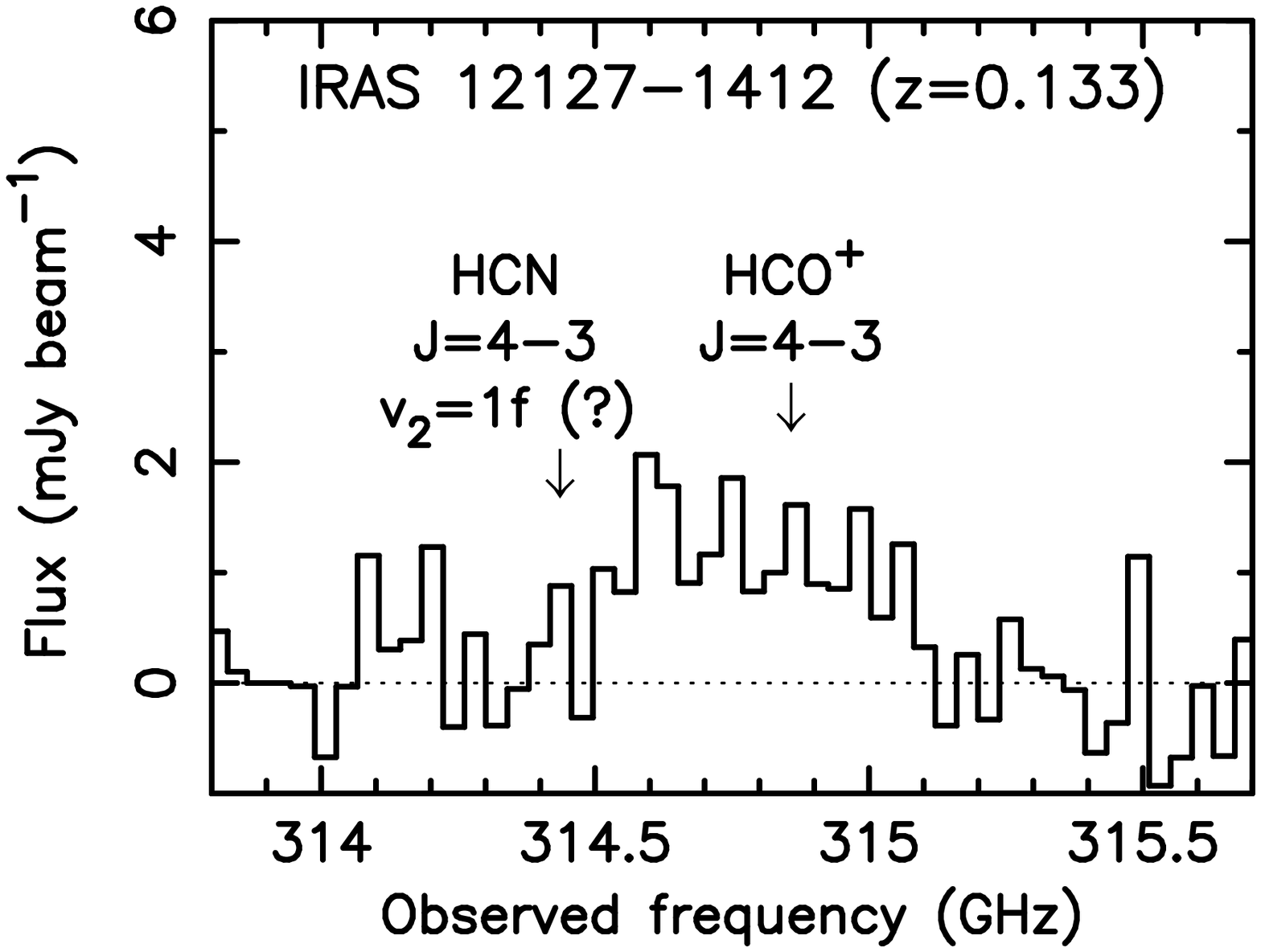} 
\includegraphics[angle=0,scale=.45]{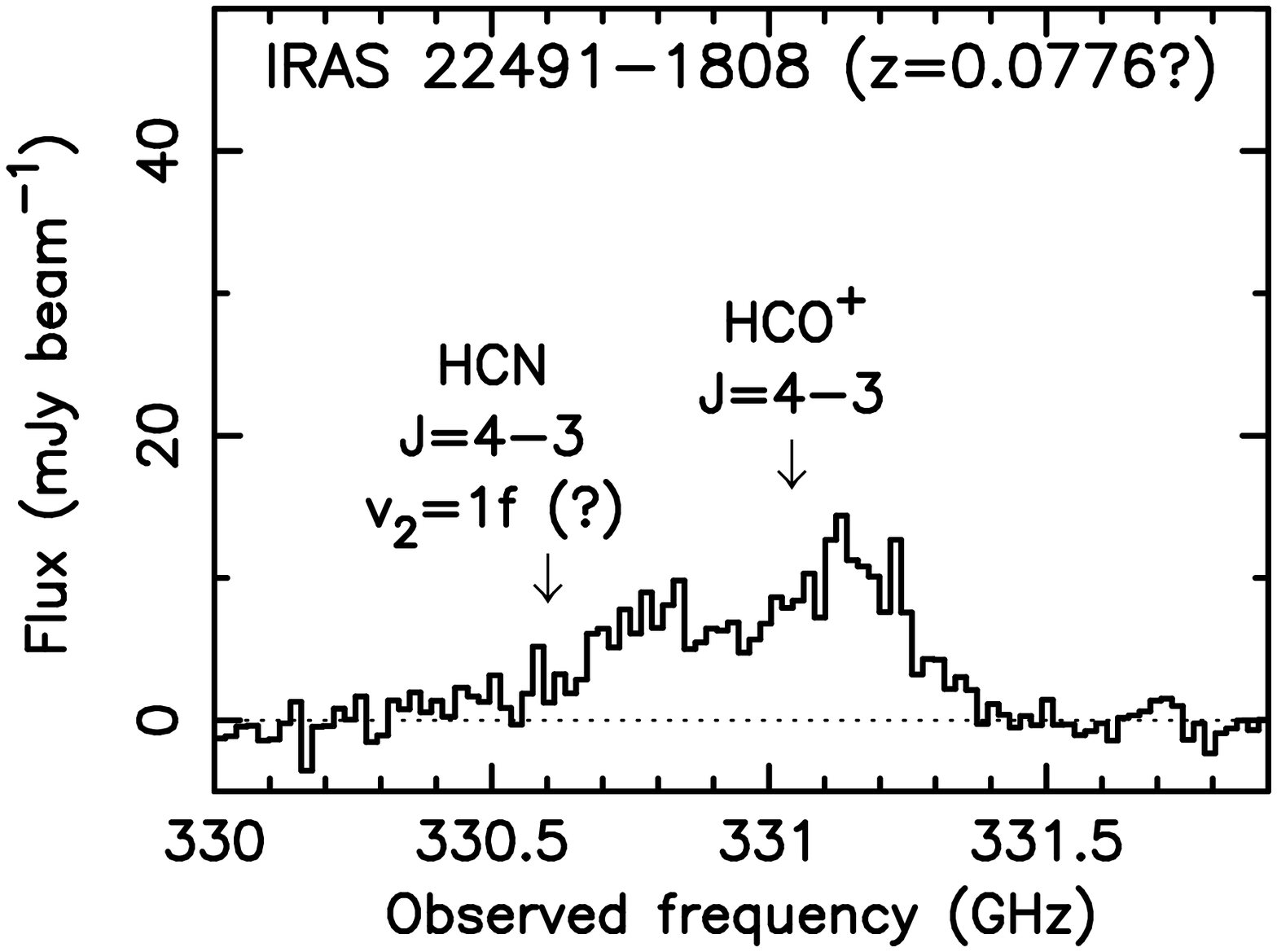} 
\caption{
Spectra around the vibrationally excited 
(v$_{2}$=1f) HCN J=4--3 emission line ($\nu_{\rm rest}$ = 356.256 GHz). 
The abscissa shows the observed frequency in [GHz], and the ordinate
shows the flux in [mJy beam$^{-1}$]. 
The expected frequencies of HCN J=4--3 (v$_{2}$=1f) and HCO$^{+}$ J=4--3 
(v=0) are shown with downward arrows. 
For IRAS 22491$-$1808, although the adopted optically derived redshift is $z=$ 
0.076 \citep{kim98}, we place the downward arrows at the redshift of 0.0776 
derived from our ALMA data of dense molecular gas tracers.
}
\end{figure}

%--- Figure 10 ---%
\begin{figure}
\begin{center}
\includegraphics[angle=0,scale=.75]{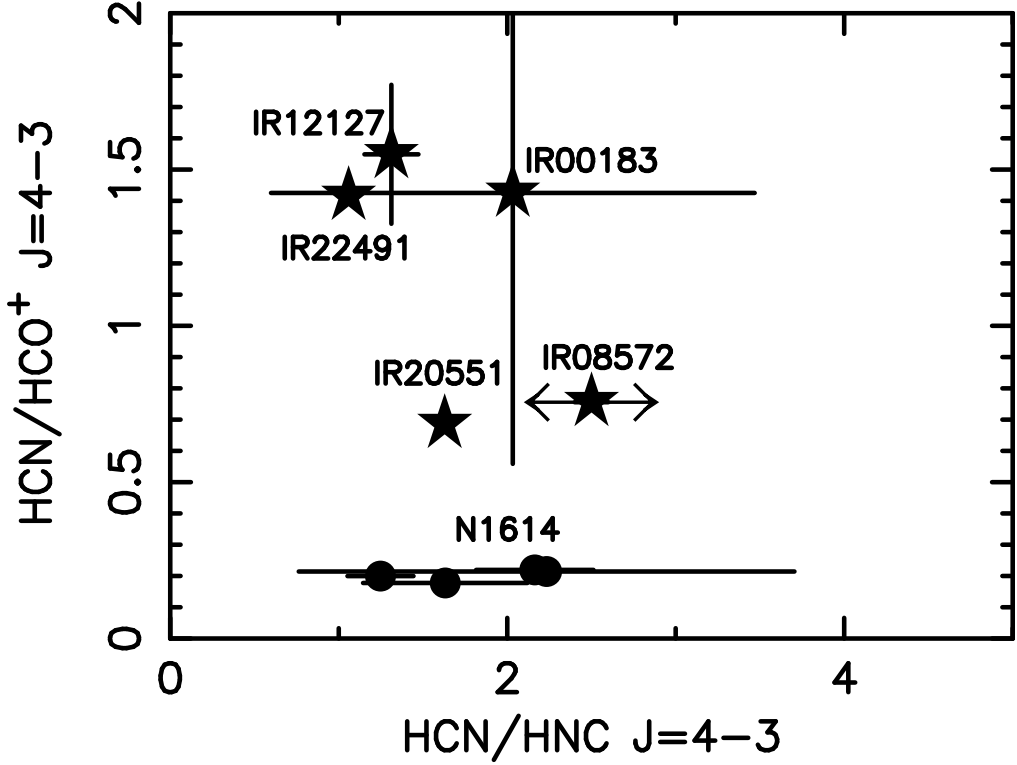} 
\end{center}
\caption{
HCN-to-HNC (abscissa) and HCN-to-HCO$^{+}$ (ordinate) flux 
ratios at J=4--3. 
The filled stars are four ULIRGs studied in this paper and 
IRAS 20551$-$4250 \citep{ima13b}. 
Since no HNC data are available for IRAS 08572$+$3915, we have 
no constraint on the HCN/HNC J=4--3 flux ratio. 
We tentatively positioned the plot at the middle of the abscissa. 
The filled circles are four data points of the starburst-dominated 
LIRG NGC 1614 \citep{ima13a}.
}
\end{figure}

%--- Figure 11 ---%
\begin{figure}
\begin{center}
\includegraphics[angle=-90,scale=.34]{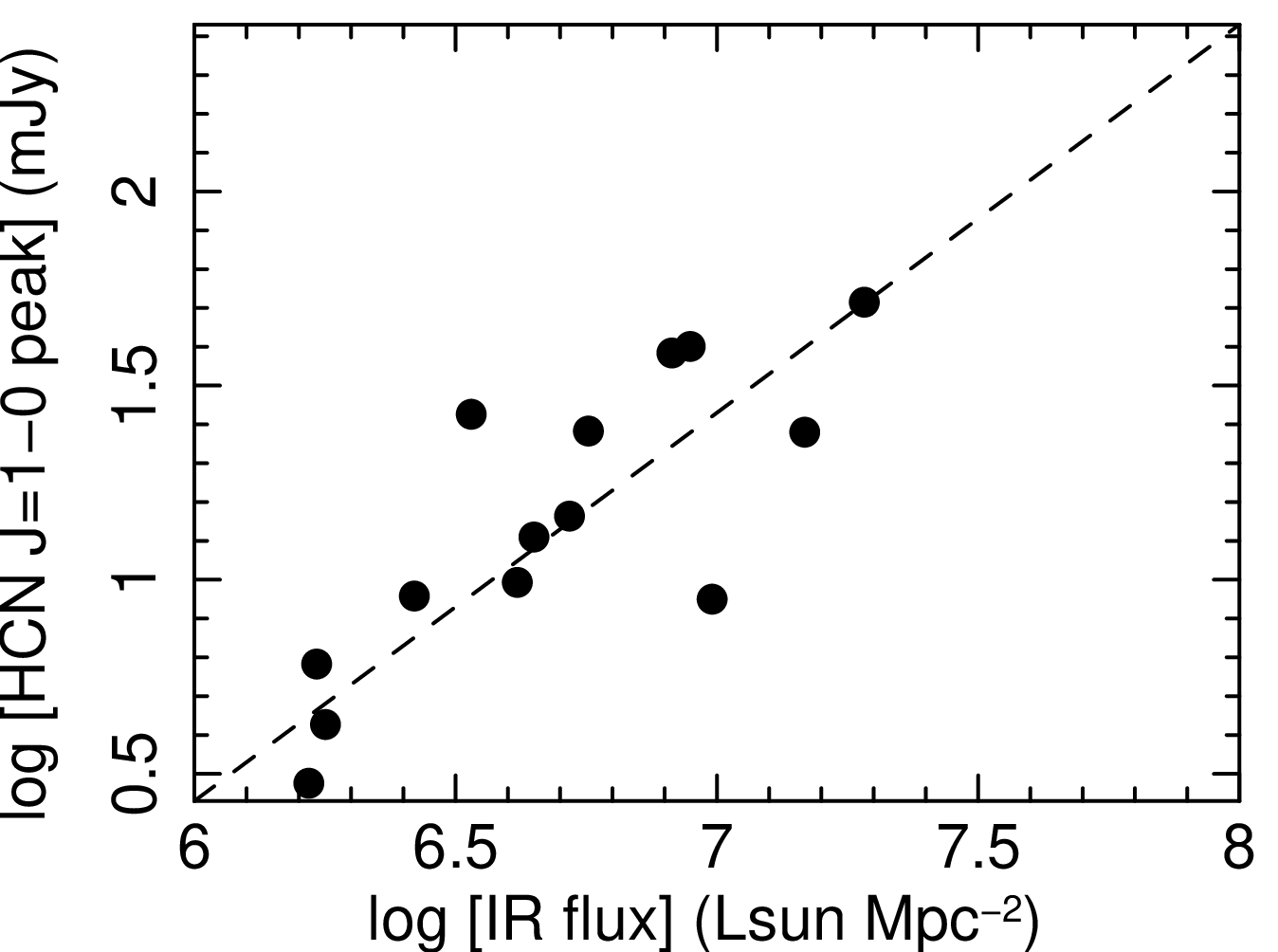} 
\includegraphics[angle=-90,scale=.34]{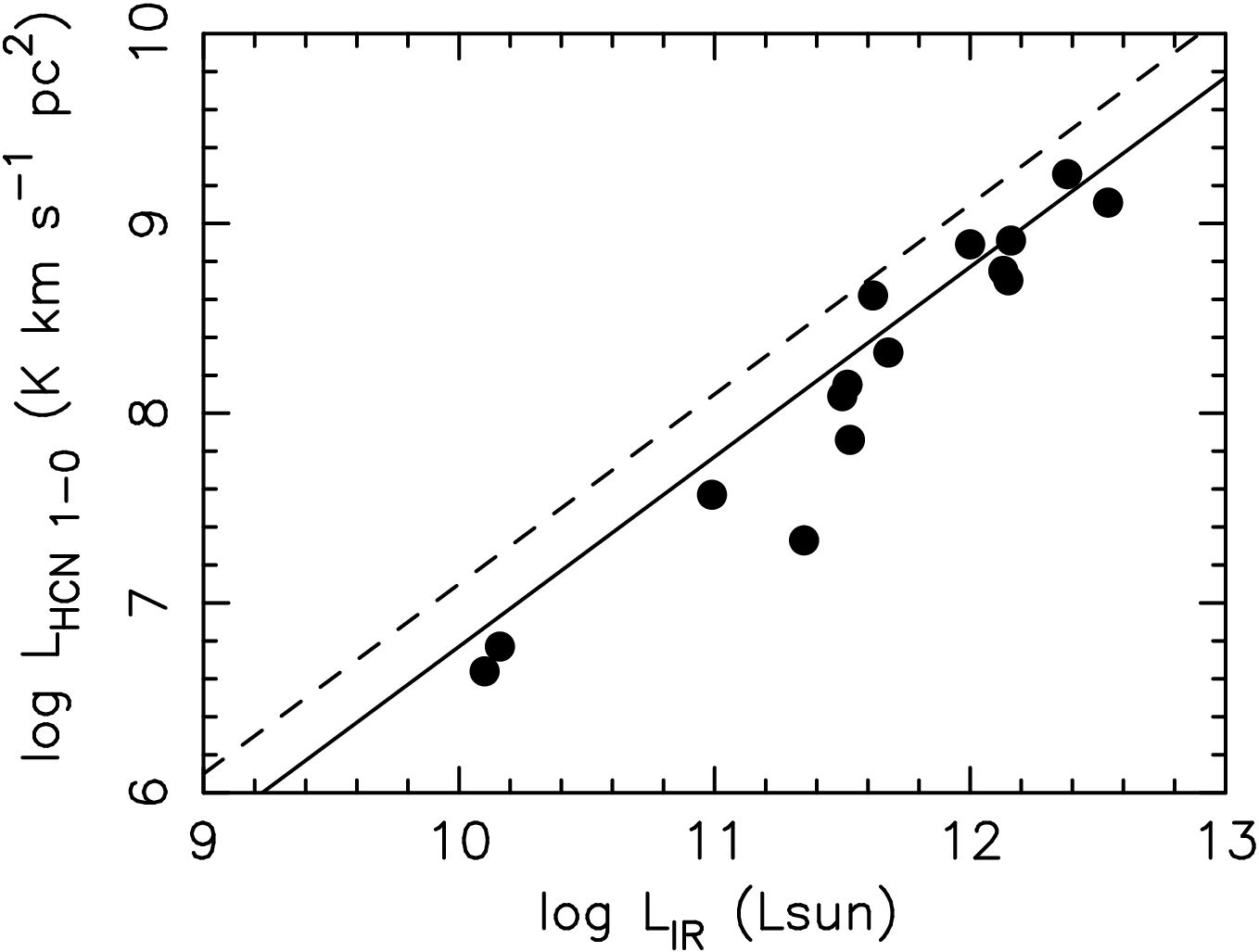} 
\end{center}
\caption{
({\it left}): Comparison of the decimal logarithm of the infrared flux 
in [L$_{\odot}$ Mpc$^{-2}$] from whole galactic regions measured with 
IRAS (abscissa) and decimal logarithm of the HCN J=1--0 {\it peak flux} 
density in [mJy] from (U)LIRG {\it nuclei} (ordinate), obtained 
from our pre-ALMA interferometric observations 
\citep{ima04,ima06b,in06,ima07b,ima09b}. 
The dashed line (log [HCN peak] = log [IR flux] $-$ 5.57) is the ratio 
adopted here. 
({\it right}): Comparison of the decimal logarithm of the infrared 
luminosity in [L$_{\odot}$] (abscissa) and decimal logarithm of the HCN 
J=1--0 {\it luminosity} in [K km s$^{-1}$ pc$^{2}$] (ordinate). 
The dashed line is the ratio for starburst galaxies 
(log L$_{\rm HCN}$ = log L$_{\rm IR}$ $-$ 2.9) derived by 
\citet{gao04b}. 
The solid line is the ratio for ULIRGs 
(log L$_{\rm HCN}$ = log L$_{\rm IR}$ $-$ 3.2) estimated by 
\citet{gra08}.
}
\end{figure}

\end{document}